\RequirePackage{fix-cm}
\documentclass[smallextended]{svjour3}       
\smartqed  
\usepackage{graphicx}
\usepackage{url}
\usepackage{amsmath}
\usepackage{amssymb}
\usepackage{graphicx}
\usepackage{rotating}
\usepackage{natbib}
\usepackage{color}
\usepackage{txfonts}
\usepackage{subfigure}
\usepackage{mathptmx}      
\def\kms{\,\mathrm{km\,s}^{-1}}

\newcommand{\HI}{\mathrm{H\,\scriptstyle I}}

\begin{document}

\title{Magnetic Fields in Spiral Galaxies
}


\author{Rainer Beck
}


\institute{R. Beck \at
              Auf dem H\"ugel 69, 53121 Bonn, Germany \\
              \email{rbeck@mpifr-bonn.mpg.de}           
}

\date{Accepted: 8 September 2015}

\maketitle

\begin{abstract}
Radio synchrotron emission, its polarization and Faraday rotation of the polarization angle
are powerful tools to study the strength and structure of magnetic fields in galaxies.
Unpolarized synchrotron emission traces isotropic turbulent fields which are strongest in
spiral arms and bars ($20-30\,\mu$G) and in central starburst regions
($50-100\,\mu$G). Such fields are dynamically important; they affect gas
flows and drive gas inflows in central regions. Polarized emission traces
ordered fields, which can be regular or anisotropic turbulent, where the latter
originates from isotropic turbulent fields by the action of compression or shear. The strongest
ordered fields ($10-15\,\mu$G) are generally found in interarm regions.
In galaxies with strong density waves, ordered fields
are also observed at the inner edges of spiral arms.
Ordered fields with spiral patterns exist in grand-design, barred
and flocculent galaxies, and in central regions. Ordered fields in
interacting galaxies have asymmetric distributions and are a
tracer of past interactions between galaxies or with the intergalactic medium.
-- Faraday rotation measures (RM) of the diffuse polarized radio emission
from galaxy disks reveal large-scale spiral patterns that can be
described by the superposition of azimuthal modes; these are
signatures of regular fields generated by mean-field dynamos.
``Magnetic arms'' between gaseous spiral arms may also be products of dynamo action,
but need a stable spiral pattern to develop.
Helically twisted field loops winding around spiral arms were found in two galaxies
so far. Large-scale field reversals, like the one found in the Milky Way, could not yet
be detected in external galaxies. In radio halos around edge-on galaxies,
ordered magnetic fields with X-shaped patterns are observed.
-- The origin and evolution of cosmic magnetic fields, in particular their first occurrence
in young galaxies and their dynamical importance during galaxy evolution, will be studied
with forthcoming radio telescopes like the Square Kilometre Array.
\end{abstract}

\section{The role of magnetic fields in spiral galaxies}
\label{intro}

Whether and how magnetic fields influence the formation and evolution of galaxies is still unknown.
Most numerical models of the interstellar medium in galaxies
ignored magnetic fields. Magneto-hydrodynamical (MHD) models of evolving galaxies are
computationally challenging and require various simplifications \citep{wang09,pakmor13,pakmor14}.

Much more is known about magnetic fields in the interstellar medium (ISM) of nearby galaxies and the
Milky Way, thanks to modern radio telescopes, the construction of sensitive and wide-band receiving
systems and the development of new methods of observation and data analysis in the radio range.
Major progress was also achieved in numerical modelling
of the magnetized ISM \citep[e.g.][]{avillez05,gissinger09,hill12,gressel13,gent13,machida13}, using
a variety of codes and assumptions. Not surprisingly, the results differ and are not always consistent
with observations, which calls for further efforts on both sides.

A widespread argument to neglect galactic magnetic fields in the past was based on assuming field strengths
of a few $\mu$G taken from Faraday rotation data in our Milky Way, which trace only a small component
of the total field. The dynamical importance of galactic fields, with typical strengths of $15\,\mu$G,
is now widely appreciated in many fields of astrophysics.
Magnetic and cosmic-ray pressures control the overall star-formation rate \citep{birnboim15}.
Magnetic fields are important for the gas dynamics of molecular clouds \citep{ade15b}.
With stronger fields fewer cloud cores are formed but with larger masses \citep{vazquez05,price08}.
Magnetic fields control the density and propagation of cosmic rays.
Together with cosmic rays they can provide the pressure to drive fast outflows of hot gas, in particular
in galaxies with high star-formation rates in the early Universe \citep{hanasz13}.
Cosmic rays propagating along the field lines in galaxy halos may heat the warm ionized gas and
explain the optical line ratios \citep{wiener13}.
Outflows from starburst galaxies may have magnetized the intergalactic medium
\citep{bertone06}.

The detection of ultrahigh-energy cosmic rays (UHECRs) reaching the Earth with the AUGER observatory
and the possibly anisotropic distribution of their arrival directions \citep{aab15} calls for a
proper model of particle propagation. As UHECR particles are deflected by large-scale regular fields
and scattered by turbulent fields, the structure and the extent of the fields in the disk and halo
of the Milky Way need to be known, but are hard to measure from our inside location.
The view onto external spiral galaxies can help.

\section{Origin of galactic magnetic fields}
\label{dynamo}

Building up magnetic fields in galaxies is a three-stage process:
(1) seeding, (2) amplifying, (3) ordering and sustaining.

(1) {\em Seed fields}\ can be ``primordial'', generated in the early Universe \citep{durrer13},
or they may originate from later epochs,
e.g. during cosmological structure formation by the Weibel instability \citep{lazar09},
from plasma fluctuations in protogalaxies \citep{schlickeiser12,schlickeiser13}, from injection
by the first stars \citep{bisnovatyi73} or jets generated by the first black holes \citep{rees05}
or from the Biermann mechanism in the first supernova remnants \citep{hanayama05}.

Indication for a weak all-pervasive intergalactic seed field as a relic from the early Universe
comes from the non-detection of GeV $\gamma$--ray emission with the FERMI satellite
from blazars, which were observed at TeV energies with the HESS observatory.
The secondary particles generated by the TeV photons may be deflected by intergalactic fields
of at least $10^{-16}$\,G strength and a high volume filling factor \citep{dolag11}.
However, fluctuations in the intergalactic plasma may also disperse the $\gamma$--ray
emission \citep{broderick12} (see also the discussion in \citet{durrer13}).
Analysis of the CMB power spectra data from PLANCK gave an upper limit of about 5\,nG on a (comoving)
scale of 1\,Mpc \citep{ade15a}.

(2) An efficient source of field {\em amplification}\ is turbulence in the gas driven by supernova
explosions \citep{ferriere96} or by spiral shocks \citep{kim06},
called the {\em small-scale dynamo}. In protogalaxies this mechanism can amplify weak
seed fields to several $\mu$G strength (the energy level of turbulence) within less than $10^8$\,yr
\citep{kulsrud97,schleicher10,abeck12,rieder15}. The resulting field is {\em turbulent}.

(3) The final and most time consuming stage is the {\em ordering}\ of the turbulent field. Numerical
models of evolving galaxies show fast field amplification and ordering with help of differential
rotation \citep{wang09,kotarba09}, possibly supported by the magneto-rotational instability
(MRI) \citep{gressel13,pakmor13}. In these models the ordered magnetic field forms spiral arm segments,
but it has frequent reversals in azimuthal and radial directions and hence is {\em anisotropic turbulent}\
(Sect.~\ref{comp}).

The most promising mechanism to {\em sustain}\ magnetic fields and generate large-scale
{\em regular fields}\ from turbulent fields in the interstellar medium of galaxies is the
$\alpha-\Omega$ dynamo \citep{beck96}. It is based on differential rotation
($\Omega$), expanding gas flows, driven by supernova explosions \citep{ferriere96,gressel13}
or cosmic rays \citep{hanasz09}, carrying magnetic fields that are twisted by the Coriolis
force ({\em $\alpha$--effect}), and on magnetic diffusivity ($\eta$) driven by turbulence.
The ``mean-field'' approximation of the $\alpha-\Omega$ dynamo equation allows analytical
solutions by mathematically separating the large-scale and small-scale parts of the velocity
field and the magnetic field. The ``mean-field'' dynamo equation for the regular field $\vec{B}$

\begin{equation}
\partial \vec{B}/ \partial t = \nabla \times (\vec{v} \times \vec{B)} + \nabla \times \alpha \vec{B} + \eta \, \nabla^2 \vec{B}
\end{equation}

\noindent describes the field amplification by the induction term
$\nabla \times (\vec{v} \times \vec{B)}$, where $\vec{v}$ is the large-scale velocity,
the gain term $\nabla \times \alpha \vec{B}$ and the loss term
$\eta \, \nabla^2 \vec{B}$. MHD modeling with high spatial
resolution showed that the scale separation of the mean-field approximation is reasonable
\citep{gent13}.

Solutions of the mean-field dynamo equation are described by modes $m$ with different
azimuthal symmetries in the disk plane (axisymmetric, bisymmetric, quadri-symmetric, etc.)
and two different vertical symmetries (even parity  or odd parity) perpendicular
to the disk plane. Several modes can be excited in the same object, where lower modes have
shorter growth times and larger amplitudes. In flat, rotating objects like
galaxy disks, the strongest mode is composed of a toroidal field of
{\em axisymmetric spiral}\ shape within the disk ($m=0$), without sign
reversals across the equatorial plane, and a weaker poloidal field
of even symmetry. Azimuthal dynamo modes can be identified
observationally from the patterns of polarization angles and Faraday rotation measures
of the diffuse polarized emission from galaxy disks (see Sect.~\ref{large}).

Mean-field models of the $\alpha-\Omega$ dynamo in galaxy disks predict
that turbulent fields are transformed into large-scale regular fields within a
few $10^9$\,yr \citep{beck94,arshakian09,rodrigues15}. Field reversals from the
early phases may survive until today \citep{moss12}. Global numerical models
of galaxies \citep{gissinger09,hanasz09,kulpa11,siejkowski14} confirmed
the basic results of the mean-field models.

The twisted field loops needed for the operation of the $\alpha-\Omega$ dynamo generate
large-scale helicity. \footnote{Magnetic helicity $H$ is defined as $H = A \cdot (\nabla \times A)$,
where $A$ is the magnetic vector potential.}
As total helicity in a system is a conserved quantity \citep{brandenburg05}, small-scale fields
with opposite helicity are generated which quench dynamo action, unless they are removed from
the system \citep{shukurov06}. Hence, moderately fast outflows are essential for effective
$\alpha-\Omega$ dynamo action, but strong outflows can suppress dynamos. This leaves a
certain range of outflow velocities for optimal dynamo action \citep{rodrigues15}.
As outflows are strongest above spiral arms, the $\alpha-\Omega$ dynamo is expected to be
more efficient {\em between}\ the spiral arms \citep{chamandy15}.

The total magnetic field $B_\mathrm{tot}$ can be presented as the quadratic sum of the regular field
$B_\mathrm{reg}$ generated by the $\alpha-\Omega$ dynamo and the component $B_\mathrm{turb}$
generated by the small-scale dynamo. Tangling of $B_\mathrm{reg}$ can produce another component,
$B_\mathrm{tan}$. $B_\mathrm{tan}$ and $B_\mathrm{turb}$ should be comparable,
because both components are driven by the same turbulent gas motions.
On the solar surface, the small-scale dynamo seems to be weak \citep{stenflo12}, but may still
operate in the solar interior.
In contrast to the Sun, stars and planets, the regime of dynamo action is fully accessible in galaxies.
The existence of the small-scale dynamo is of fundamental importance for the evolution of magnetic
fields in galaxies. Without a small-scale dynamo the seed fields in protogalaxies would
be much weaker and the generation of large-scale regular fields would take much longer.
Turbulent and tangled fields can be distinguished with (future) high-resolution observations.

\section{Observational tools}

Magnetic fields need illumination to be detectable. {\em Polarized emission}\ at optical, infrared,
submillimeter and radio wavelengths holds the clue to measure magnetic fields in galaxies.
Huge progress has been achieved in the last decade.

The various methods to measure magnetic fields are briefly discussed in the following.
The basic concepts are presented in more detail in \citet{klein15}.

\subsection{Basic magnetic field components}
\label{comp}

Table~\ref{table:components} summarizes the different field components and the methods used to observe them.

\begin{table}[h!]
\caption{Magnetic field components and methods to observe them. The symbols $\perp$ and $\parallel$ indicate
field components perpendicular and parallel to the line of sight, respectively.
}
\centering
\begin{tabular}{l l l}
\hline\hline
Field component	& Notation & Observational signatures\\
\hline
Total field perpendicular & $B_\mathrm{tot,\perp}^{\,2}=\,\,\,B_\mathrm{turb,\perp}^{\,2}+B_\mathrm{reg,\perp}^{\,2}$
& Total synchrotron intensity\\
\,\,\,to the line of sight & & \\
Turbulent or tangled field	& $B_\mathrm{turb,\perp}^{\,2}=\,\,\,B_\mathrm{iso,\perp}^{\,2}+B_\mathrm{aniso,\perp}^{\,2}$
& Total synchrotron emission,\\
\,\,\,perpendicular to the line of sight& & \,\,\,partly polarized\\
Isotropic turbulent or tangled & $B_\mathrm{iso,\perp}\,\,\,(=\sqrt{2/3}\,B_\mathrm{iso})$
& Unpolarized synchr. intensity,\\
\,\,\,field perpendicular & & \,\,\,beam depolarization,\\
\,\,\,to the line of sight & & \,\,\,Faraday depolarization\\
Isotropic turbulent or tangled & $B_\mathrm{iso,\parallel}\,\,\,(=\sqrt{1/3}\,B_\mathrm{iso})$
& Faraday depolarization\\
\,\,\,field along the line of sight\\
Ordered field perpendicular & $B_\mathrm{ord,\perp}^{\,2}=\,\,\,B_\mathrm{aniso,\perp}^{\,2}+B_\mathrm{reg,\perp}^{\,2}$
& Intensity and vectors of radio, \\
\,\,\,to the line of sight & & \,\,\,optical, IR \& submm pol.\\
Anisotropic turbulent or & $B_\mathrm{aniso,\perp}$ & Intensity and vectors of radio,\\
\,\,\,tangled field perpendicular & & \,\,\,optical, IR \& submm pol.,\\
\,\,\,to the line of sight & & \,\,\,Faraday depolarization\\
Regular field perpendicular & $B_\mathrm{reg,\perp}$ & Intensity and vectors of radio, \\
\,\,\,to the line of sight & & \,\,\,optical, IR \& submm pol.,\\
\,\,\, & & \,\,\,Goldreich-Kylafis effect\\
Regular field along the & $B_\mathrm{reg,\parallel}$ & Faraday rotation + depol.,\\
\,\,\,line of sight & & \,\,\,longitudinal Zeeman effect\\
\hline
\label{table:components}
\end{tabular}
\end{table}
\normalsize

The total magnetic field is separated into a {\em regular}\ and a {\em turbulent}\ component. A regular field has
a well-defined direction within the telescope beam, while a turbulent field frequently reverses its direction
within the telescope beam. In other words, the coherence scale of regular fields is much larger than the
turbulent scale. Turbulent fields can be isotropic turbulent (i.e. the same dispersion in all three spatial
dimensions) or anisotropic turbulent (i.e. different dispersions). Anisotropic turbulent fields originate
from isotropic turbulent fields by the action of compressing or shearing gas flows.

The power spectra of turbulent fields are generally assumed to be of Kolmogorov type, with the largest scale
given by the scale driving turbulence (e.g. supernova remnants in galaxies) and the smallest scale given
by magnetic diffusivity. The power spectra of regular fields generated by the mean-field dynamo show a
broad maximum over a range of scales which develops with time \citep{moss12}.

Present-day radio polarization observations with limited
spatial resolution cannot distinguish turbulent fields generated by turbulent gas flows from tangled fields
generated from regular fields by small-scale gas motions; the components of both types of fields in the sky
plane give rise to unpolarized synchrotron emission.

Anisotropic turbulent, anisotropic tangled and regular field components contribute to the {\em ordered}\ field,
observable in polarized synchrotron emission.
Polarization angles are ambiguous by $180^\circ$, so that polarized emission is not sensitive to field reversals.
Faraday rotation and the longitudinal Zeeman effect are sensitive to the field {\em direction}\ and hence
trace regular fields.

\subsection{Optical, infrared and submm polarization}

\begin{figure*}[t]
\vspace*{2mm}
\begin{center}
\includegraphics[width=8cm]{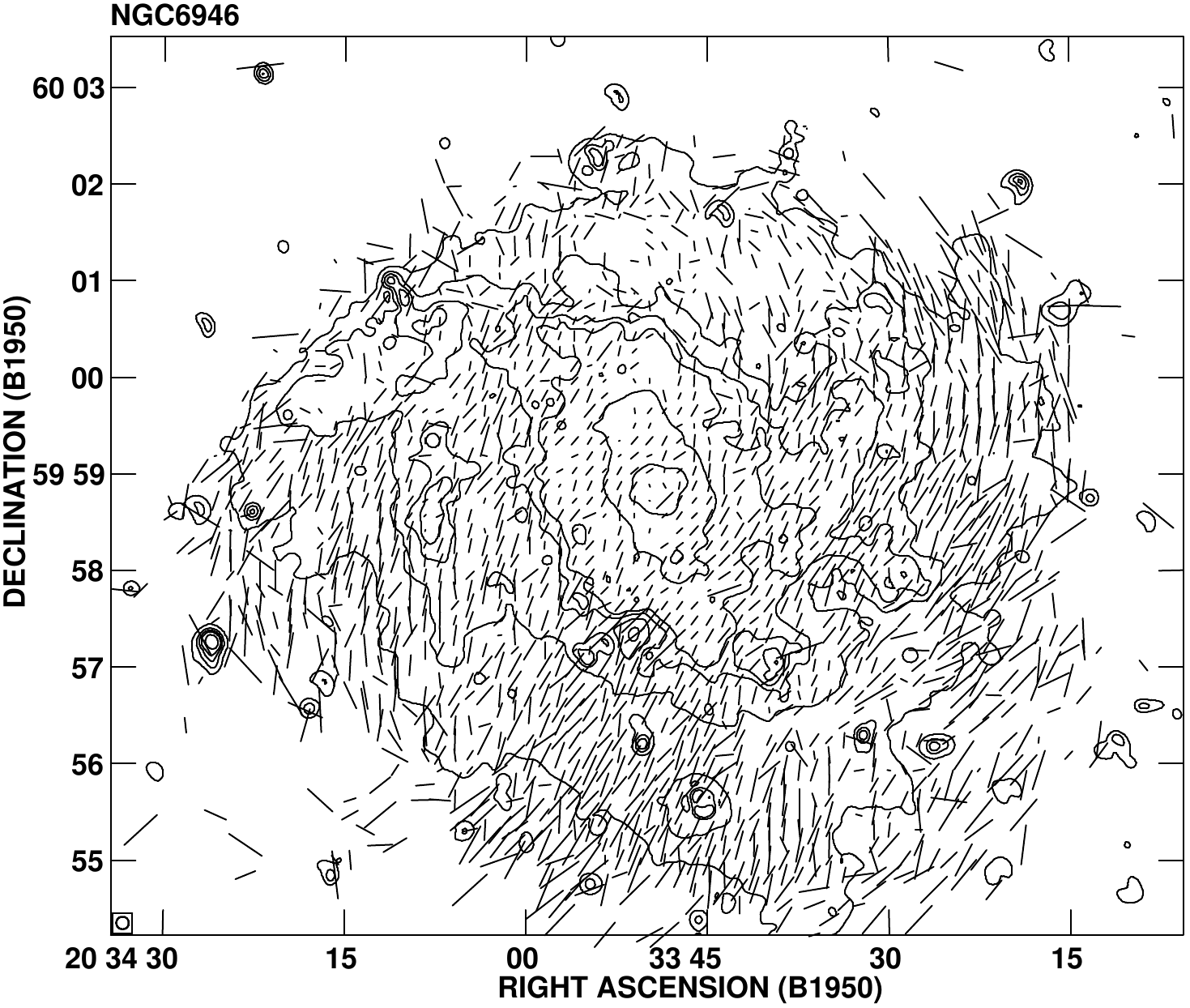}
\caption{Optical emission (contours) and polarization E--vectors of NGC~6946,
observed with the polarimeter of the Landessternwarte Heidelberg at the MPIA Calar Alto
1-m telescope. The optical E--vectors are along the spiral arms in the eastern and western regions
(compare with the radio polarization B--vectors in Fig.~\ref{fig:n6946}).
Polarization due to light scattering is probably occurring in the southern part (from \cite{fendt98}).}
\label{fig:n6946opt}
\end{center}
\end{figure*}

Elongated dust grains can be oriented with their major axis
perpendicular to the interstellar magnetic field lines by paramagnetic alignment (the
{\em Davis-Greenstein effect}) or, more efficiently, by radiative torque
alignment \citep{hoang08,hoang14}. When particles on the line of sight to a star
are oriented with their major axis perpendicular to the line of sight
(and the field is oriented in the same plane), the different levels of
extinction along the major and the minor axis leads to polarization of the starlight,
with the E--vectors pointing parallel to the component of the
interstellar magnetic field perpendicular to the line of sight  (Fig.~\ref{fig:n6946opt}).
The interpretation of optical and near-infrared polarization observations
of individual stars or diffuse starlight is based on this mechanism.
Extinction is most efficient for grains of sizes similar to the
wavelength. These small particles are only aligned in the medium between
molecular clouds, not in the dense clouds themselves \citep{cho05}.

The detailed physics of the alignment is complicated and not fully understood.
The degree of polarization p generated in a volume element along the line of
sight of extent $L$ is proportional to ($B_\mathrm{tot,\perp}^{\,\,2}\,L$),
but also depends on the magnetic properties, density and temperature of the grains
\citep{ellis78}.

Starlight polarization yields the orientation of large-scale magnetic fields
in the Milky Way (Sect.~\ref{MW}), and in the galaxies M~51
\citep{scarrott87}, NGC~6946 (Fig.~\ref{fig:n6946opt}) and the Small Magellanic Cloud \citep{lobo15}.
The major shortcoming when applying this method to extended sources like gas clouds
or galaxies is that light can also be polarized by scattering, a contamination
that is unrelated to magnetic fields and difficult to subtract.

At far-infrared and submillimeter wavelengths, elongated dust grains emit linearly polarized waves,
without contributions by polarized scattered light. The B--vector is parallel
to the magnetic field. The field structure can be mapped in gas clouds of the Milky Way,
e.g. in the massive star formation site W~51 \citep{tang09} and in galaxies, e.g. in the
halo of the starburst galaxy M~82 \citep{greaves00}.


Magnetic force (total field strength $B$) and kinetic force by turbulent gas motions (turbulent velocity $\mathrm{v_{turb}}$ and density $\rho$) are in competition in the interstellar medium.
In a strong field the field lines are straight and the dispersion of polarization angles $\sigma$
is small. According to \cite{chandra53}, the total field strength $B_\mathrm{tot,\perp}$ in the sky
plane can be estimated from:
\begin{equation}
B_\mathrm{tot,\perp} \, \simeq \, ({4\over3} \, \pi \rho)^{1/2} \,\, \mathrm{v_{turb}} \, / \sigma \, .
\end{equation}

\noindent Application to starlight polarization in the Milky Way yielded field strengths of about
7\,$\mu$G, in agreement with other methods (Sect.~\ref{MW}).
The Chandrasekhar--Fermi method was improved by correcting for the observational errors in polarization
angle and signal integration \citep{hildebrand09, houde09, houde11}. It can also be applied to radio
polarization angles, like in M~51 \citep{houde13}.

\subsection{Zeeman effect}

The Zeeman effect is the most direct method of remote sensing of magnetic fields.
In the presence of a magnetic field $B_{\parallel}$ along the line of sight,
a spectral line is split into two circularly polarized components of opposite sign
(longitudinal Zeeman effect). The frequency shift is 2.8\,MHz/Gauss for the $\HI$ line and
larger for molecular lines like OH, CN or H$_2$O \citep{heiles05}.
The Zeeman effect is sensitive to regular fields, but turbulent fields can also be
measured from the dispersion of the circularly polarized signal at a large number of
locations on the source \citep{watson01}.
The interpretation of Zeeman splitting is hampered by its weakness, requiring high line
intensities and careful correction of instrumental polarization. Most results have been obtained
for the $\HI$ line that traces clouds of diffuse (warm) gas in the Milky Way
\citep{crutcher10} and for the OH line for starburst galaxies \citep{robishaw08}.


Magnetic fields perpendicular to the line of sight generate two shifted lines in addition to the
main unshifted line, all linearly polarized (transversal Zeeman effect). These lines cannot be
resolved for observations in the Milky Way and external galaxies and no net polarization is observed
under symmetric conditions. Detection of linearly polarized lines becomes possible for unequal
populations of the different sublevels, a gradient in optical depth or velocity, or an anisotropic
velocity field \citep{goldreich81}. Depending on the line ratios, the orientation of linear
polarization can be parallel or perpendicular to the magnetic field orientation. This
{\em Goldreich--Kylafis effect}\ was detected in molecular clouds, star-forming regions, outflows
of young stellar systems and supernova remnants of the Milky Way. It was also observed in the ISM
of the galaxy M~33 \citep{li11}, where it is consistent with the field orientations along the spiral
arms as measured by polarized radio continuum emission \citep{taba08}.
The $\pm90^\circ$ ambiguity needs to be solved with help of additional continuum polarization data.

\subsection{Total radio synchrotron emission and the equipartition assumption}
\label{eq}

Radio continuum emission from galaxies is a mixture of thermal and nonthermal (synchrotron) components.
The average thermal fractions are typically 10\% and 25\% at 20\,cm and 6\,cm wavelengths, respectively.
At long wavelengths thermal emission is negligible, but thermal absorption may reduce the synchrotron
intensity in star-forming regions. A first-order separation of the thermal and nonthermal components can
be performed by assuming a uniform spectral index of $\alpha=0.8-1.0$ for the nonthermal intensity
($I \propto \nu^{-\alpha}$) and using the spectral index of the thermal emission in an optically thin plasma
of $\alpha=0.1$. However, synchrotron intensity depends on energy and age of cosmic-ray electrons (CREs)
propagating from the spiral arms and losing energy by synchrotron radiation and inverse Compton scattering
with photons, so that the synchrotron spectrum is expected to be flatter (smaller $\alpha$) in spiral arms
than in interarm regions, outer disk and halo. The advanced method to separate the two components introduced
by \citet{taba07} uses an extinction-corrected H$\alpha$ image as a template for the thermal radio emission
and was first applied to the galaxy M~33. Simplified separation methods use an H$\alpha$ image \citep{heesen14}
or a combination of H$\alpha$ and $24\,\mu$m infrared images as a thermal template that is subtracted from the
total emission \citep{basu13}.

The intensity of total synchrotron emission (examples in Figs.~\ref{fig:m31},
\ref{fig:n1097}, \ref{fig:n1097central}, \ref{fig:n891}--\ref{fig:n4631} and \ref{fig:n4258}--\ref{fig:n4449},
all of which show mostly synchrotron emission) is a measure of the number
density of cosmic-ray electrons (CREs) in the relevant energy range and of the strength of the total
magnetic field component $B_\mathrm{tot,\perp}$ perpendicular to the line of sight (i.e. in the sky plane).
Synchrotron emission at a frequency $\nu$ is emitted by a range of CREs with average energy $E$ (in GeV)
in a field of strength $B_\mathrm{tot,\perp}$ (in $\mu$G):
\begin{equation}
\nu \, \simeq \, 16\, \mathrm{MHz} \,\, E^2 \,\, B_\mathrm{tot,\perp} \, .
\label{sync}
\end{equation}


Equipartition is expected between the energy densities of the total cosmic rays (dominated by protons) and the
total magnetic field. The cosmic-ray energy density is determined by integrating over their energy spectrum.
This allows us to calculate the total magnetic field strength from the synchrotron intensity
\citep{beck+krause05, arbutina12}. The synchrotron intensity $I_\mathrm{syn}$ scales with the total magnetic
field strength as $I_\mathrm{syn} \propto B_\mathrm{tot,\perp}^{\,\, 3 + \alpha}$. Vice versa,
$B_\mathrm{tot,\perp}$ scales as:
\begin{equation}
B_\mathrm{tot,\perp} \propto (I_\mathrm{syn} \, (K_0 + 1) \, / \, L)^{\,\, 1/(3 + \alpha)} \, ,
\label{equi}
\end{equation}

\noindent where $\alpha$ is the synchrotron spectral index ($I_\mathrm{syn} \propto \nu^{-\alpha}$)
and $L$ is the effective pathlength through the source. $K_0$ is the ratio of number densities of CR
protons and electrons. $K_0\simeq100$ is a reasonable assumption in the star forming regions of the
disk \citep{bell78}. For an electron-positron plasma, e.g. in jets of radio galaxies, $K_0=0$.
The input parameters, especially the pathlength $L$ and the ratio $K_0$ are not well constrained.
Changing one of these values by a factor of e.g. $2$ changes the field strength only weakly by
$2^{1/(3 + \alpha)}\simeq1.2$.

Issues with the equipartition estimate are:

\noindent (1) Eq.~\ref{equi} can be applied only for steep radio spectra with $\alpha > 0.5$. For flatter spectra,
the integration over the energy spectrum of the cosmic rays has to be restricted to a limited energy interval.\\
(2) Due to the highly nonlinear dependence of $I_\mathrm{syn}$ on $B_\mathrm{tot,\perp}$,
the average equipartition value $B_\mathrm{tot,\perp}$ derived from synchrotron intensity is biased
towards high field strengths and hence is an overestimate if $B_\mathrm{tot}$ varies along the line
of sight or across the telescope beam \citep{beck03}.
For $\alpha=1$ and the equipartition case, the overestimation factor $g$ of the total field is
\citep[see Appendix A in][]{stepanov14}:
\begin{equation}
g = (<B_\mathrm{tot,\perp}^{\,\,4}>)^{1/4} \, / <B_\mathrm{tot,\perp}> \,\, =
(1 + {8\over3} Q^2 + {8\over9} Q^4)^{1/4} \, ,
\label{over2}
\end{equation}

\noindent where $< \,\,\, >$ indicates the averages along the line of sight and the beam, and
$Q=(<\delta B_\mathrm{tot,\perp}^2>)^{1/2}\, / <B_\mathrm{tot,\perp}>$ is the amplitude of the
field fluctuations relative to the mean field.
For strong fluctuations of $Q=1$, the overestimation factor is 1.46.\\
(3) If energy losses of CREs are significant, especially in starburst regions or massive spiral
arms, the equipartition values are lower limits and are underestimated. The ratio $K$ increases
because energy losses of aging CREs are much more severe than those of cosmic-ray protons. Using the
standard value $K_0$ underestimates the total magnetic field by a factor of $(K/K_0)^{1/(3 + \alpha)}$
in such regions \citep{beck+krause05}.
The same applies to the outer disk and halo of a galaxy, where the emitting CREs propagated far away
from the places of origin and energy losses are significant.\\
(4) In dense gas, e.g. in starburst regions, secondary positrons and electrons may be responsible
for most of the radio emission via pion decay. Notably, the ratio of protons to secondary
electrons is also about 100 for typical radio wavelengths \citep{lacki13}.\\
(5) Energy equipartition needs time to develop and hence is not valid on small scales in time
and space. A correlation analysis of radio continuum images from the Milky Way and the
nearby galaxy M~33 by \cite{stepanov14} showed that equipartition does not hold on scales smaller
than about 1\,kpc, which is understandable in view of the typical propagation length of CREs
away from their sources (Sect.~\ref{scalelength}). The equipartition field strength is underestimated
in regions around the CRE sources and overestimated in regions far away from the sources.

Arguments for the validity of equipartition on large (kpc) scales come from the joint analysis
of radio continuum and $\gamma$–-ray data, allowing an independent determination of magnetic
field strengths, e.g. in the Milky Way \citep{strong00}, the Large Magellanic Cloud (LMC)
\citep{mao12} and in M~82 \citep{yoast13} \footnote{but probably not in NGC~253 \citep{yoast14}}.
Furthermore, the estimates of strengths of large-scale
ordered fields derived from Faraday rotation data are similar to those from equipartition in several
galaxies \citep[see Table~2 in][]{vaneck15}.

\subsection{Polarized radio synchrotron emission}
\label{pol}

{\em Linearly polarized synchrotron emission}\ (examples in Figs.~\ref{fig:m51}, \ref{fig:m83a},
\ref{fig:n6946}, \ref{fig:ic342central} and \ref{fig:n4535}) emerges from CREs in ordered
fields in the sky plane. As polarization angles are ambiguous by $180^\circ$, they cannot
distinguish {\em regular fields}, defined to have a constant direction within the
telescope beam, from {\em anisotropic turbulent or tangled fields}. Unpolarized synchrotron
emission indicates {\em isotropic turbulent or tangled fields}\ that have random directions in 3D
and have been amplified by turbulent gas flows (see Sect.~\ref{comp}).

The intrinsic degree of linear polarization $p_0$ of synchrotron emission in a perfectly
regular field is:
\begin{equation}
p_0 = (1 + \alpha) \, / ({5\over3} \, + \alpha) \, ,
\label{intrinsic}
\end{equation}

\noindent where $\alpha$ is the spectral index of the synchrotron emission.
In spiral galaxies, typical values are $\alpha=0.8-1.0$, so that $p_0=73-75\%$.

The observed degree of polarization is smaller than $p_0$ due to the contribution from unpolarized thermal
emission, wavelength-dependent Faraday depolarization (Sect.~\ref{dp}) and wavelength-independent
``beam depolarization'' that is discussed in the following.

Turbulent magnetic fields in galaxies preserve their direction over a coherence scale that depends
on the properties of turbulence. \footnote{The degrees of polarization
expected from compressed or sheared fields are discussed in the Appendix of \citet{beck05c}.}
In the case of isotropic turbulent fields with a constant coherence length $d$,
the observable degree of synchrotron polarization is (see \citet{sokoloff98} for a detailed discussion):
\begin{equation}
p \, = \, p_0 \,\, N^{-1/2} \, ,
\label{beam}
\end{equation}

\noindent where $N \simeq D^2 L f / d^3$ is the number of synchrotron-emitting turbulent cells
with diameter $d$ and filling factor $f$ within the volume defined by the telescope beam with the linear
size $D$ at the galaxy's distance and pathlength $L$ through the source. Eq.~(\ref{beam}) is valid
only at short radio wavelengths, where Faraday depolarization is small.

Typical degrees of synchrotron polarization observed in nearby galaxies between 3\,cm and 6\,cm wavelengths
with typical spatial resolutions of $D\simeq500$\,pc are $1-5\%$ in spiral arms (where isotropic
turbulent fields dominate),
e.g. in NGC~6946 \citep{beck07} and in M~33 \citep{taba08}. With $L\simeq1000$\,pc and $f\simeq0.5$
for the turbulent field coupled to the warm diffuse gas, we get $d\simeq50$\,pc,
consistent with estimates from Faraday depolarization at long wavelengths (Eq.~\ref{dispersion}).

If the medium is pervaded by an isotropic turbulent field $B_\mathrm{iso}$
plus an ordered field $B_\mathrm{ord}$ (regular and/or anisotropic turbulent) that has a constant
orientation in the volume observed by the telescope beam,
the wavelength-independent polarization (for the case of equipartition between the energy
densities of magnetic field and cosmic rays) amounts to \citep{sokoloff98}:

\begin{equation}
p \, = \, p_0 \, (1 + 3.5 \, q^2) \, / \, (1 + 4.5 \, q^2 + 2.5 \, q^4) \, ,
\end{equation}

\noindent where $q = B_\mathrm{iso,\perp} \,  / \, B_\mathrm{ord,\perp}$ and
$B_\mathrm{iso,\perp} = \sqrt{2/3}\,\, B_\mathrm{iso}$.

\subsection{Faraday rotation and Faraday Synthesis}
\label{rm}

The polarization plane is rotated in a magnetized thermal plasma by
{\em Faraday rotation}. As the rotation angle is sensitive to the
sign of the field direction, only regular fields give rise to
Faraday rotation, while the Faraday rotation contributions from
turbulent fields largely cancel along the line of sight.
Measurements of Faraday rotation from multi-wavelength
observations (examples in Figs.~\ref{fig:m31rm} and \ref{fig:ic342rm})
yield the strength and direction of the average regular field component along
the line of sight. If Faraday rotation is low (in galaxies
typically at wavelengths shorter than a few centimeters), the
B--vector of polarized emission gives the intrinsic field
orientation in the sky plane, so that the magnetic pattern can be
mapped directly (Sect.~\ref{structure}).

The Faraday {\em rotation measure $RM$}\ is defined as the slope of the observed variation
of the polarization rotation angle $\chi$ with the square of wavelength $\lambda$:
\begin{equation}
\chi = \chi_0 \, + \, RM \, \lambda^2 \, ,
\end{equation}

\noindent where $\chi_0$ is the intrinsic polarization angle. $\lambda$ is measured in meters and RM
in rad\,m$^{-2}$. RM is constant if $\chi$ is a linear function
of $\lambda^2$, e.g. if one or more Faraday-rotating regions are located in front of the emitting region
({\em Faraday screen}).

A nonlinear variation of $\chi$ with $\lambda^2$ and hence a variation of RM with $\lambda^2$
occurs in case of:\\
(1) emission and rotation in the same region where the distribution of electrons or regular magnetic
field strength along the line of sight is not symmetric, or field reversals occur, or the field is
helical \citep{sokoloff98},\\
(2) Faraday depolarization (Sect.~\ref{dp}),\\
(3) several distinct emitting and rotating regions with different contributions to Faraday rotation
are located within the beam or along the line of sight or within the source (internal structure).\\
In these cases, RM needs to be replaced by the {\em Faraday depth $\Phi$} \citep{burn66}:

\begin{equation}
\chi = \chi_0 \, +\Phi \, \lambda^2 \, \, , \,\,\, \mathrm{where} \,\,\,
\Phi = 0.812 \, \int\limits_{source}^{observer}{B_\mathrm{reg,\parallel} \, n_\mathrm{e} \, dl} \, ,
\end{equation}

\noindent where $\Phi$ is measured in rad\,m$^{-2}$, the line-of-sight magnetic field
$B_\mathrm{reg,\parallel}$ in $\mu$G, the thermal electron density $n_\mathrm{e}$ in cm$^{-3}$
and the line of sight $l$ in pc.

\begin{figure*}[t]
\vspace*{2mm}
\begin{center}
\subfigure[Sketch of different components along the line of sight. Some of them are emitting ($Ei$), are
Faraday-rotating ($Ri$), or both.]{
    \includegraphics[scale=0.28]{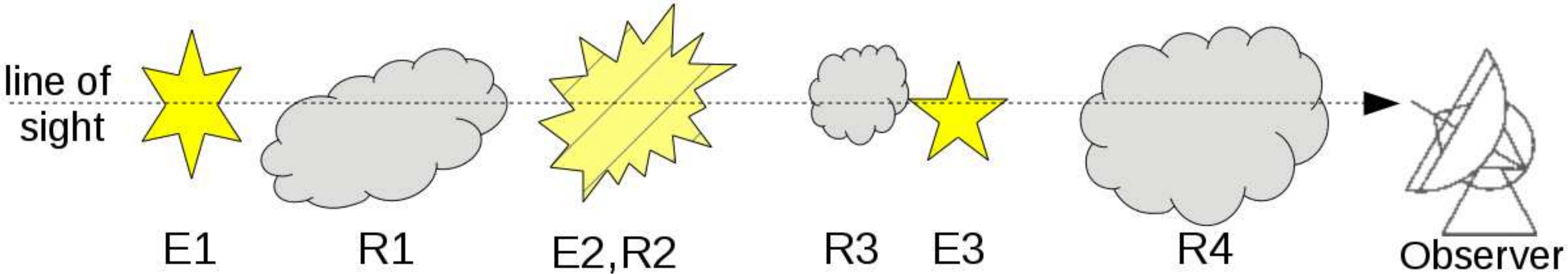}
    \label{LOS}
}
\subfigure[Resulting Faraday spectrum $F(\Phi)$, where $\Phi$ is the Faraday depth, of the components
depicted in~\ref{LOS}.]{
    \includegraphics[scale=0.4]{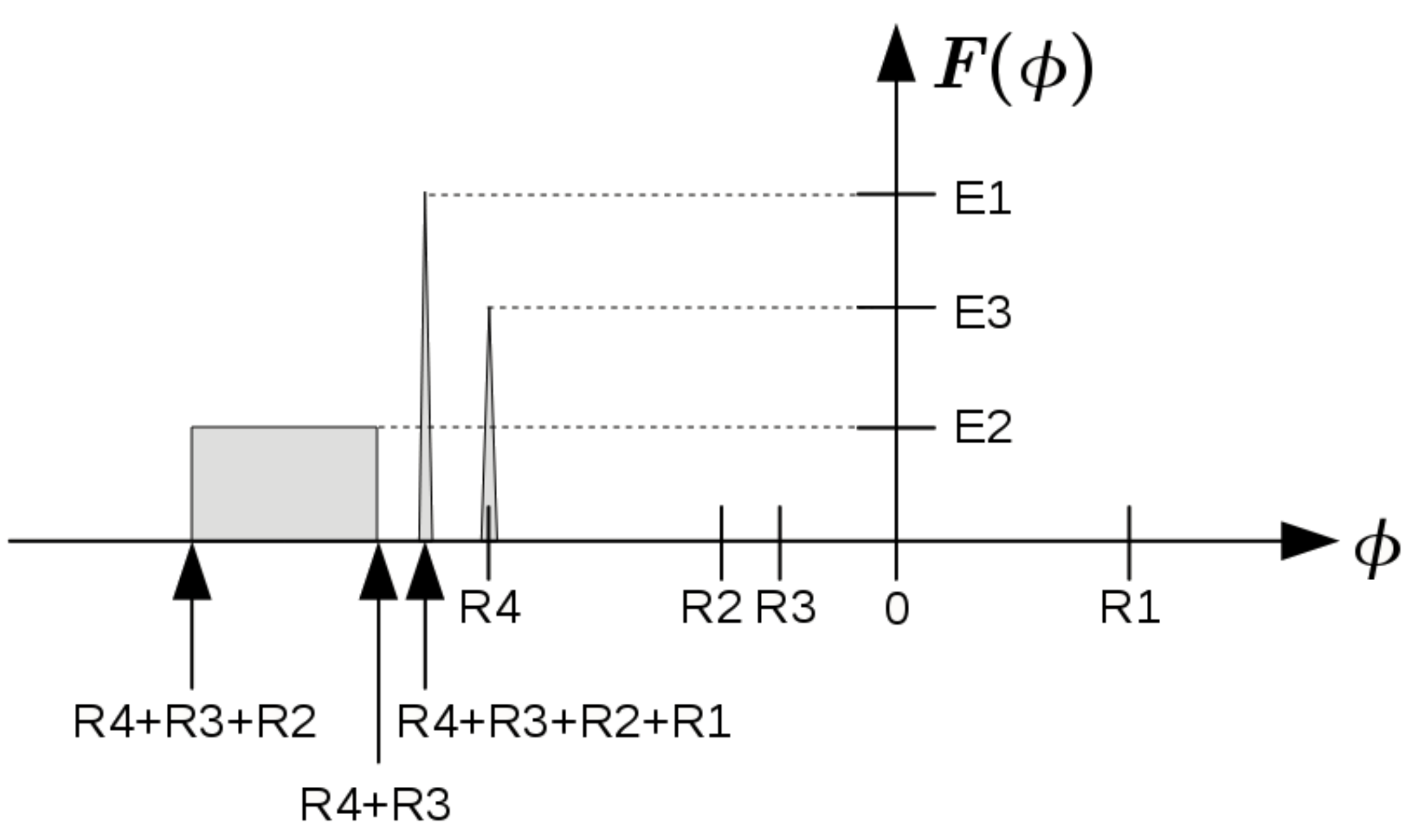}
    \label{LOS-spectrum}
}
\caption{Example for a set of different polarization-emitting and rotating components along the line of
sight and the resulting Faraday spectrum $F(\Phi)$. $E1$ and $E3$ are point sources,
$E2,R2$ is an extended emitting and rotating region. $R1$, $R3$, and $R4$ are Faraday screens that are
not emitting. Faraday depolarization effects are neglected here for simplicity (from \citet{giess13}).}
\label{fig:rmsynth}
\end{center}
\end{figure*}

To measure the contribution of various sources to $\Phi$, wide-band, multi-channel spectro--polarimetric
radio data are needed. These are Fourier-transformed by a software tool called
{\em RM Synthesis} \citep{brentjens05,heald15} or {\em Faraday Synthesis}
\citep[following the terminology proposed by][]{sun15},
to obtain the {\em Faraday spectrum}\ $F(\Phi)$ (previously called {\em Faraday dispersion function})
at each position of the radio image. $F(\Phi)$ shows the intensity of polarized emission and its polarization
angle as a function of $\Phi$ (Fig.~\ref{fig:rmsynth}).

If the medium has a relatively simple structure, Faraday spectra at
several positions can reveal the 3D structure of the magnetized interstellar medium
({\em Faraday tomography}). For example, regular fields, field reversals and turbulent fields can
be recognized from their signatures in Faraday spectra \citep{bell11,frick11,beck12}. Helical fields
can also imprint characteristic features in the Faraday spectrum \citep{brandenburg14,horellou14}.
Realistic galaxies have complicated Faraday spectra, as revealed from simplified galaxy
models \citep{ideguchi14}.

Crucial to successful application of Faraday Synthesis is that the observations cover a wide range
in $\lambda^2$, giving high resolution in Faraday spectra.
Even for a wide frequency coverage, two components with a difference in intrinsic polarization angle
of about $90^\circ$ cannot be properly recovered by Faraday Synthesis, so that model fitting to
the data in Stokes Q and U is needed \citep{farnsworth11}.


A {\em grid of $RM$ measurements}\ of polarized background sources is another powerful tool to
measure magnetic field patterns in galaxies \citep{stepanov08,mao12}. At least 10 RM values from
sources behind a galaxy's disk are needed to recognize a simple large-scale field pattern
if the Galactic foreground contribution is constant and the background sources have no intrinsic
contributions to RM.


Measuring the strength of the regular field from RM needs additional information about the thermal
electron density, e.g. from the arrival times of pulsar signals that are delayed in a cloud of ionized gas
proportional to the {\em dispersion measure} $DM=\int{n_\mathrm{e} \, dl}$. The ratio

\begin{equation}
RM / DM = 0.812
< B_\mathrm{reg,\parallel} \, n_\mathrm{e} > / < n_\mathrm{e} > \,
\label{dmrm}
\end{equation}

\noindent allows us to compute the field strength $B_\mathrm{reg,\parallel}$ (in $\mu$G) if the fluctuations
in field strength and in electron density are {\em uncorrelated}.
The value for $B_\mathrm{reg,\parallel}$ derived
from pulsar data is an underestimate if small-scale fluctuations in field strength and in electron
density are anticorrelated, as expected for local pressure equilibrium, while it is an overestimate if the
fluctations are correlated, as expected for compression by shock fronts or turbulence \citep{beck03}.

\subsection{Faraday depolarization}
\label{dp}

%

In a single region containing CREs, thermal electrons and purely {\em regular magnetic fields},
wavelength-dependent Faraday depolarization occurs because the polarization planes of waves from
the far side of the emitting layer are more rotated than those from the near side. This effect is
called {\em differential Faraday rotation}. For one single layer with a symmetric distribution of
thermal electron density and field strength along the line of sight the degree of polarization is
reduced to \citep{burn66,sokoloff98}:
\begin{equation}
p \, = \, p_0 \,\, | sin (2\,RM\,\lambda^2) / (2\,RM\,\lambda^2) | \, ,
\label{diff}
\end{equation}

\noindent where RM is the observed rotation measure from the source, which is half of the Faraday depth $\Phi$
through the whole layer. $p$ varies periodically with the square of wavelength.

Applying Faraday Synthesis (Sect.~\ref{rm}) to polarization data of a region that is emitting and
Faraday-rotating reveals an extended component in the Faraday spectrum (Fig.~\ref{fig:rmsynth}).
However, regions broader than
$\Phi_\mathrm{max} \simeq \pi / \lambda_\mathrm{min}^2$ cannot be recovered, where $\lambda_\mathrm{min}$
is the minimum wavelength of the observations \citep{brentjens05}; only two ``horns'' remain visible at
the edges of the structure in the Faraday spectrum $F(\Phi)$ \citep{beck12}. This problem is similar
to the missing short baselines in synthesis imaging.

For multiple emitting + rotating layers, the Faraday spectrum contains several extended components and
Eq.~(\ref{diff}) is no longer applicable.
\citet{shneider14} extended Burn's model to a multi-layer (e.g. galaxy disk + halo) medium.

Faraday rotation in {\em helical fields}\ has a completely different behaviour and may lead to an increase
of the degree of polarization with increasing wavelength (``re-polarization'') in certain wavelength ranges
\citep{sokoloff98,brandenburg14,horellou14}.


{\em Turbulent fields}\ also cause wavelength-dependent depolarization, called {\em Faraday dispersion}\
\citep{sokoloff98,arshakian11}. For an emitting and Faraday-rotating region (internal dispersion)
the degree of polarization is reduced to:
\begin{equation}
p \, = \, p_0 \,\, (1 - exp(-S)) / S \, ,
\end{equation}

\noindent where $S = 2 \sigma_\mathrm{RM}^2\,\lambda^4$. $\sigma_\mathrm{RM}$ is the dispersion in
rotation measure and depends on the turbulent field strength along the line of sight, the turbulence
scale, the thermal electron density, and the pathlength through the medium.

Depolarization by external dispersion occurs in a turbulent Faraday-rotating (but not emitting) medium in the foreground
if the diameter of the telescope beam at the distance of the screen is larger than the turbulence scale:
\begin{equation}
p \, = \, p_0 \,\, exp(-S) \, .
\end{equation}

An important effect of Faraday dispersion is that the interstellar medium becomes ``Faraday thick''
for polarized radio emission beyond a certain wavelength, depending on $\sigma_\mathrm{RM}$, and only a front
layer remains visible in polarized intensity. Typical values for galaxy disks are
$\sigma_\mathrm{RM}= 20-100$\,rad\,m$^{-2}$, while star-forming regions can have dispersions of
$\sigma_\mathrm{RM}\simeq 1000$\,rad\,m$^{-2}$ \citep{arshakian11}. The maximum in the spectrum
of polarized intensity falls into the wavelength range 2--20\,cm.

In a random-walk approach, we may write
$\sigma_\mathrm{RM} = 0.812 \, B_\mathrm{iso,\parallel} \,\, n_\mathrm{e} \, d \, N_\parallel^{1/2}$
\citep{beck07}, where $B_\mathrm{iso,\parallel} = \sqrt{1/3}\,\,B_\mathrm{iso}$ is the strength of the
isotropic turbulent field, $n_\mathrm{e}$ is the electron density within a cell of size $d$, and
$N_\parallel = L  \, f / d$
is the number of cells along the line of sight $L$ with a volume filling factor $f$. The average electron
density along the line of sight is $<n_\mathrm{e}> = n_\mathrm{e} / f$, so that we get:
\begin{equation}
\sigma_\mathrm{RM} = 0.812 \, B_\mathrm{iso,\parallel} \, <n_\mathrm{e}> \, \sqrt{L \, d / f}  \, .
\label{dispersion}
\end{equation}

The average value for galaxy disks of $\sigma_\mathrm{RM} \simeq 50$\,rad\,m$^{-2}$ is consistent
with typical values for the warm diffuse ISM of $<n_\mathrm{e}> \simeq 0.03$\,cm$^{ -3}$ and $f \simeq 0.5$
\citep[Fig.~10 in][]{berkhuijsen06}, $B_\mathrm{iso} \simeq 10\,\mu$G, $L \simeq 1000$\,pc, and $d \simeq 50$\,pc.

The dispersion $\sigma_\mathrm{RM}$ leads to a ``Faraday forest'' of $N$ components in the Faraday spectrum.
If $N$ is not large, the components are possibly resolvable with very high Faraday resolution, hence a
wide $\lambda^2$ span of the observations \citep{beck12,bernet12}.

\section{Results}

\subsection{Magnetic field strengths}
\label{strength}

Strength is the fundamental quantity to estimate the dynamical importance of magnetic fields.
The equipartition assumption (Sect.~\ref{eq}) is the only presently applicable method to measure the
total field strength. Zeeman measurements will become possible in spiral galaxies with future radio
telescopes, but will trace fields on small scales only.

The average equipartition strength (Eq.~\ref{equi}) of {\em total fields}\ (corrected for inclination)
for a sample of 74 spiral galaxies is $B_\mathrm{tot}=9\pm 2\,\mu$G \citep{niklas95}. The average
strength of 21 bright galaxies observed between 2000 and 2010 is $B_\mathrm{tot}=17\pm3\,\mu$G \citep{fletcher10}.
Gas-rich spiral galaxies with high star-formation rates, like M~51 (Fig.~\ref{fig:m51}), M~83 (Fig.~\ref{fig:m83a})
and NGC~6946 (Fig.~\ref{fig:n6946}), have total field strengths of $20-30\,\mu$G in their spiral arms.
The strongest total fields of $50-100\,\mu$G are found in starburst galaxies like M~82 \citep{adebahr13},
in the ``Antennae'' NGC~4038/9 \citep{chyzy04}, in nuclear starburst regions like in NGC~253
\citep{heesen11a}, and in barred galaxies \citep{beck05c}.
In such galaxies, energy losses of CREs can be strong, so that
the equipartition values are underestimates (Sect.~\ref{eq}).
Radio-faint galaxies like M~31 (Fig.~\ref{fig:m31}) and M~33 have weaker total magnetic fields (about $6\,\mu$G).
The similarity to the values derived for our Milky Way with various methods (Sect.~\ref{MW}) gives confidence
that the equipartition assumption is valid.

Field strengths of $0.5-18$\,mG were detected in starburst galaxies by the Zeeman effect in the OH megamaser
emission line at 18\,cm wavelength \citep{robishaw08}. These values refer to highly compressed gas clouds
and are not typical of the diffuse interstellar medium.

The strength of {\em ordered fields}\ in the sky plane as observed by polarized synchrotron emission
varies strongly between galaxies, from 10--15\,$\mu$G in M~51 and the magnetic arms of NGC~6946
(Fig.~\ref{fig:n6946}) to about $5\,\mu$G in the star-forming ring of M~31 (Fig.~\ref{fig:m31}).
The average strength of the ordered fields of 21 bright galaxies observed since 2000 is
$B_\mathrm{ord,\perp}=5\,\mu$G with a standard deviation of $3\,\mu$G and an average ratio
$B_\mathrm{ord,\perp}/B_\mathrm{tot}$ of 0.3 \citep{fletcher10}.

The only deep observation of an Sa galaxy, M~104, with a prominent dust ring, revealed ordered,
partly regular magnetic fields \citep{krause06}. Spiral galaxies of type S0 and elliptical
galaxies without an active nucleus have very little star formation and hence hardly
produce CREs that could emit synchrotron emission.
Irregular starburst galaxies show only spots of ordered fields \citep{heesen11b}.
No ordered fields could be detected so far in dwarf irregular galaxies \citep{chyzy11}.

\subsection{Magnetic energy densities}
\label{energy}

\begin{figure*}[t]
\vspace*{7mm}
\begin{center}
\includegraphics[width=9cm]{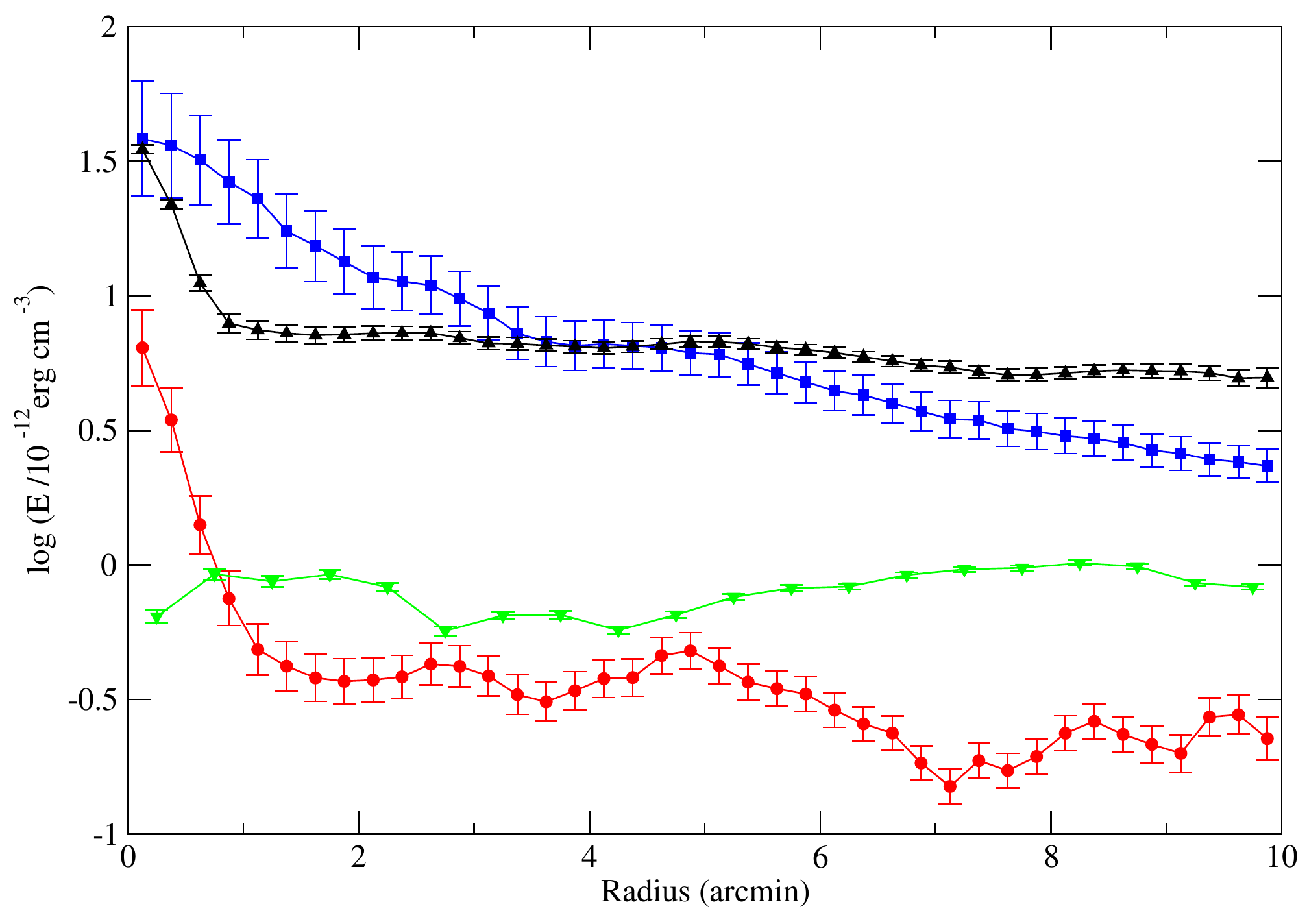}
\caption{Radial variation of the energy densities in IC~342, determined from observations
of synchrotron and thermal radio continuum and the CO and $\HI$ line emissions:
for the total magnetic field $B_\mathrm{tot}^2/8\pi$ (black triangles), identical to that for the total
cosmic rays, for the ordered magnetic field $B_\mathrm{ord}^2/8\pi$ (green triangles), for the
turbulent motion of the neutral gas $E_\mathrm{turb}=0.5 \, \rho_n \, \mathrm{v_{turb}}^2$ (blue squares),
where $\mathrm{v_{turb}}=7$~km/s, and for the thermal energy of the warm ionized gas
$E_\mathrm{th}=1.5 \, <n_\mathrm{e}> \, k \, T_\mathrm{e}$ (red circles), where $T_\mathrm{e}=10^4$~K.
The error bars include only errors due to rms noise in the images from which the energy densities are derived.
No systematic errors are included, e.g. those imposed by a radial variation of thermal gas temperature,
filling factor or turbulent gas velocity, nor errors due to deviations from the equipartition assumption
(from \citet{beck15}).}
\label{fig:ic342_energies}
\end{center}
\end{figure*}

The relative importance of various competing forces in the interstellar medium can be
estimated by comparing the corresponding {\em energy densities}. The mean energy densities of the total
(mostly turbulent) magnetic field and the total cosmic rays, averaged in rings of about 1\,kpc width,
are $\simeq10^{-11}$\,erg~cm$^{-3}$ in NGC~6946 \citep{beck07}, M~63, M~83, NGC~4736 \citep{basu13}
and IC~342 (Fig.~\ref{fig:ic342_energies}), and $\simeq10^{-12}$\,erg~cm$^{-3}$ in M~33 \citep{taba08}.
In all these galaxies the average magnetic energy density is similar to the average kinetic energy
density of the turbulent motions of the neutral gas across the star-forming disk (assuming a constant
turbulent velocity of $10\kms$) and about 10 times larger than that of the warm ionized gas
(but still 500--1000~times smaller than the energy density of the general rotation of the neutral gas).
This result is similar to that derived for the Milky Way \citep{cox05}.
The ISM is a \emph{low--$\beta$ plasma}\ (where $\beta$ is the ratio of thermal to magnetic energy densities).
Magnetic fields are dynamically important.

According to Fig.~\ref{fig:ic342_energies}, $E_\mathrm{turb}/E_\mathrm{th} \gg 1$, i.e. turbulence is
\emph{supersonic}. Supersonic turbulence leads to shocks and hence dissipation of kinetic energy into heat,
until turbulence becomes transonic and the system reaches an equilibrium state. The observation of supersonic
turbulence may indicate that a quasi-steady state is never reached. However, observational bias effects
need to be considered. The energy density of hot gas in the ISM, neglected in Fig.~\ref{fig:ic342_energies},
is similar or somewhat larger than that of warm gas, depending on its volume filling factor \citep{ferriere01},
so that its inclusion would not change the above result significantly.
Another bias is the assumption of a constant gas temperature $T_\mathrm{e}$. If cosmic rays
propagate via the streaming instability (Sect.~\ref{scalelength}), they may heat the gas and increase
$T_\mathrm{e}$ in the outer disk \citep{wiener13}.

Another important message from Fig.~\ref{fig:ic342_energies} (and similarly from the other galaxies
studied so far) is that the magnetic energy density decreases radially more slowly
than the kinetic energy density and dominates in the outer galaxy disk, i.e. the Alfv{\'e}n velocity
v$_\mathrm{A} \propto B_\mathrm{tot}/\sqrt{\rho}$ is larger than the turbulent velocity and increases with
radius, which may cause magnetic effects on the rotation curve of the gas \citep{elstner14}.
Note that the ratio of magnetic to kinetic energy may be even higher in the outer galaxy disk than
shown in Fig.~\ref{fig:ic342_energies}, because the equipartition field strengths are underestimates
due to energy losses of the CREs (Sect.~\ref{eq}). Furthermore, the turbulent velocity
tends to decrease with radius \citep{tamburro09}, which would also enhance the ratio of magnetic
to kinetic energy.

Possible reasons why equipartition between magnetic and kinetic energy density does not hold
within galaxies are:\\
(1) energy or pressure balance is valid between the magnetic field and the sum of pressures
related to all gas components (kinetic and thermal) \citep{basu13};\\
(2) the efficiency of the small-scale dynamo increases from the inner to the outer galaxy;\\
(3) the magneto-rotational instability (MRI) leads to super-equipartition fields
\citep[Fig.~7 in][]{gressel13};\\
(4) the equipartition field strengths are underestimated in the inner galaxy and overestimated
in the outer galaxy due to CRE propagation (see Sect.~\ref{eq}), hence flattening the profile of
the magnetic energy density.

Scenario (4) is supported by the analysis of radio and IR images \citep{berkhuijsen13}
and the cross-correlation based on wavelet transforms that breaks down below a certain scale
that can be interpreted as the propagation length of CREs \citep{taba13b}.
CRE propagation in IC~342 seems to be exceptionally fast (Sect.~\ref{scalelength}).

The energy density of the {\em ordered}\ magnetic field (Fig.~\ref{fig:ic342_energies}) even increases
radially. This field may be assumed to be mostly regular, so that we can invoke the $\alpha-\Omega$ dynamo
that also operates in the outer disk of galaxies \citep{mikhailov14}. Although the star-formation activity
is low in the outer disk, the magneto-rotational instability (MRI) may serve as the source of turbulence
required for $\alpha-\Omega$ dynamo action \citep{sellwood99,gressel13}. Alternatively, regular fields
generated in the inner disk could be transported outwards by the joint action of a dynamo and turbulent
diffusivity \citep{moss98c}.

\subsection{Cosmic-ray propagation}
\label{scalelength}

The measurable extent $L_\mathrm{syn}$ of synchrotron disks or halos (see Sect.~\ref{halo}) gives us information
about the extent of the total magnetic field and the propagation of CREs. Under energy equipartition conditions,
the radial exponential scale length of the total field in a disk of mildly inclined galaxies (or the
vertical exponential scale height in a halo of almost edge-on galaxies)
is $L_\mathrm{B} \ge (3+\alpha)\,L_\mathrm{syn}$, where $\alpha\simeq 0.9$ is the synchrotron spectral index.
The scale length $L_\mathrm{CR}$ (or scale height) of the total cosmic rays is half of this value.
These are lower limits because the CREs lose their energy with distance from the star-forming disk, so that
the equipartition assumption yields too small values for the field strength (Sect.~\ref{eq}).
The scale lengths of synchrotron disks of typically $L_\mathrm{syn}\simeq3-5$\,kpc at 20\,cm wavelength
\citep{beck07,basu13,mulcahy14} yield $L_\mathrm{B} \simeq 12-19$\,kpc and $L_\mathrm{CR} \simeq 6-9.5$\,kpc.
The nearby galaxy IC~342 (Fig.~\ref{fig:ic342}) reveals an exceptionally large synchrotron
scale length of about 16\,kpc \citep{beck15}.

The radial scale length and its frequency dependence give us a general idea about the speed and type
of CRE propagation.
At wavelengths of $\ge1$\,m, synchrotron and inverse Compton losses are weaker than at shorter
wavelengths, so that CREs can diffuse further outwards from their locations of origin. This is
clearly seen in M~31 and M~33 \citep{berkhuijsen13}. The LOFAR image of the galaxy M~51 at
2\,m wavelength reveals a steepening of the radial distribution of synchrotron emission, located at
about 10\,kpc radius, beyond the sharp decrease in the star-formation rate, indicating a propagation
length of a few kpc \citep{mulcahy14}.
The frequency dependence of the scale length favours diffusive propagation in the cases of M~51
and several other galaxies \citep{basu13,berkhuijsen13,mulcahy14} and fast convective transport
in IC~342 \citep{beck15}.


The galactic fields may extend further out into intergalactic space than visible in radio synchrotron
images. A large radial scale length means that magnetic fields may affect the global rotation
of the gas in the outer parts of spiral galaxies, possibly explaining some fraction of the flattening
of the rotation curves \citep{battaner07,elstner14}.

\subsection{Relations between gas, star formation and magnetic fields}
\label{RIC}


Radio continuum and infrared emissions from galaxies are closely correlated.

Firstly, a {\em global}\ correlation exists between the galaxy-integrated luminosity of the total
radio continuum emission at around 20\,cm wavelengths (frequencies of around 1\,GHz), which is mostly
of synchrotron origin, and the infrared (IR) luminosity of star-forming galaxies. This is one of the
tightest correlations known in astronomy. The correlation extends over five orders of magnitude
\citep{bell03}, is slightly nonlinear in log-log scale \citep{bell03,pierini03} with an exponent
of $1.09\pm0.05$ \citep{basu15} and is valid in starburst galaxies to redshifts of at least
four \citep{seymour08}, without evidence for evolution with redshift up to 1.2 \citep{basu15}.
As the correlation between radio thermal and IR luminosity should be strictly linear,
the exponent of the nonthermal (synchrotron)--IR correlation is steeper than 1.1.
A breakdown of synchrotron emission and of its correlation with IR is expected beyond a
critical redshift when inverse Compton loss of the CREs dominates
synchrotron loss; this critical redshift will give us information about the field
evolution in young galaxies \citep{schleicher13}.

\begin{figure*}[t]
\vspace*{2mm}
\begin{center}
\includegraphics[width=9cm]{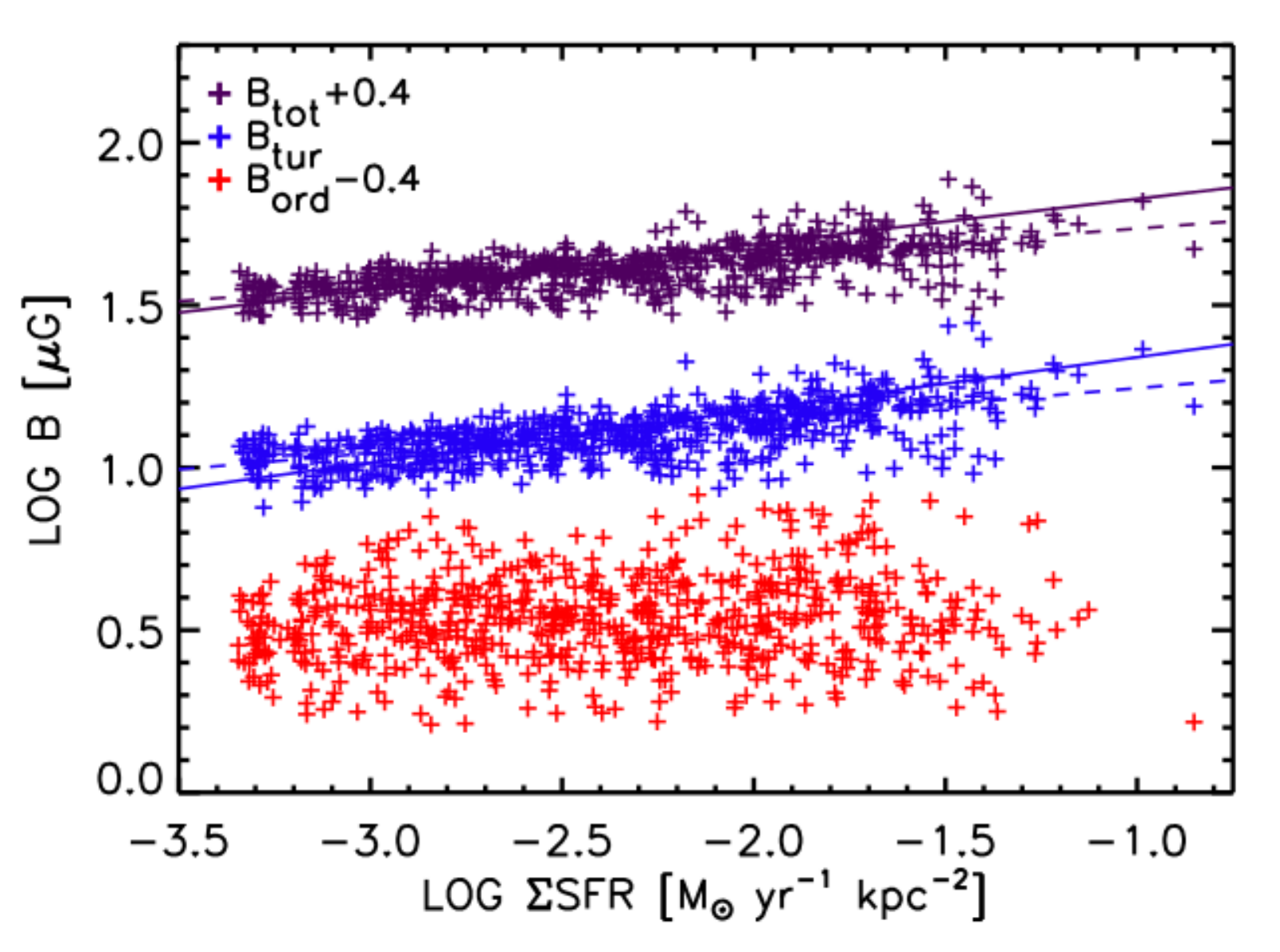}
\caption{Correlation between the strength of the total, turbulent and
ordered fields and the star-formation rate per
area (determined from the 24~$\mu$m infrared intensities) within the
galaxy NGC~6946. The points for $B_\mathrm{tot}$ and $B_\mathrm{ord}$ are
shifted by +0.4 and -0.4, respectively, to fit into the same figure.
The solid lines are bisector fits with slopes of $0.14\pm0.01$ for $B_\mathrm{tot}$
and $0.16\pm0.01$ for $B_\mathrm{turb}$, the dashed lines ordinary least-squares fits
with slopes $0.089\pm0.004$ and $0.100\pm0.005$
(from \citet{taba13a} and priv. comm.).}
\label{fig:n6946corr}
\end{center}
\end{figure*}

In galaxies with strong magnetic fields, e.g. in starburst galaxies, the CREs may lose their energy by
synchrotron loss within the galaxy disk (``lepton calorimeter model'', \citet{voelk89}), where the correlation
is linear and due to the joint generation of CREs and dust-heating UV photons in star-forming regions,
independent of the magnetic field. In Milky Way-type galaxies, magnetic fields are crucial to understand the
correlation. \citet{niklas97} proposed an ``equipartition model'', where total (mostly turbulent)
magnetic fields, cosmic rays, gas density and star formation are related, and predicted a nonlinear
synchrotron--IR correlation with an exponent of about 1.3, in good agreement with observations.
The more detailed analysis by \citet{lacki10} identified multiple feedback mechanisms that are not yet fully
understood.

Secondly, the total radio and IR intensities {\em within}\ galaxies are also highly
correlated, but with a smaller exponent, due to CRE propagation \citep{berkhuijsen13}. CRE propagation
is also responsible for the flattening of the profile of cosmic-ray energy density in
Fig.~\ref{fig:ic342_energies}. The exponent of the correlation was found to be different in the central
region, spiral arms and interarm regions in several nearby galaxies \citep{dumas11,basu12},
and it also differs between galaxies due to differences in the diffusion length of CREs
\citep{berkhuijsen13,taba13b}.

The magnetic field and its structure play an important role in understanding the correlation
\citep{taba13b,taba13a}. The synchrotron--IR correlation can be presented as a correlation between
total field strength $B_\mathrm{tot}$ and star-formation rate surface density $\Sigma_\mathrm{SFR}$
with an exponent of $0.18\pm0.01$ within NGC~4254 \citep{chyzy08}, $0.14\pm0.01$ within NGC~6946
(Fig.~\ref{fig:n6946corr}), and $0.30\pm0.02$ (globally and locally) for a sample of 17 spiral galaxies
\citep{heesen14}. A very similar exponent is expected from theoretical considerations
\citep{schleicher13}.

The Kennicutt-Schmidt law $\Sigma_\mathrm{SFR} \propto \Sigma_\mathrm{gas}^{\,\,\,N}$ is valid within
galaxies, with $N\simeq1.0$ for the molecular gas and $N\simeq1.5$ for the atomic gas \citep{bigiel08}.
\footnote{The nonlinearity of the Kennicutt-Schmidt law can be interpreted as a dependence of the
star-formation efficiency (SFE) on the stellar mass surface density \citep{shi11}. Interestingly,
effects of the magnetic field on the SFE have not been investigated so far.}
Then the above relation $B_\mathrm{tot} \propto \Sigma_\mathrm{SFR}^{\,\,\,0.30}$ corresponds to
$B_\mathrm{tot} \propto \Sigma_\mathrm{gas}^{\,\,\,0.30}$ if molecular gas dominates, e.g. in the inner
regions of most spiral galaxies, and $B_\mathrm{tot} \propto \Sigma_\mathrm{gas}^{\,\,\,0.45}$ in the outer
regions that are dominated by atomic gas.

Turbulent fields in spiral arms are probably generated by turbulent gas motions related to star
formation activity (Sect.~\ref{dynamo}). The small-scale dynamo predicts equipartition between magnetic
and kinetic energy densities ($B_\mathrm{tot} \propto \rho^{0.5}$, assuming v$_\mathrm{turb}$=const),
which is supported by the observed average values of a galaxy sample \citep{niklas97}.
However, within a galaxy the ratio between these energy densities decreases (Fig.~\ref{fig:ic342_energies}),
which is consistent with an exponent of $<0.5$ found from the $\Sigma_\mathrm{SFR}$ data (see above).
Possible reasons why equipartition between magnetic and kinetic energy density holds on average between
galaxies, but not within galaxies, were discussed in Sect.~\ref{energy}.

The correlation between the radio continuum luminosity and the {\em molecular gas}\
(traced by its CO line emission) is similarly tight as the radio--IR correlation and is also
nonlinear, with a slope of $1.31\pm0.09$ \citep{liu10}.
The correlation between the synchrotron radio continuum and CO intensities,
observed with 60\,pc resolution within the spiral galaxy M~51
\citep{schinnerer13}, appears to be tighter than the radio--IR correlation
and may be the fundamental one. Much of the molecular gas in the spiral arms
of M~51 is not directly associated with ongoing massive star formation, implying that star formation
cannot be the sole cause for the radio--CO relation. Either the synchrotron emission emerges from
secondary CREs that are produced in the interaction of cosmic-ray protons with the dense molecular
material \citep{murgia05}, or the synchrotron emission is from primary CREs in strong magnetic fields
coupled to the dense gas \citep{niklas97}, in connection with fast diffusion of CREs along
the spiral arms. High-resolution observations of nearby galaxies (e.g. with SKA and
ALMA) can provide the necessary information on the coupling between magnetic fields and gas on small
scales, to test which of the proposed origins for the correlation is indeed true.

In contrast to the turbulent field, the ordered field is either uncorrelated with the star-formation rate
(Fig.~\ref{fig:n6946corr}) or anticorrelated in interarm regions, where the star-formation rate is low
\citep{frick01}.

\subsection{Spiral fields}
\label{structure}


The most striking result from the observation of ordered (regular and/or anisotropic) fields
with help of polarized synchrotron emission is that spiral patterns are found in almost every
spiral galaxy \citep{beck+wielebinski13}, in galaxies with a star-forming ring
\citep{chyzy+buta08}, in flocculent galaxies without gaseous spiral arms \citep{soida02},
in central regions of galaxies (Fig.~\ref{fig:ic342central}), and in circum-nuclear gas rings of
barred galaxies (Fig.~\ref{fig:n1097central}).

\begin{figure*}[t]
\vspace*{2mm}
\begin{center}
\includegraphics[width=9cm]{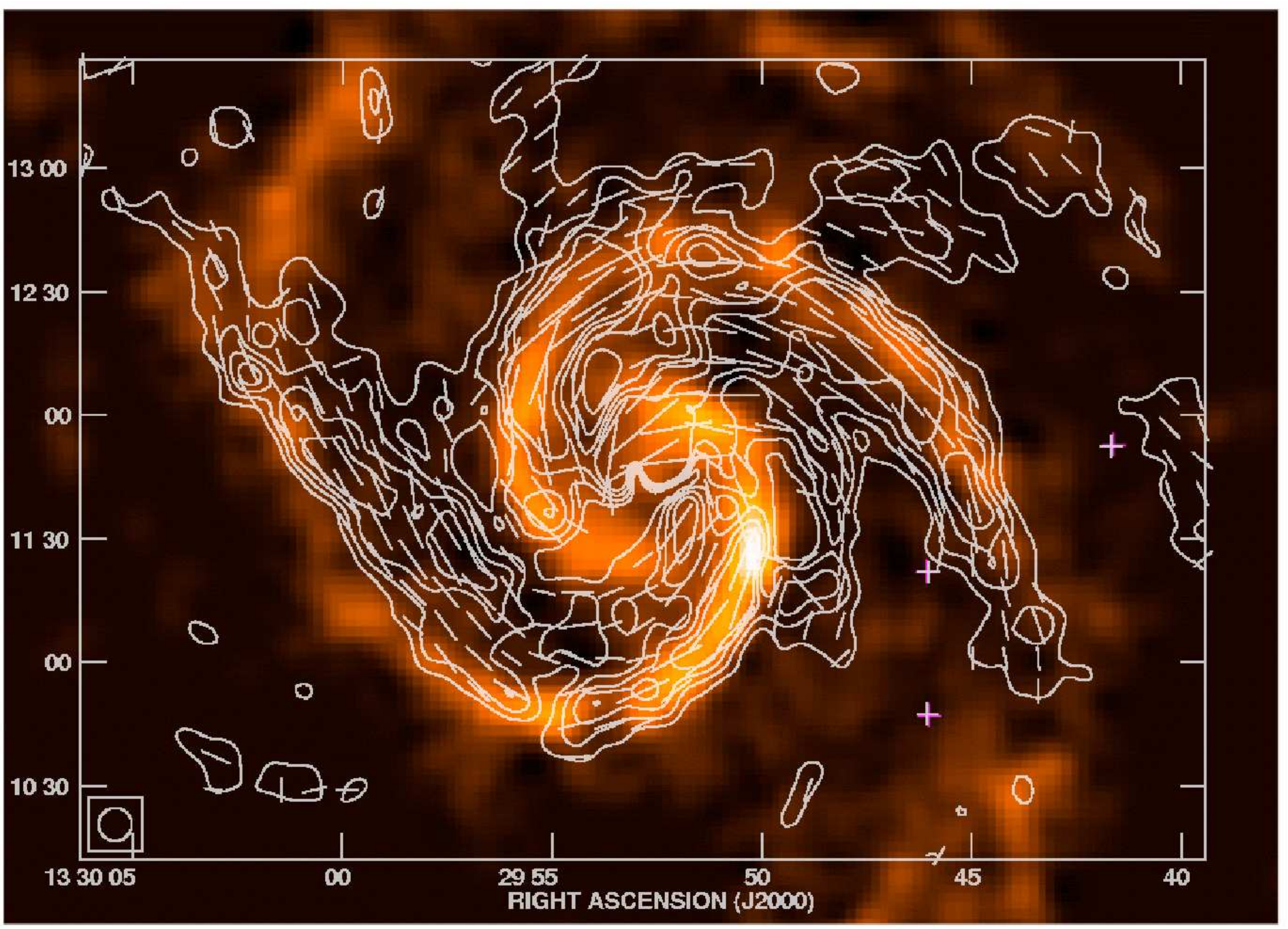}
\caption{Polarized radio emission (contours) and B--vectors of M~51,
combined from observations at 6\,cm wavelength with the VLA and Effelsberg telescopes
at $8^{\prime\prime}$ resolution (from \citet{fletcher11}). The background colour image shows
the CO line emission from molecular gas (from \citet{helfer03}).}
\label{fig:m51}
\end{center}
\end{figure*}

In galaxies with strong density waves like M~51 (Fig.~\ref{fig:m51}) and M~83
(Fig.~\ref{fig:m83a}) enhanced ordered fields occur at the inner
edges of the inner optical arms where the cold molecular gas is densest
\citep{patrikeev06}. These are probably anisotropic turbulent fields
generated by the compression of the density wave. An analysis of dispersions
of the radio polarization angles at 6\,cm wavelength in M~51 shows that the ratio of the
correlation lengths parallel and perpendicular to the local ordered
magnetic field is about two \citep{houde13}.

\begin{figure*}[t]
\vspace*{2mm}
\begin{center}
\includegraphics[width=9cm]{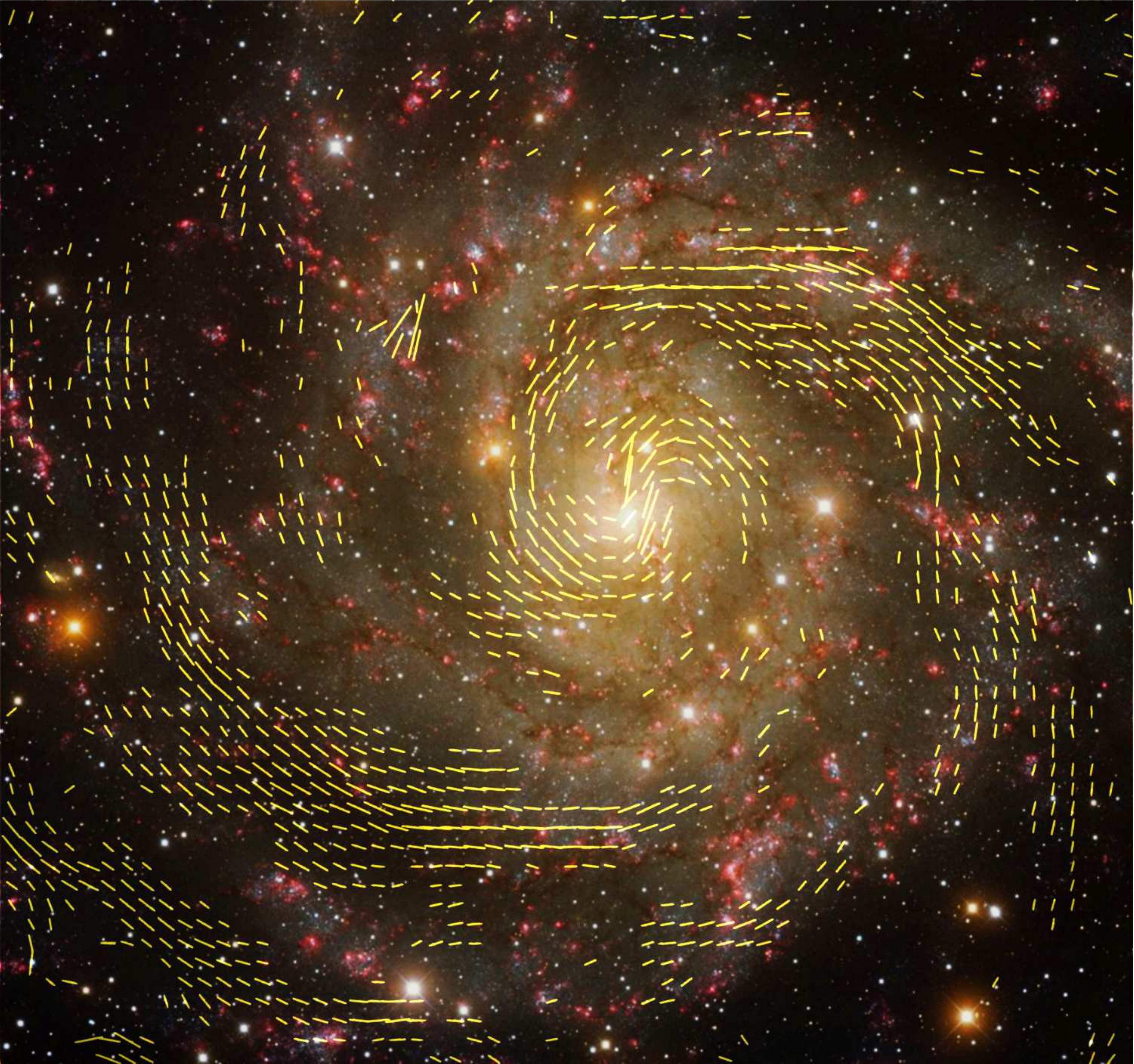}
\caption{Polarization B--vectors of IC~342, combined from observations at 6\,cm wavelength with the
VLA and Effelsberg telescopes and smoothed to $25^{\prime\prime}$ resolution  (from \citet{beck15}),
overlaid on a colour image from the Kitt Peak Observatory (credit: T.A. Rector, University of
Alaska Anchorage, and H. Schweiker, WIYN and NOAO/AURA/NSF).
A region of $16^{\prime} \times 16^{\prime}$ (about $16 \times 16$\,kpc) is shown.
(Copyright: MPIfR Bonn)}
\label{fig:ic342}
\end{center}
\end{figure*}

In galaxies without strong density waves the ordered fields are not coinciding
with the gaseous or the optical spiral arms.
The typical degree of radio polarization at small wavelengths is only a few percent
within the spiral arms, hence the field in the spiral arms must be mostly tangled or
randomly oriented within the telescope beam on scales of a few 10\,pc (Eq.~\ref{beam}).

Ordered fields are strongest ($10-15\,\mu$G) in the regions {\em
between}\ the spiral arms and oriented parallel to the
adjacent spiral arms. In some galaxies they form {\em magnetic arms},
like in NGC~6946 (Fig.~\ref{fig:n6946}),
with exceptionally high degrees of polarization up to 50\%. These
are probably generated by the $\alpha-\Omega$ dynamo (Sect.~\ref{ma}).

The ordered magnetic field in the galaxy IC~342 reveals several spiral polarization arms of different
origins (Fig.~\ref{fig:ic342}). In contrast to NGC~6946, there is only a rudimentary magnetic arm in
an interarm region in the north-west, probably because of weaker dynamo action in IC~342
(Sect.~\ref{ma}). A narrow polarization arm of about 300\,pc width, displaced inwards with
respect to the inner arm east of the central region by about 200\,pc, indicates that magnetic
fields are compressed by a density wave, like in M~51. A broad polarization arm of 300--500\,pc
width around the northern optical arm shows systematic variations in polarized
emission, polarization angles and Faraday rotation measures on a scale of about 2\,kpc, indicative
of a helically twisted flux tube generated by the Parker instability (Sect.~\ref{loop}). Several
broad polarization arms in the outer galaxy are coincident with spiral arms in the total neutral gas.

At wavelengths of around 20\,cm, most of the polarized emission from
the far side of the disk and halo is Faraday-depolarized and the
emission from the front side dominates. A striking asymmetry of the
polarized emission occurs along the major axis of 12 spiral galaxies
with inclinations of less than about $60^{\circ}$.
The emission is always much weaker
around the kinematically receding side (positive radial velocities)
of the major axis \citep{urbanik97,braun10,vollmer13}. This asymmetry is still
visible at 11\,cm wavelength, but disappears at smaller wavelengths where the
polarized emission from the far side becomes observable. In strongly inclined
galaxies, both sides of the major axis become Faraday-depolarized at
around 20\,cm. Modeling of a combination of disk and halo fields,
as predicted by $\alpha-\Omega$ dynamo models (Sect.~2), can explain
the asymmetry \citep{braun10}.

At even longer wavelengths, Faraday effects depolarize the synchrotron
emission almost completely. With help of Faraday Synthesis applied to 90\,cm data
from the Westerbork Synthesis Radio Telescope,
an extremely low average degree of polarization of $0.21\pm0.05\%$ was
measured in the star-forming ``ring'' of M~31 \citep{giess13}.
No polarized emission could be detected from M~51 with the Low Frequency Array (LOFAR)
at around 2\,m wavelength -- total Faraday depolarization occurs at such long wavelengths
\citep{mulcahy14}.

\subsection{Spiral pitch angles}
\label{pitch}

\begin{figure*}[t]
\vspace*{2mm}
\begin{minipage}[t]{5.5cm}
\begin{center}
\includegraphics[width=5.25cm]{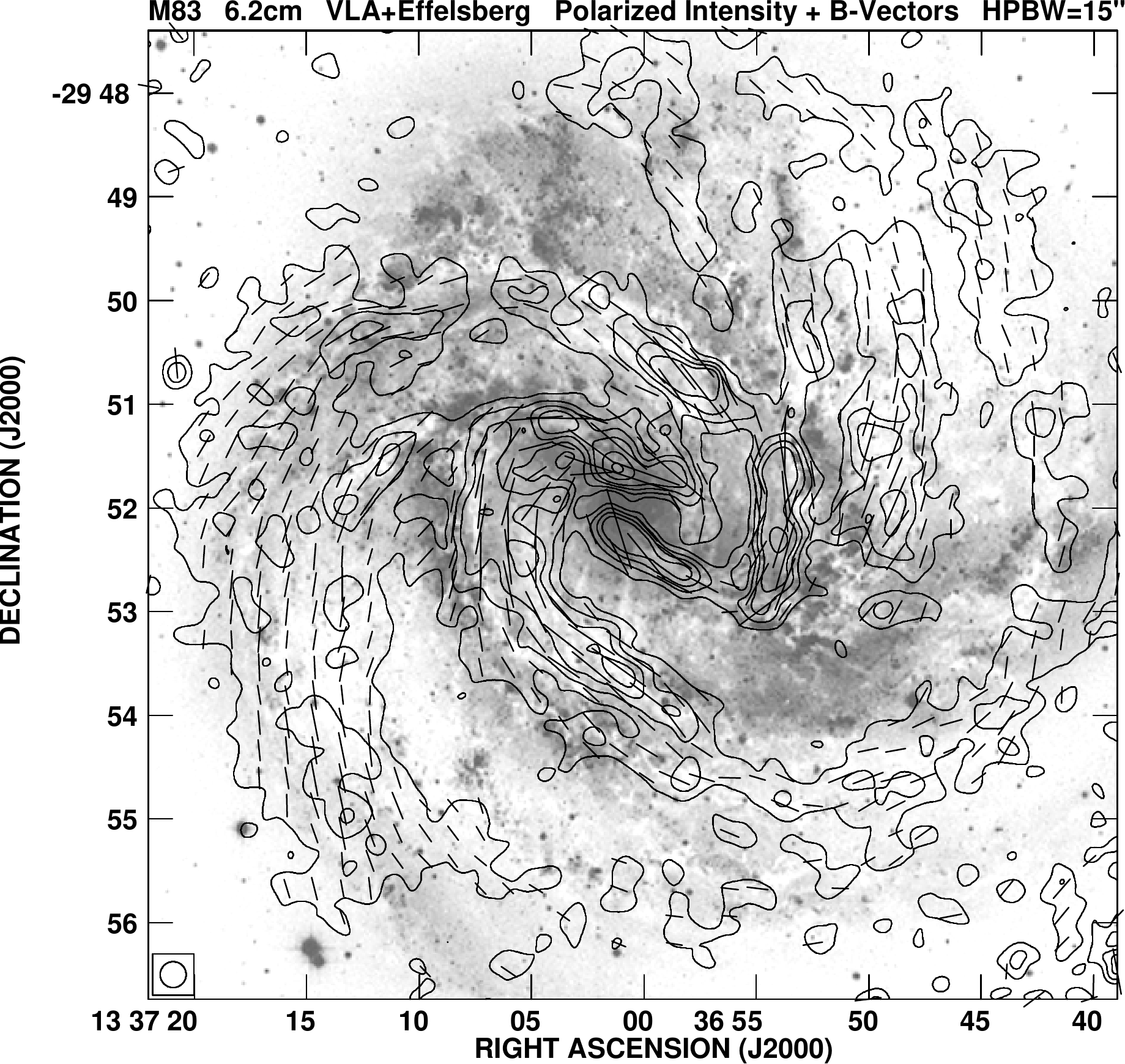}
\caption{Polarized radio emission (contours) and B--vectors of M~83, combined from
observations at 6\,cm wavelength with the VLA and Effelsberg telescopes and smoothed
to $15^{\prime\prime}$ resolution (from \citet{frick16}),
overlaid onto an optical image from Dave Malin, Anglo Australian Observatory.
(Copyright: MPIfR Bonn and AAO)}
\label{fig:m83a}
\end{center}
\end{minipage}\hfill
\begin{minipage}[t]{6cm}
\begin{center}
\includegraphics[width=5.75cm]{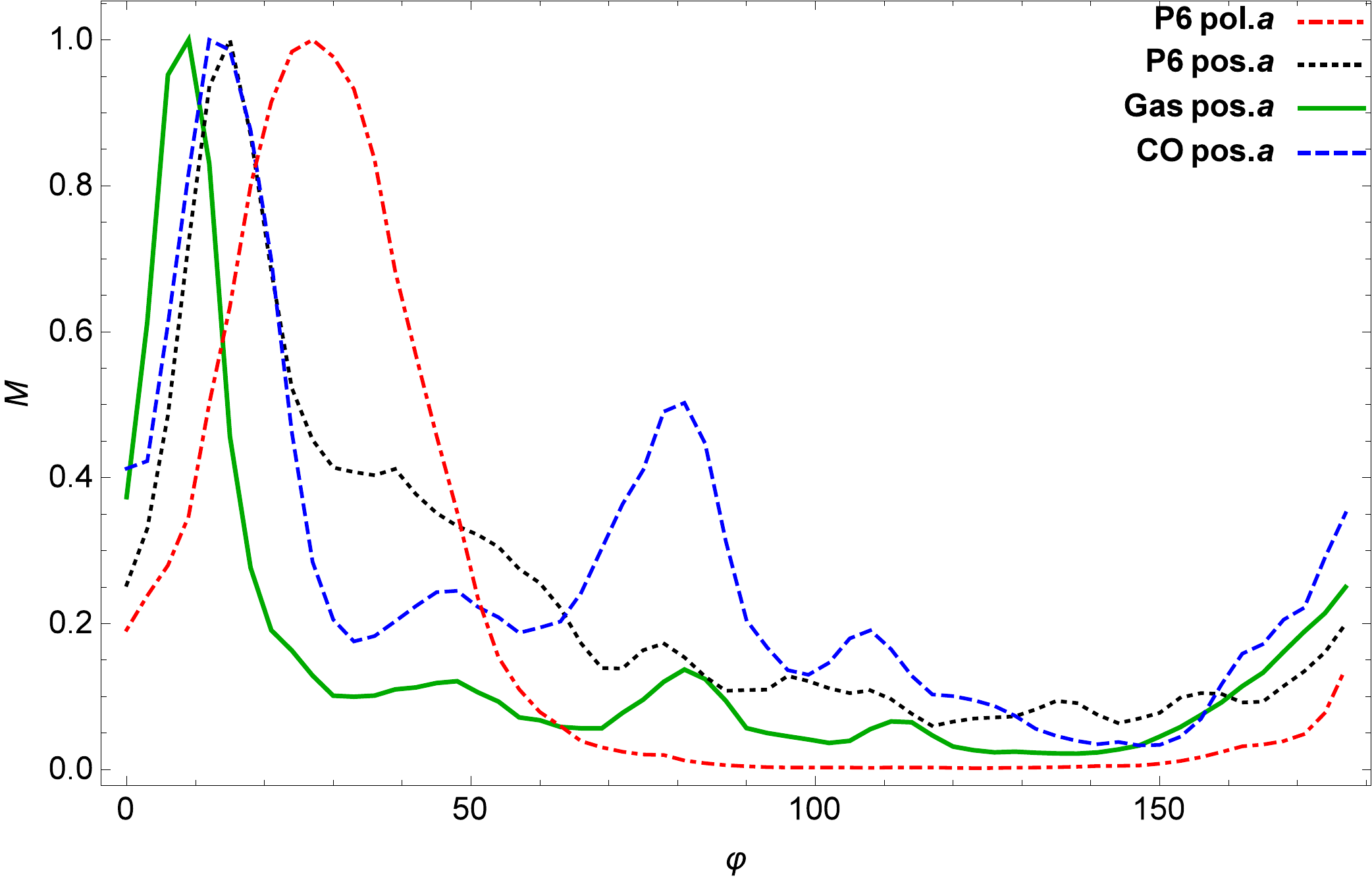}
\caption{The power spectra of pitch angles of structures at the wavelet-filtered
scale $1^{\prime}$ (about 2.6\,kpc) in M~83, normalized to unit maximum: the pitch angles
of the structures of polarized intensity at 6\,cm wavelength (black dotted),
total neutral gas (green) and molecular gas CO (blue dashed). The spectrum of the
magnetic pitch angles is shown with the red dot-dashed curve
(from \citet{frick16}).}
\label{fig:m83b}
\end{center}
\end{minipage}
\end{figure*}

\begin{figure*}[t]
\vspace*{5mm}
\begin{minipage}[t]{5.75cm}
\begin{center}
\centerline{\includegraphics[width=5.5cm]{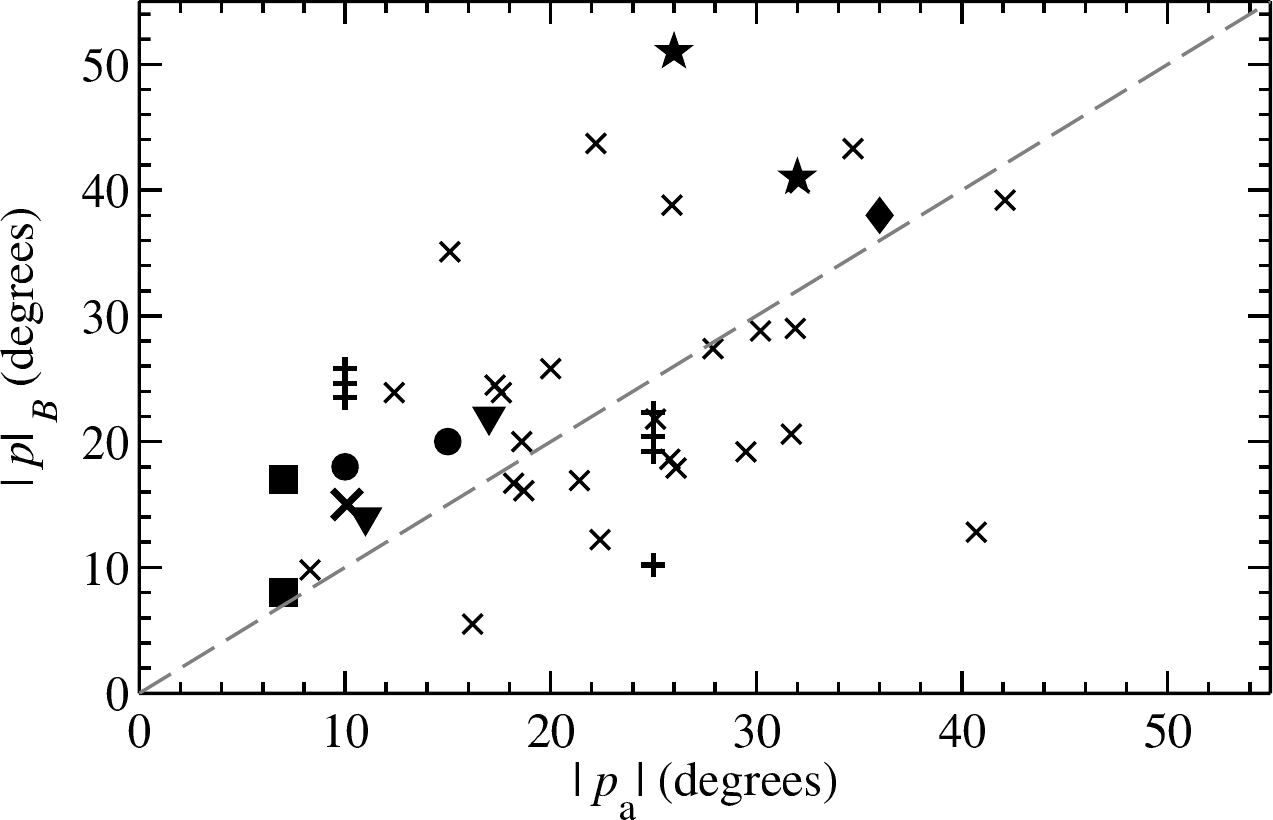}}
\caption{Average magnetic pitch angles $|p_\mathrm{B}|$ and average pitch angles of gaseous spiral
structures $|p_\mathrm{a}|$ in several radial regions of 12 galaxies. The dashed line corresponds to
$|p_\mathrm{B}| = |p_\mathrm{a}|$ (from \citet{vaneck15}).}
\label{fig:pitch}
\end{center}
\end{minipage}\hfill
\begin{minipage}[t]{5.75cm}
\begin{center}
\centerline{\includegraphics[width=5.5cm]{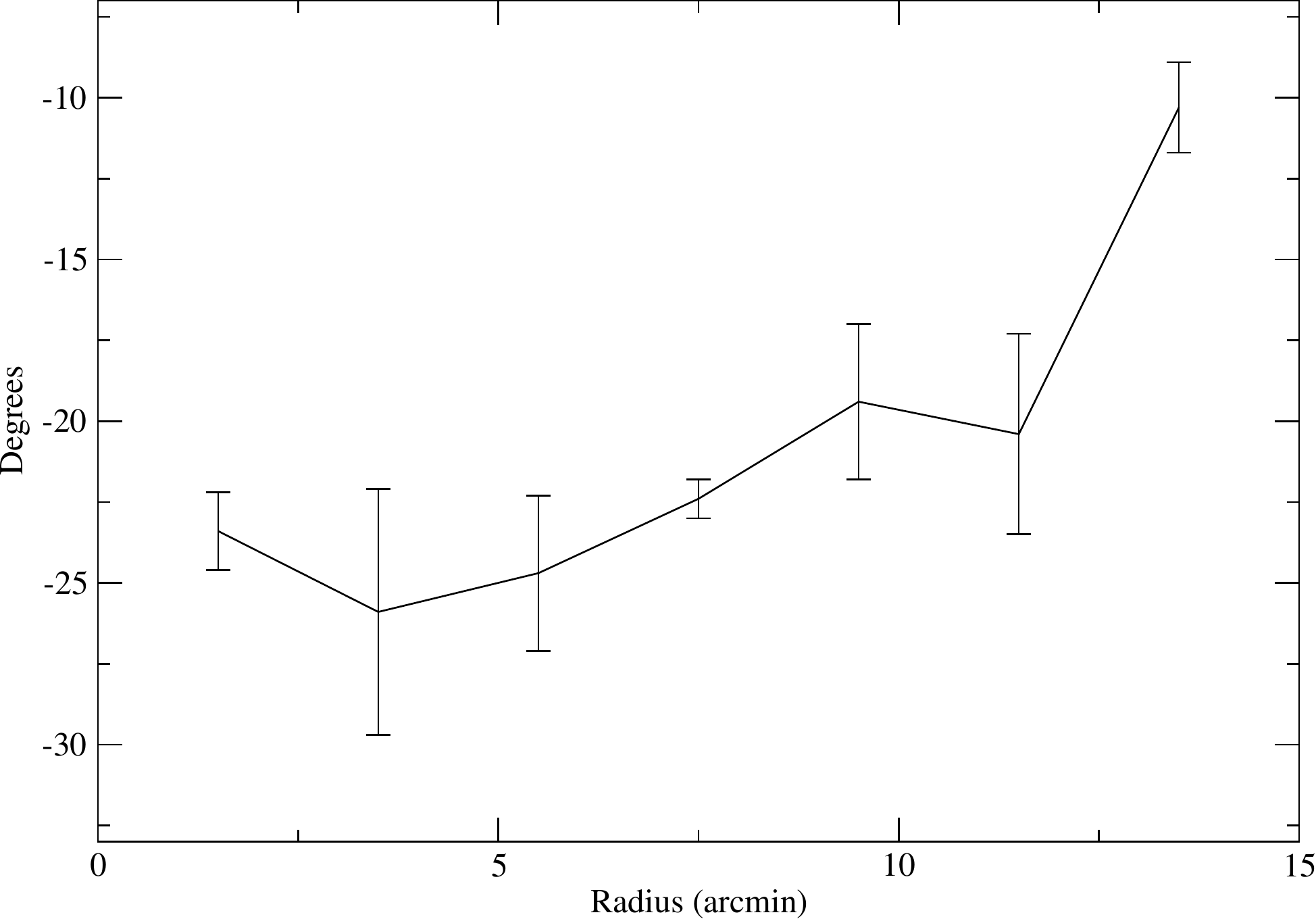}}
\caption{Azimuthally averaged magnetic pitch angle $p_\mathrm{B}$ as a function of radius (in arcminutes)
in the plane of IC~342. $p_\mathrm{B}$ is negative in this galaxy, because the spiral pattern
(see Fig.~\ref{fig:ic342}) turns outward in the clockwise sense (from \citet{beck15}).}
\label{fig:pitchradial}
\end{center}
\end{minipage}
\end{figure*}

In mean-field dynamo theory, the average pitch angle of the spiral magnetic field $p_\mathrm{B}$ and
the average pitch angle of the gaseous spiral arms $p_\mathrm{a}$ are not related, while they are observed
to be roughly similar in a limited sample of galaxies (Fig.~\ref{fig:pitch}). Within galaxies, too,
the magnetic pitch angles are similar to the pitch angles of the gaseous spiral arm structures and
of those of the polarized intensity, tracing the structures of the ordered field (e.g. Fig.~\ref{fig:ic342}).
In IC~342 the magnetic pitch angle $|p_\mathrm{B}|$ decreases with radius (Fig.~\ref{fig:pitchradial}),
following the pitch angle of the gaseous spiral structures \citep{beck15}.

The formation of spiral arms and spiral magnetic patterns seems to be related, as expected e.g. for
MHD density waves \citep{lou99}. Almost identical pitch angles would indicate field alignment with the
gas structures by shear or compression in density waves.
However, based on anisotropic wavelet transformations, a systematic shift was detected in the
barred galaxy M~83 (Fig.~\ref{fig:m83a}), in the sense that $|p_\mathrm{B}|$ is larger than
$|p_\mathrm{a}|$ by about $20^{\circ}$ (Fig.~\ref{fig:m83b}). Similar results were found in
other galaxies. Such differences can be regarded as a signature of $\alpha-\Omega$ dynamo action
(Sect.~\ref{test}).

In the thin-disk approximation of the $\alpha-\Omega$ dynamo, the magnetic pitch angle
is given by $|p_\mathrm{B}| = (R_{\alpha}/R_{\Omega})^{1/2}$, where $R_{\alpha}$ and $R_{\Omega}$ are the dynamo
numbers \citep{shukurov05}. Simplified estimates for the dynamo numbers and a flat rotation curve,
which is valid beyond a radius of about a few kpc in most spiral galaxies, give $|p_\mathrm{B}| = d/H$,
where $d$ is the turbulence scale (about 50\,pc, see \citet{fletcher11}), and $H$ is the scale height
of the ionized gas. The scale heights of $\HI$ disks in spiral galaxies increase radially
\citep{bagetakos11}. One may speculate that the disk of ionized gas also flares
\citep[e.g. assumed in the model by][]{gressel13}, which would explain the decrease of $|p_\mathrm{B}|$
observed in many galaxies. If so, the radial variation of $|p_\mathrm{B}|$ in IC~342
(Fig.~\ref{fig:pitchradial}) indicates that the scale height $H$ is approximately constant
until $12^{\prime}$ (about 12\,kpc) radius and then increases.
Alternatively, the magnetic pitch angle may be affected by gas flows, for instance by outflows
that become weaker towards the outer disk and can decrease the effective dynamo number \citep{shukurov06}.

\subsection{Regular fields}
\label{large}

Spiral fields can be generated by compression in spiral arms, by shear in interarm regions or by
dynamo action (Sect.~\ref{dynamo}). Measuring Faraday rotation (Sect.~\ref{rm}) is crucial to distinguish
between these mechanisms. Large-scale sinusoidal patterns of Faraday rotation measures (RM) along azimuthal
angle in the galaxy plane ({\em modes}) are signatures of regular fields generated by the $\alpha-\Omega$
dynamo and can be identified in RM derived from polarized emission from the galaxy disks at
several frequencies \citep{krause90} or in RM data of polarized background sources
\citep{stepanov08}. If several modes are superimposed, a Fourier analysis of the RM variation
with azimuthal angle is needed. The resolution and sensitivity of present-day radio observations is
sufficient to identify 2--3 modes.

\begin{figure*}[t]
\vspace*{2mm}
\begin{center}
\includegraphics[width=10.5cm]{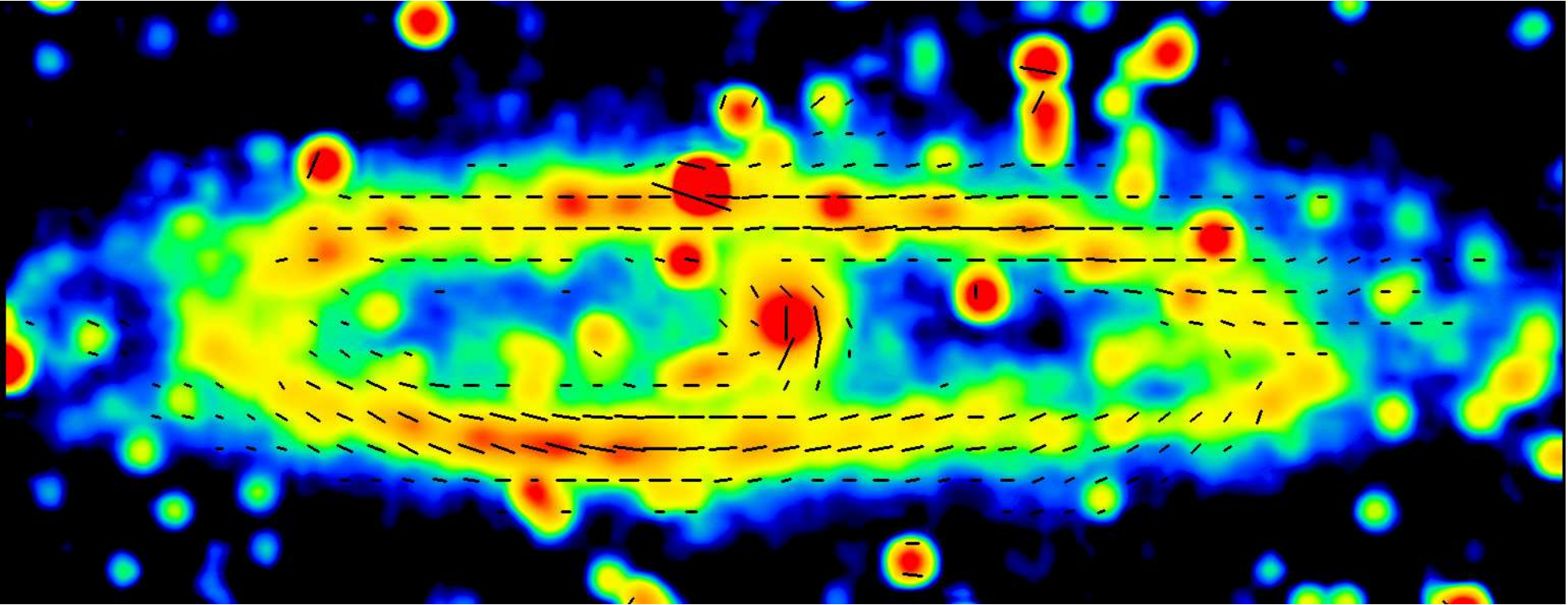}
\caption{Total radio emission (colour) and B--vectors (corrected for Faraday rotation
in the foreground of the Milky Way) in the Andromeda galaxy (M~31), observed at
6\,cm wavelength with the Effelsberg telescope at $3^{\prime}$ resolution (from \citet{giess12}).
(Copyright: MPIfR Bonn)}
\label{fig:m31}
\end{center}
\end{figure*}

\begin{figure*}[t]
\vspace*{2mm}
\begin{center}
\includegraphics[width=10.5cm]{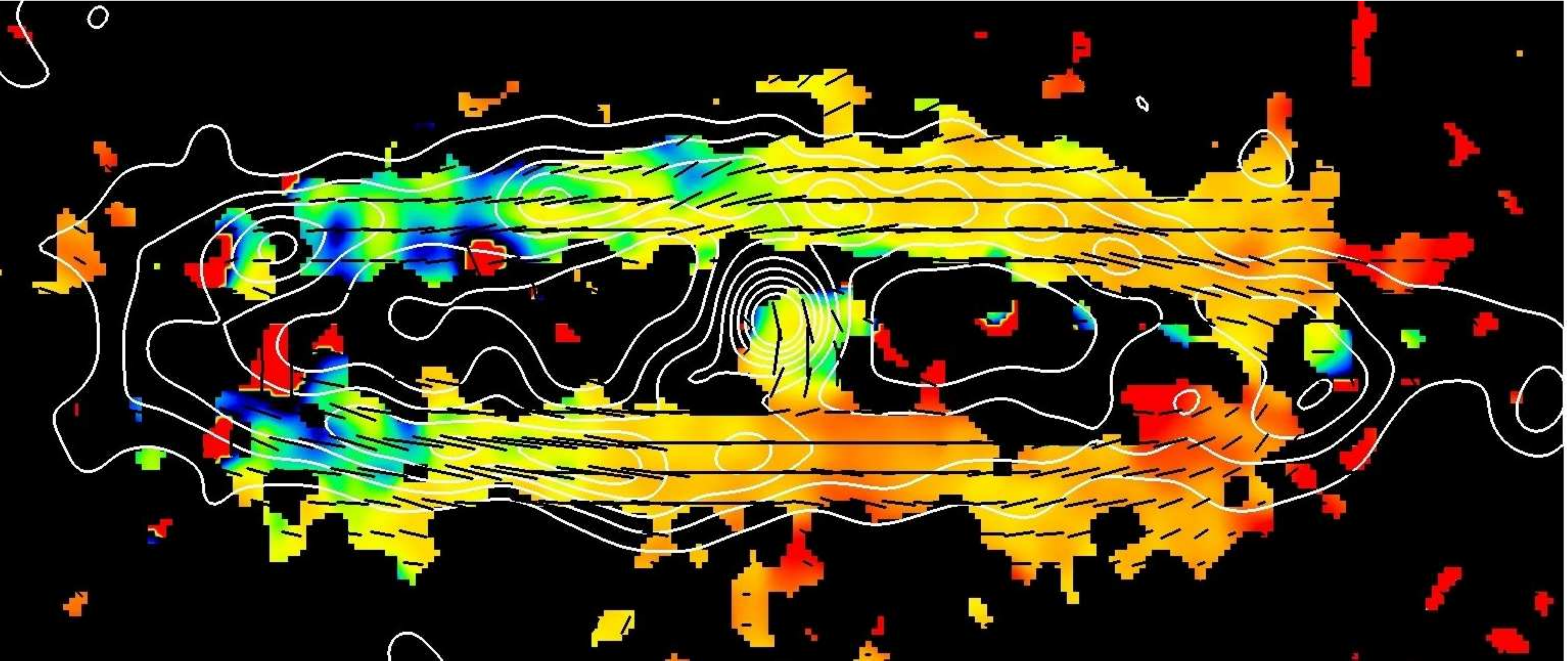}
\caption{Faraday rotation measures between 6\,cm and 11\,cm wavelengths (colour, from blue=-175\,rad\,m$^{-2}$ to
red=+25\,rad\,m$^{-2}$), total radio emission at 6\,cm (contours) and B--vectors (corrected
for Faraday rotation) in the Andromeda galaxy (M~31), observed with the Effelsberg telescope
at $5^{\prime}$ resolution (from \citet{berkhuijsen03}). (Copyright: MPIfR Bonn)}
\label{fig:m31rm}
\end{center}
\end{figure*}

The disks of about a dozen nearby spiral galaxies reveal large-scale RM patterns. The Andromeda
galaxy M~31 (Fig.~\ref{fig:m31}) is the prototype of a dynamo-generated axisymmetric spiral disk
field, with a striking sinusoidal RM variation along the star-forming ring (Fig.~\ref{fig:m31rm}),
which is a strong indication of an axisymmetric spiral field (mode $m = 0$) \citep{fletcher04}.
Other candidates for a dominating axisymmetric disk field are the nearby spirals IC~342
\citep{sokoloff92,beck15} and NGC~253 \citep{heesen09b}. The axisymmetric field in the irregular Large
Magellanic Cloud (LMC) is almost azimuthal (i.e. small pitch angles) \citep{gaensler05,mao12}. Dominating
bisymmetric spiral fields (mode $m = 1$) are rare, as predicted by dynamo models, but possibly exist
in M~81 \citep{krause89b,sokoloff92}. Faraday rotation in NGC~6946 and in other similar
galaxies with magnetic arms can be described by a superposition of two azimuthal modes
($m = 0$ and $m = 2$) with about equal amplitudes, where the quadrisymmetric ($m = 2$) mode
is phase shifted with respect to the density wave \citep{beck07}. For several other galaxies,
three modes ($m = 0$, 1 and 2) are necessary to describe the data (Table~\ref{table:modes}).

\begin{table}[h!]
\caption{Decomposition of regular magnetic fields in galaxies into azimuthal modes of order $m$.
Columns $2-4$ give the approximate amplitudes relative to the dominating mode. A dash indicates
modes that were not sought, whereas a zero indicates that no signature of this mode
was found. Modes of order higher than 2 cannot be detected with present-day telescopes.
This is an updated version of a table in \citet{fletcher10}.}
\centering
\begin{tabular}{l l l l l}
\hline\hline
Galaxy 	& $m=0$ &  $m=1$ & $m=2$ & Reference\\
\hline
M~31        & 1 & 0   & 0   & \citet{fletcher04}\\
M~33        & 1 & 1   & 0.5 & \citet{taba08}\\
M~51 (disk) & 1 & 0   & 0.5 & \citet{fletcher11}\\
M~51 (halo) & 0 & 1   & 0   & \citet{fletcher11}\\
M~81        & 0.5 & 1 & --  & \citet{sokoloff92}\\
NGC~253     & 1 & --  & --  & \citet{heesen09b}\\
NGC~1097    & 1 & 1   & 1   & \citet{beck05c}\\
NGC~1365    & 1 & 1   & 1   & \citet{beck05c}\\
NGC~4254    & 1 & 0.5 & --  & \citet{chyzy08}\\
NGC~4414    & 1 & 0.5 & 0.5 & \citet{soida02}\\
NGC~6946    & 1?& --  & 1?  & \citet{ehle93,rohde99}\\
IC~342      & 1 & --  & --  & \citet{sokoloff92,beck15}\\
LMC         & 1 & --  & --  & \citet{gaensler05,mao12}\\
\hline
\label{table:modes}
\end{tabular}
\end{table}

The spiral pattern of magnetic fields cannot be solely the
result of $\alpha-\Omega$ dynamo action. If the beautiful spiral pattern
of M~51 seen in radio polarization (Fig.~\ref{fig:m51}) were only
due to a regular field in the disk, its line-of sight component
should generate a conspicuous large-scale pattern in RM, but this is
not observed \citep{fletcher11}. This means that a large amount
of the ordered field is {\em anisotropic turbulent}\ and probably generated by
compression and shear of the non-axisymmetric gas flows in the
density-wave potential. The anisotropic turbulent field is strongest at the
positions of the prominent dust lanes on the inner edge of the inner
gas spiral arms, due to compression of turbulent fields in the
density-wave shock. A regular field (composed of modes $m = 0$ and $m = 2$)
also exists in the disk of M~51, but is much weaker than the
anisotropic turbulent field \citep{fletcher11}. Notably, RMs between 18\,cm
and 22\,cm wavelengths, tracing only the polarized emission from regions nearest to the observer,
show a large-scale pattern that indicates a regular field in the halo
of M~51 \citep{heald+09,fletcher11,mao15}.

The large-scale regular field in the nearby galaxy IC~342 is similarly weak as in M~51.
It is only visible in the pattern of RMs obtained from the low-resolution Effelsberg images
\citep{beck15}, while RMs derived from the higher-resolution VLA images (Fig.~\ref{fig:ic342rm})
are about 10\,times larger. The field direction changes along the northern spiral arm
(Sect.~\ref{loop}).

The central regions of M~31 (Fig.~\ref{fig:m31}) and IC~342 (Fig.~\ref{fig:ic342central}) host
regular spiral fields that are disconnected from the disk fields \citep{giess14,beck15}.
As the direction of the radial field component points outwards, opposite to that of the disk
field, two separate dynamos seem to operate in these galaxies.

In the disks of many other galaxies no clear pattern of Faraday
rotation was found. Either several high-order modes are
superimposed and cannot be distinguished with the limited
sensitivity and resolution of present-day telescopes, or the
timescale for the generation of large-scale modes is longer than the
galaxy's lifetime \citep{arshakian09}. Also field injection by strong
star-formation activity may perturb the generation of a large-scale
regular field \citep{moss12}.

While the azimuthal symmetries of regular magnetic fields have been well observed in many
galaxies (Table~\ref{table:modes}), the vertical symmetry (even or odd) is much harder to
determine. The field of odd-symmetry modes reverses its sign above and below the galactic plane.
The symmetry type is best visible in strongly inclined galaxies, via the RM signs above and
below the galaxy plane. In mildly inclined galaxies the RMs of diffuse polarized emission
from even and odd-symmetry fields differ by a factor of about two, which is hard to measure,
while the RMs of background sources are close to zero for even symmetry
(because the RM contributions from the two halves cancel) but large for odd symmetry.
Background RMs in the area of the Large Magellanic Cloud (LMC) indicate an even-symmetry
field \citep{mao12}. Indications for even-symmetry patterns were found in NGC~891
\citep{krause09} and NGC~5775 \citep{soida11}.


Spectro--polarimetric data of spiral galaxies are still rare and the application of Faraday Synthesis
(Sect.~\ref{rm}) has just started \citep[e.g.][]{giess13,mao15}. The Faraday spectrum reflects the
distribution of regular magnetic fields, thermal gas and CREs and in principle allows tomography of
the ISM in the disk and the halo.
However, Faraday spectra do not have a simple correspondence to physical properties and cannot provide
a straight-forward interpretation, as demonstrated by the models of \citet{ideguchi14}.

\subsection{Testing dynamo models}
\label{test}

Observations provide several qualitative evidences for the action of dynamos in spiral galaxies:\\
(1) The tight radio--IR correlation (Sect.~\ref{RIC}) and the similarity of kinetic and
magnetic energy densities (Fig.~\ref{fig:ic342_energies}) indicate field amplification by turbulent
gas motions generated by star-forming processes \citep{taba13a,schleicher13}.\\
(2) The magnetic spiral patterns observed in all spiral galaxies so far indicate a general decoupling
between magnetic fields and the (almost circular) gas flow due to magnetic diffusivity. The magnetic
pitch angle deviates systematically from that of the spiral structures (e.g. in M~83,
Fig.~\ref{fig:m83b}), as predicted by the $\alpha-\Omega$ dynamo. Magnetic spiral patterns also exist
in flocculent galaxies (Sect.~\ref{irr}). At present, no other model can explain the magnetic spiral
patterns in the various types of galaxies.\\
(3) Large-scale regular fields are observed in all spiral galaxies so far (Table~\ref{table:modes}),
an important result giving a strong hint to the action of $\alpha-\Omega$ dynamos. No alternative model
exists so far.\\
(4) By comparing the signs of the RM distribution and the velocity field
on both sides of a galaxy's major axis, the inward and outward
directions of the radial component of the axisymmetric spiral field
can be distinguished \citep{krause98}. Dynamo models predict that
both signs should have the same probability, which is confirmed by
observations. The axisymmetric fields of M~31, IC~342, NGC~253, and
the axisymmetric field component in NGC~6946 point inwards, while
those of NGC~891, NGC~4254 and NGC~5775 \citep{krause09}, and the axisymmetric
component of the disk field in M~51 \citep{fletcher11} point outwards.

A major effort of quantitative comparisons between observable quantities and predictions was performed
by \citet{vaneck15}. Identifying the precise form of the connection between magnetic and other
galactic properties is not straightforward because the parameters can combine to produce
non-trivial scalings. For example, different physical mechanisms have been proposed to provide
saturation of the field growth in dynamo models, like the balance between magnetic and kinetic
(turbulent) forces, the balance between Coriolis and Lorentz forces, or outflows of hot gas carrying
small-scale magnetic fields out of the disk in order to preserve the balance of magnetic helicity.
These non-linear saturation mechanisms depend differently on galactic parameters, such as star-formation
rate, gas density, rotational velocity, and rotational shear.

From their sample of 20 well-observed galaxies, \citet{vaneck15} found a statistically significant
relation of the total magnetic field strength with the surface density of molecular gas and
surface density of the star-formation rate, confirming earlier results (Sect.~\ref{RIC}).
The only other significant relation was found between the magnetic pitch angle and the strength
of the axisymmetric spiral ($m=0$) component of the regular field: a more tightly wound field
has a stronger ASS component. However, IC~342 does not fit into this relation; has a tightly wound
spiral field (Fig.~\ref{fig:pitchradial}), but only a weak axisymmetric component \citep{beck15}.

In conclusion, dynamos certainly operate in all spiral galaxies, but other processes are amplifying
and shaping the field as well. MHD density waves are compressing and aligning the field along the gaseous
spiral arms. Non-axisymmetric gas flows around spiral arms and bars are shearing field lines.
Parker instabilities form helically twisted field loops that are winding around spiral arms.
To measure the importance of these competing or cooperating effects, the fundamental
scaling relations between the properties of magnetic fields and the other galactic parameters need
to be determined, based on radio observations of a large number of galaxies with high angular resolution.

\subsection{Magnetic arms}
\label{ma}

\begin{figure*}[t]
\vspace*{2mm}
\begin{center}
\includegraphics[width=8cm]{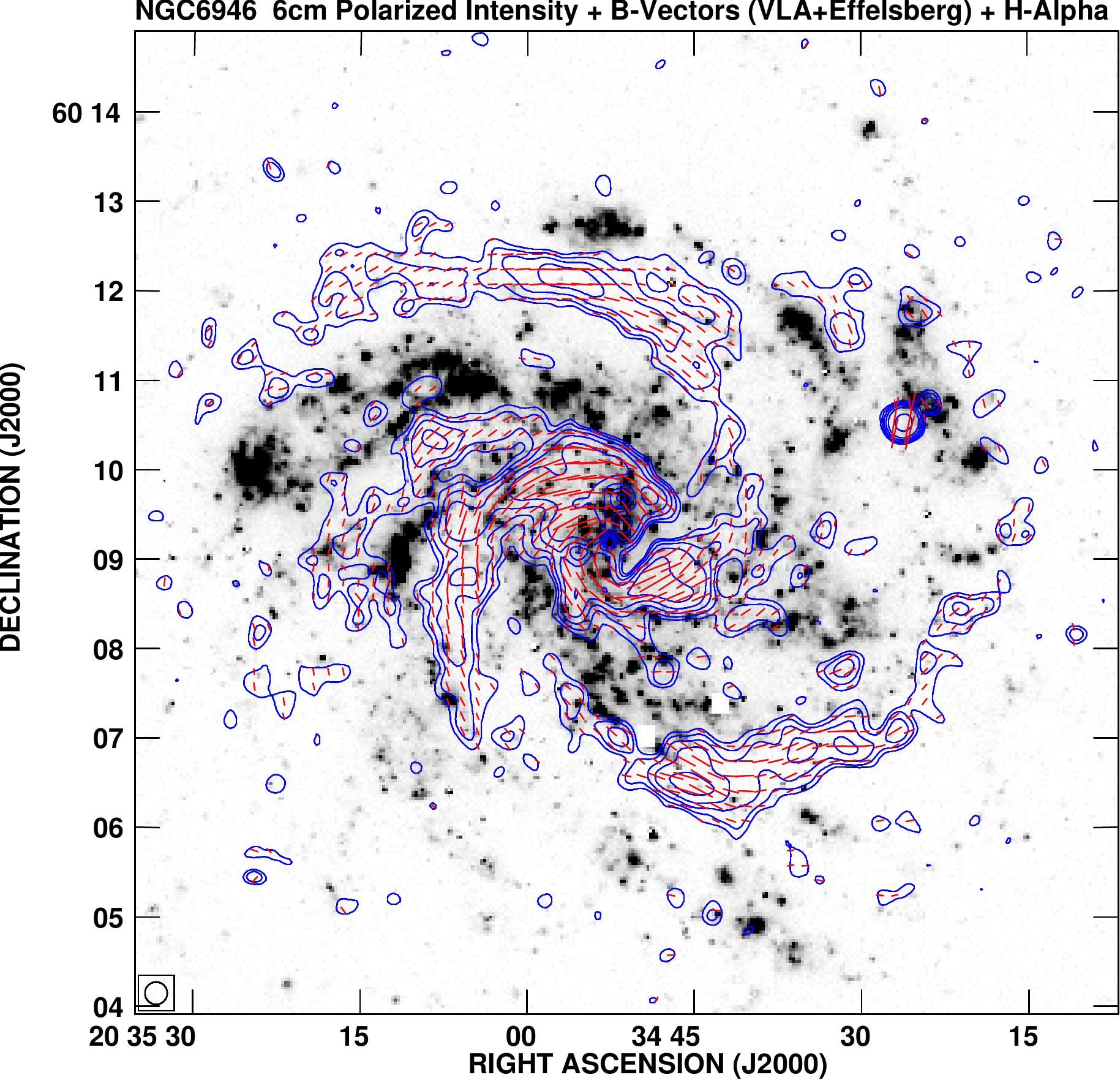}
\caption{Polarized radio emission (contours) and B--vectors of
NCC~6946, combined from observations at 6\,cm wavelength with the
VLA and Effelsberg telescopes and smoothed to $15^{\prime\prime}$ resolution
(from \citet{beck07}), overlaid onto an H$\alpha$ image from Anne Ferguson.
(Copyright: MPIfR Bonn)}
\label{fig:n6946}
\end{center}
\end{figure*}

Surprisingly, in most spiral galaxies observed so far the highest polarized intensities
(i.e. the strongest ordered fields) are
detected {\em between}\ the optical arms, filling a large fraction of the interarm space, sometimes
concentrated in magnetic arms, as in NGC~6946 (Fig.~\ref{fig:n6946}). The southern-sky spiral galaxy
NGC~2997 hosts compressed magnetic fields at the inner edges of material arms, as well as one well-developed
magnetic arm \citep{han99}. The strong density waves in NGC~2997 lead to high degrees of polarization
at 6\,cm wavelength of typically 25\% at the inner edge of the northern arm, similar to those
in M~51 \citep{fletcher11}. The degree of polarization of the magnetic arm of 40\% and its length of at least
10\,kpc are similar to the values in NGC~6946 and are much higher than in IC~342 \citep{beck15}.
Like IC~342, M~51 has a short, rudimentary magnetic arm with a low degree of polarization \citep{fletcher11}.
NGC~1566 shows signatures of a magnetic arm in the south-east \citep{ehle96},
but the angular resolution of these radio observations was too coarse to clearly detect magnetic arms.
In large barred galaxies, ordered magnetic fields also fill most of the interarm space, e.g. in NGC~2442
\citep{harnett04}, NGC~1097, and NGC~1365 \citep{beck05c}, but do not form well-defined magnetic arms.

Several mechanisms have been proposed to explain the high degree of field order in interarm regions:\\
(1) Magnetic field ropes as a result of a magnetic buoyancy instability in a turbulent high--$\beta$ plasma
\citep{kleeorin90}; however, the ISM of galaxies is a low--$\beta$ plasma,
meaning that the thermal pressure is lower than the magnetic pressure (Sect.~\ref{energy});\\
(2) slow MHD density waves in the rigidly rotating inner region of a galaxy \citep{lou99,poedts02}; however,
slow MHD density waves in 3D can be subject to the Parker and shearing instabilities \citep{foglizzo95};\\
(3) more efficient action of the mean-field dynamo between the optical
arms due to lower turbulent velocity in interarm regions \citep{moss98a,shukurov98};
however, the observed turbulent velocity is {\em \emph{not}}\ lower in the interarm regions
(e.g. in IC~342, \citet{crosthwaite01});\\
(4) introduction of a relaxation time of the magnetic response in the dynamo equation, leading to a phase
shift between the material and magnetic spiral arms \citep{chamandy13a,chamandy13b}; however, the resulting
magnetic arms are restricted to a relatively small region around the corotation radius and have a much smaller
pitch angle than the gaseous arms;\\
(5) drift of magnetic fields with respect to the gaseous arms in a non-axisymmetric gas flow caused by a spiral
perturbation \citep{otmian02} or by a bar \citep{kulpa11};\\
(6) weakening of the mean-field dynamo in the material arms by continuous injection and amplification of
turbulent fields by supernova shock fronts \citep{moss13,moss15};\\
(7) weakening of the mean-field dynamo in the material arms by star-formation-driven outflows
\citep{chamandy15};\\
(8) magnetic arms as a transient phenomenon during the evolution of galactic magnetic fields,
possibly related to the short lifetimes of spiral patterns seen in numerical simulations
\citep[e.g.][]{dobbs14}.

Models (5) -- (8) are promising. Still, these models are simplified and consider either
gravitational perturbations or dynamo action.
Self-consistent MHD models of galaxies including the gravitational potential
with spiral perturbations and mean-field dynamo action are still missing.

The origin of the rudimentary magnetic arm in IC~342 (Fig.~\ref{fig:ic342}) deserves a detailed discussion.
The absence of long magnetic arms in IC~342 is surprising in view of the apparent similarities between
IC~342 and NGC~6946, like the rotation curve and the star-formation surface density $\Sigma_\mathrm{SFR}$ \citep{calzetti10}.
The evolution of a spiral magnetic pattern in IC~342 may be hampered by the possible tidal interaction
with the Local Group \citep{buta99} and/or by the complex spiral pattern in the gas of IC~342, described by a
superposition of a two- and four-armed spiral pattern with different pattern speeds \citep{meidt09}, which may
lead to an unstable spiral pattern. The mean-field dynamo needs a least a few rotation
periods to build up a regular field \citep[e.g.][]{moss12}. A short lifetime of a stable pattern may not allow
the formation of magnetic arms.

The spiral pattern in the gas of M~51 is also distorted and short-lived, so that only a weak large-scale
regular field and a rudimentary magnetic arm could develop. On the other hand, NGC~6946 has a two-armed
spiral pattern with a well-defined pattern speed \citep{fathi07}, the spiral field extends smoothly
into the central region \citep{beck07} and the large-scale regular field is strong \citep{ehle93}.
In summary, {\em \emph{the existence of magnetic arms may indicate a stable spiral arm pattern over
several galactic rotation periods,}} which may not be given in IC~342.

Most models of spiral gravitational perturbations in stars and gas have neglected the effect of magnetic
fields so far. On the other hand,
the only dynamo model for the amplification and ordering of magnetic fields including the
spiral perturbations of gas density and gas velocity is the kinematical model by \citet{otmian02}, in which
the back reaction of the field onto the gas flow was not included. The MHD model of \citet{pakmor13}
includes self-gravity and spiral perturbations, but no mean-field dynamo action. There is urgent need for
a synergy between these approaches to achieve a comprehensive description of the evolving magnetized
ISM in galaxies.


\subsection{Helically twisted Parker loops}
\label{loop}

\begin{figure*}[t]
\vspace*{5mm}
\begin{center}
\includegraphics[width=7cm]{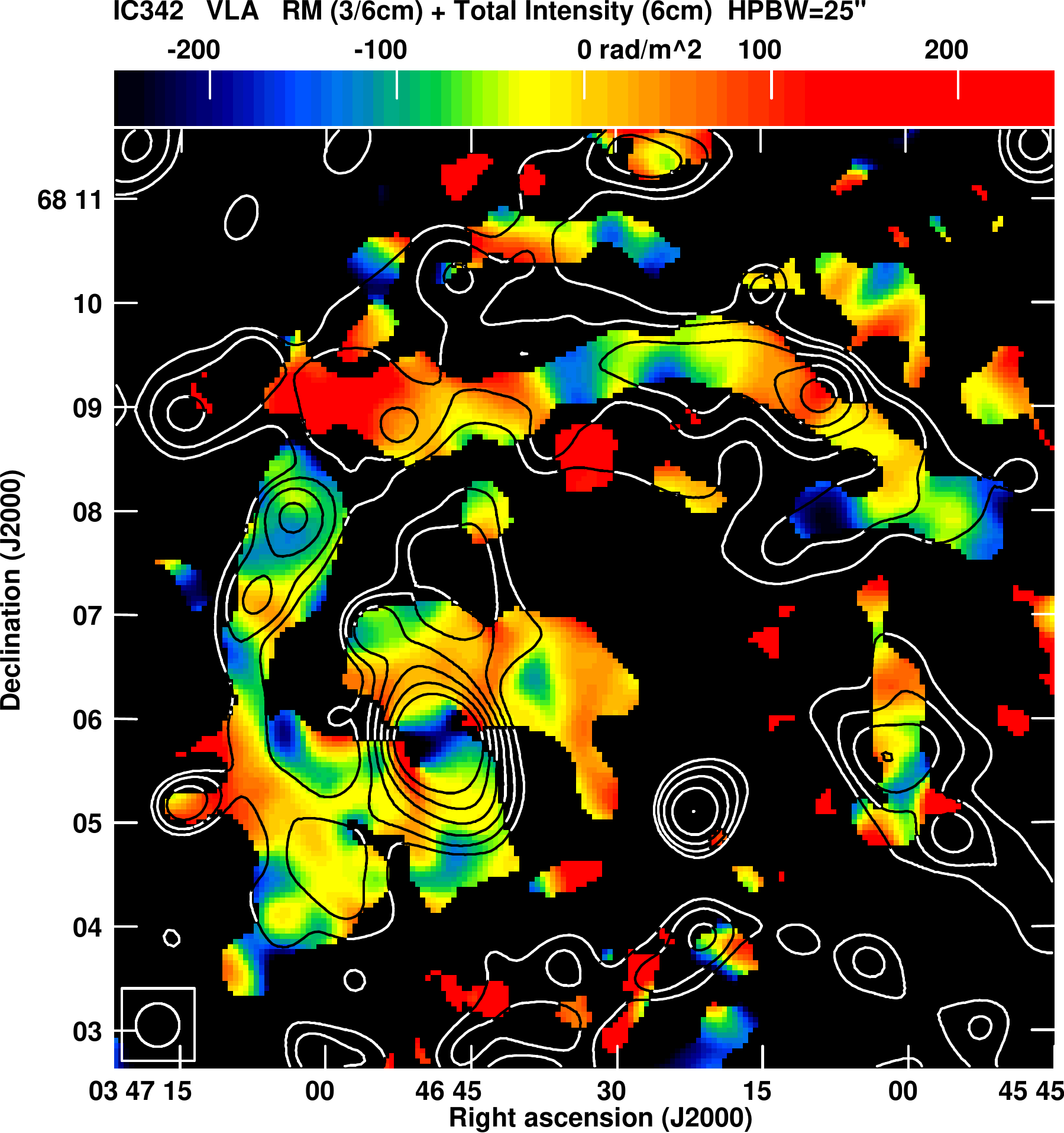}
\caption{Faraday rotation measures (RM) between 3.5\,cm and 6\,cm wavelengths (colour) and total emission
at 6\,cm (contours) in the central and northern regions of IC~342, observed with the VLA
at $25^{\prime\prime}$ resolution. RM is shown only at points where the polarized intensities
at both wavelengths exceeds 3\,times the rms noise (from \citet{beck15}).
}
\label{fig:ic342rm}
\end{center}
\end{figure*}

IC~342 is one of the best studied nearby spiral galaxies. Its small distance and bright radio disk allows
detailed studies of the magnetic field that are still impossible in most other galaxies.

Faraday rotation measures (Fig.~\ref{fig:ic342rm}) are found to vary periodically
along the most prominent northern radio spiral arm. The average distance between the extrema corresponds to about
2.1\,kpc. This feature can be interpreted as a magnetic loop (Parker instability) extending out
of the galaxy's disk and out of the sky plane, giving rise to a periodic pattern in RM.
A regular field bending out of the plane should lead to minima in polarized intensity (tracing the field
component in the sky plane) at locations where $|RM|$ (tracing the field component along the line of sight)
is at its maximum. Minima of polarized intensity occur roughly at the locations of the
extrema in RM, with a similar distance between the maxima or minima of about 2.2\,kpc.

This is the second indication of a Parker loop in the magnetic field of a nearby galaxy,
after M~31 \citep{beck89}.
The numerical models by \citet{kim02} (who assume $\beta=1$) predict a wavelength of the most unstable
symmetric mode between $4\,\pi\,H$ and $17\,H$, where $H$ is the scale height of the gas. The peak-to-peak
wavelength of about 4\,kpc measured in IC~342 corresponds to $H\approx 230-320$\,pc, which is larger than
typical scale heights of $\HI$ gas disks of spiral galaxies \citep{bagetakos11}. The reason for this
discrepancy may be that the assumption of $\beta=1$ is incorrect, because observations indicate $\beta<1$
(Sect.~\ref{energy}).

The magnetic field in the northern arm of IC~342 diverts not only in the vertical direction, but also in
the disk plane. The ridge line of the polarization spiral arm as well as the magnetic pitch angle
(Fig.~\ref{fig:ic342}) oscillate around the northern arm with similar periodicity. This gives
indication for a large-scale, helically twisted flux tube, as predicted by models of the Parker loops
under the Coriolis force \citep{shibata91,hanasz02}.

In the galaxy NGC~7479 (Fig.~\ref{fig:n7479}), two jets serve as bright polarized background sources.
High-resolution observations revealed several reversals in RM on $1-2$\,kpc scale, originating in the
foreground disk of the galaxy \citep{laine08}, which may represent another case of a helically twisted
field loop.

\subsection{Large-scale field reversals}
\label{reversals}

\begin{figure*}[t]
\vspace*{2mm}
\begin{center}
\includegraphics[width=8cm]{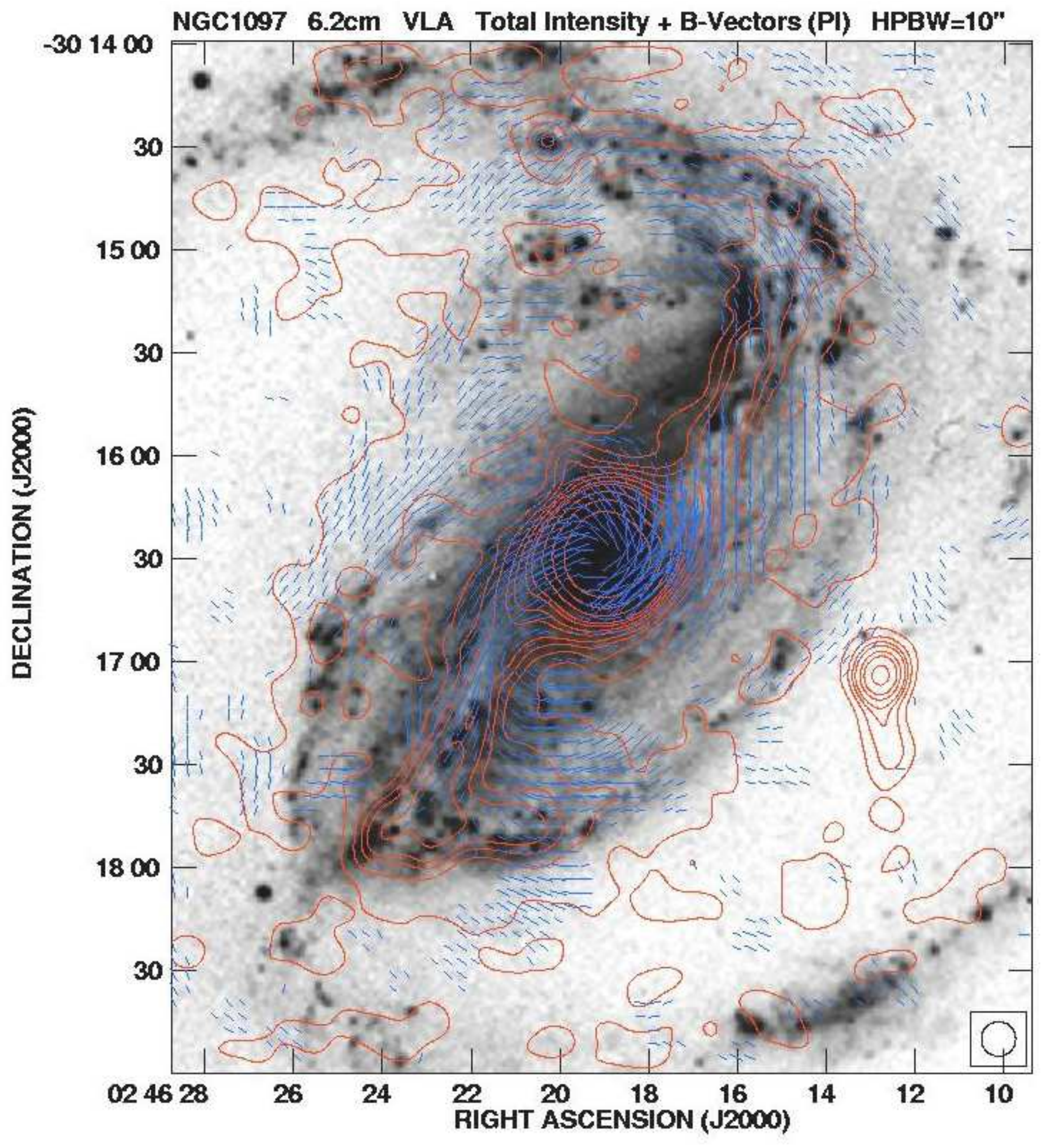}
\caption{Total radio emission (contours) and B--vectors of the
barred galaxy NGC~1097, observed at 6\,cm wavelength with the VLA
and smoothed to $10^{\prime\prime}$ resolution (from \citet{beck05c}).
The background optical image is from Halton Arp.}
\label{fig:n1097}
\end{center}
\end{figure*}

A large-scale field reversal exists between the central region
and the disk in M~31 \citep{giess14} and in IC~342 \citep{beck15}.
A large-scale field reversal at about constant azimuthal angle across the whole galaxy disk
was found in NGC~4414 \citep{soida02}. To fulfil the divergence-free condition,
this reversal cannot occur in the same volume, but e.g. between disk and halo.
Indication for a large-scale field reversal between disk and halo was also found
in M~51 \citep{berkhuijsen97,fletcher11,mao15}. New observations and applying Faraday Synthesis
(Sect.~\ref{rm}) are needed.

Large-scale field reversals between spiral arms, like that observed in the Milky Way
(Fig.~\ref{fig:Galaxy2}), have not been detected in external galaxies, although
high-resolution images of Faraday rotation are available for many spiral galaxies.
Possible reasons are discussed in Sect.~\ref{MW}.

\subsection{Barred galaxies}
\label{bar}

\begin{figure*}[t]
\vspace*{2mm}
\begin{minipage}[t]{6.25cm}
\begin{center}
\includegraphics[width=6cm]{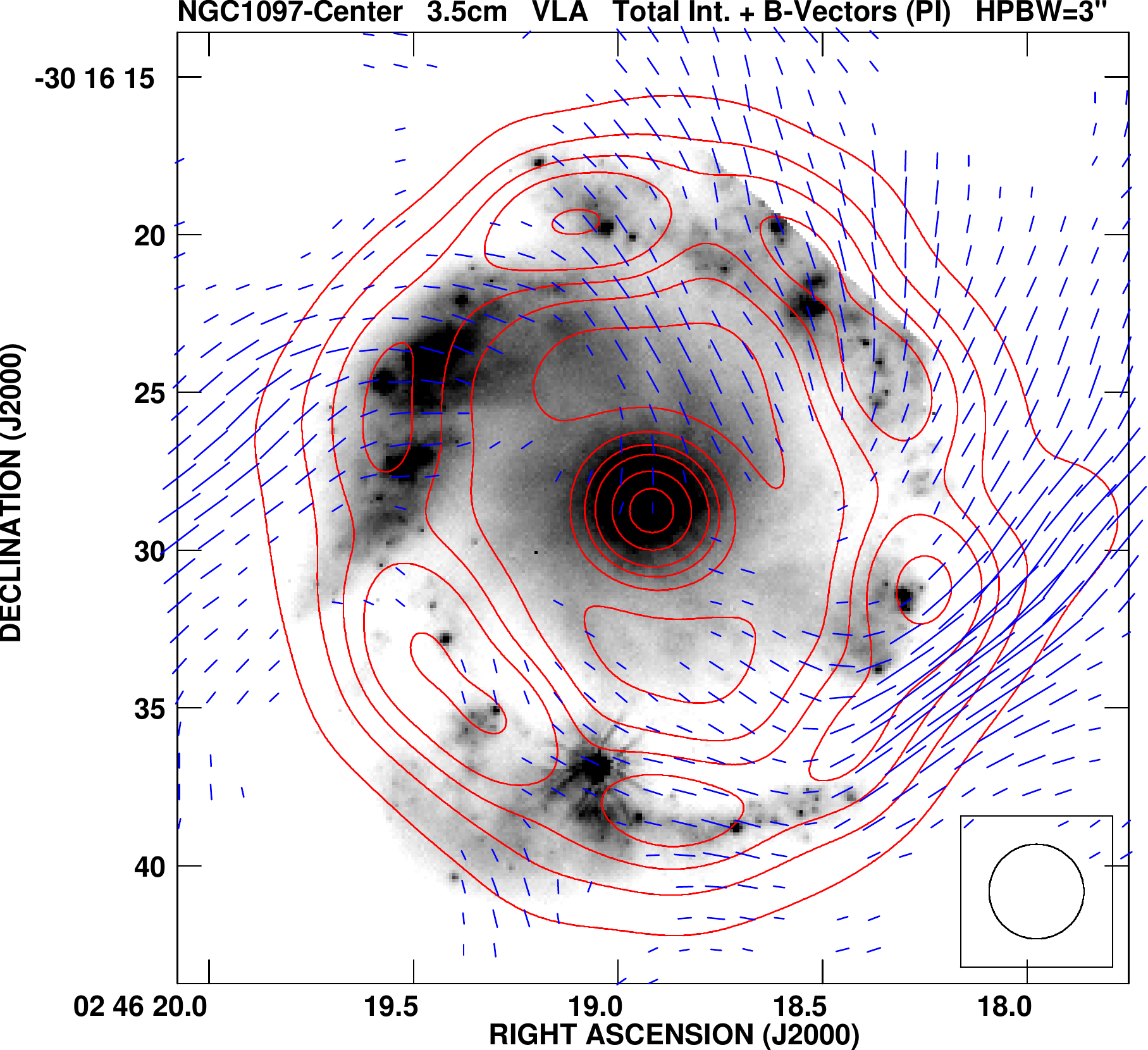}
\caption{Total radio emission (contours) and B--vectors in the
circumnuclear ring of the barred galaxy NGC~1097, observed at 3.5\,cm
wavelength with the VLA at $3^{\prime\prime}$ resolution. The background
optical image is from the HST (from \citet{beck05c}).}
\label{fig:n1097central}
\end{center}
\end{minipage}\hfill
\begin{minipage}[t]{5.25cm}
\begin{center}
\centerline{\includegraphics[width=5cm]{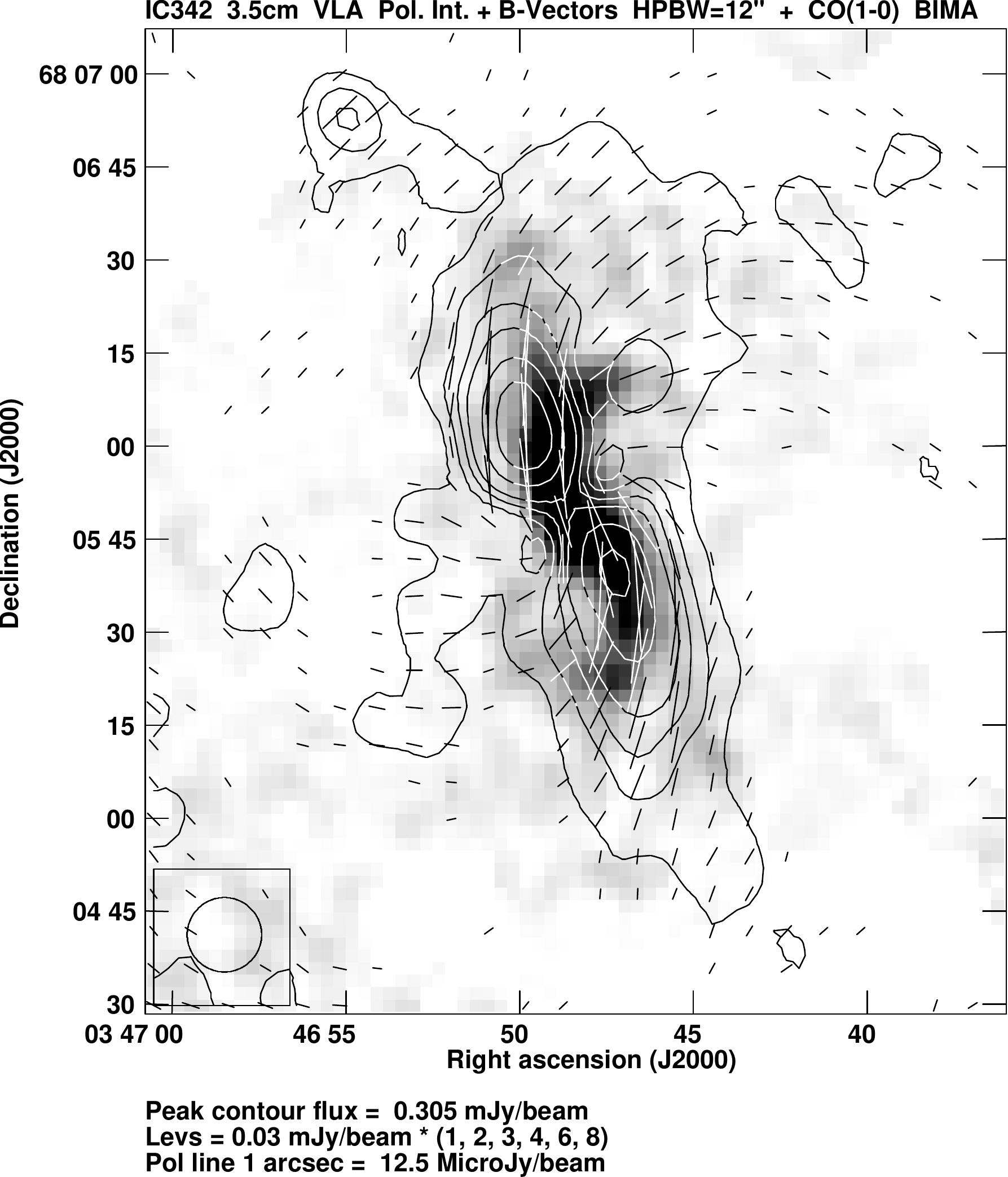}}
\caption{Polarized intensity and B--vectors in the central region
of IC~342, observed at 3.5\,cm wavelength with the VLA at $12^{\prime\prime}$ resolution,
overlaid on a greyscale presentation of the CO(1-0) emission at 2.6\,mm from the
BIMA survey \citep{helfer03}.}
\label{fig:ic342central}
\end{center}
\end{minipage}
\end{figure*}

Galaxies with massive bars have non-axisymmetic gas flows that interact with magnetic fields.
The magnetic field lines in NGC~1097, one of the closest and brightest barred galaxies (Fig.~\ref{fig:n1097}),
follow a pattern that resembles the flow of cold gas in a bar potential. As the gas rotates faster
than the bar pattern of a galaxy, a shock occurs in the cold gas, which has a
small sound speed, while the flow of warm, diffuse gas is only
slightly compressed but sheared. The ordered field is also hardly
compressed, probably coupled to the diffuse gas and strong enough to
affect its flow \citep{beck05c}. The ordered field is also strong in
the upstream region (south of the center in Fig.~\ref{fig:n1097}),
oriented almost perpendicularly to the bar and aligned with the dust
filaments seen on the optical image in the background.
The polarization pattern in barred galaxies can be used as a tracer
of shearing gas flows in the sky plane and complements spectroscopic
measurements of radial velocities.

The central regions of barred galaxies are often sites of ongoing
intense star formation and strong magnetic fields that can affect
gas flows. NGC~1097 hosts a bright ring with about 1.5\,kpc
diameter and an active nucleus in its center
(Fig.~\ref{fig:n1097central}). The ordered field in the ring
has a spiral pattern and extends to the nucleus. The
orientation of the innermost spiral field agrees with that of the
spiral dust filaments visible on optical images. Magnetic stress in
the circumnuclear ring due to the strong total magnetic field (about
50\,$\mu$G) can drive gas inflow \citep{balbus98} at a rate of
several solar masses per year, which is sufficient to fuel the activity of
the nucleus \citep{beck05c}. MHD modeling confirmed that magnetic fields
can strongly enhance the mass gas inflow rate \citep{kim12}.

The central region of IC~342 hosts a bar of dust and cold gas with a wealth of molecular lines.
In terms of size, dynamical mass, molecular mass, and star-formation rate, the nucleus of IC~342
is a potential twin of the Galactic centre \citep{meier14}. While the central bar in total radio
intensity coincides with the central bar in the CO line emission from molecular gas, the polarized
emission reveals a double-lobe structure that is displaced from the CO bar
(Fig.~\ref{fig:ic342central}).

\subsection{Magnetic halos}
\label{halo}

\begin{figure*}[t]
\vspace*{2mm}
\begin{minipage}[t]{5cm}
\begin{center}
\includegraphics[width=4.75cm]{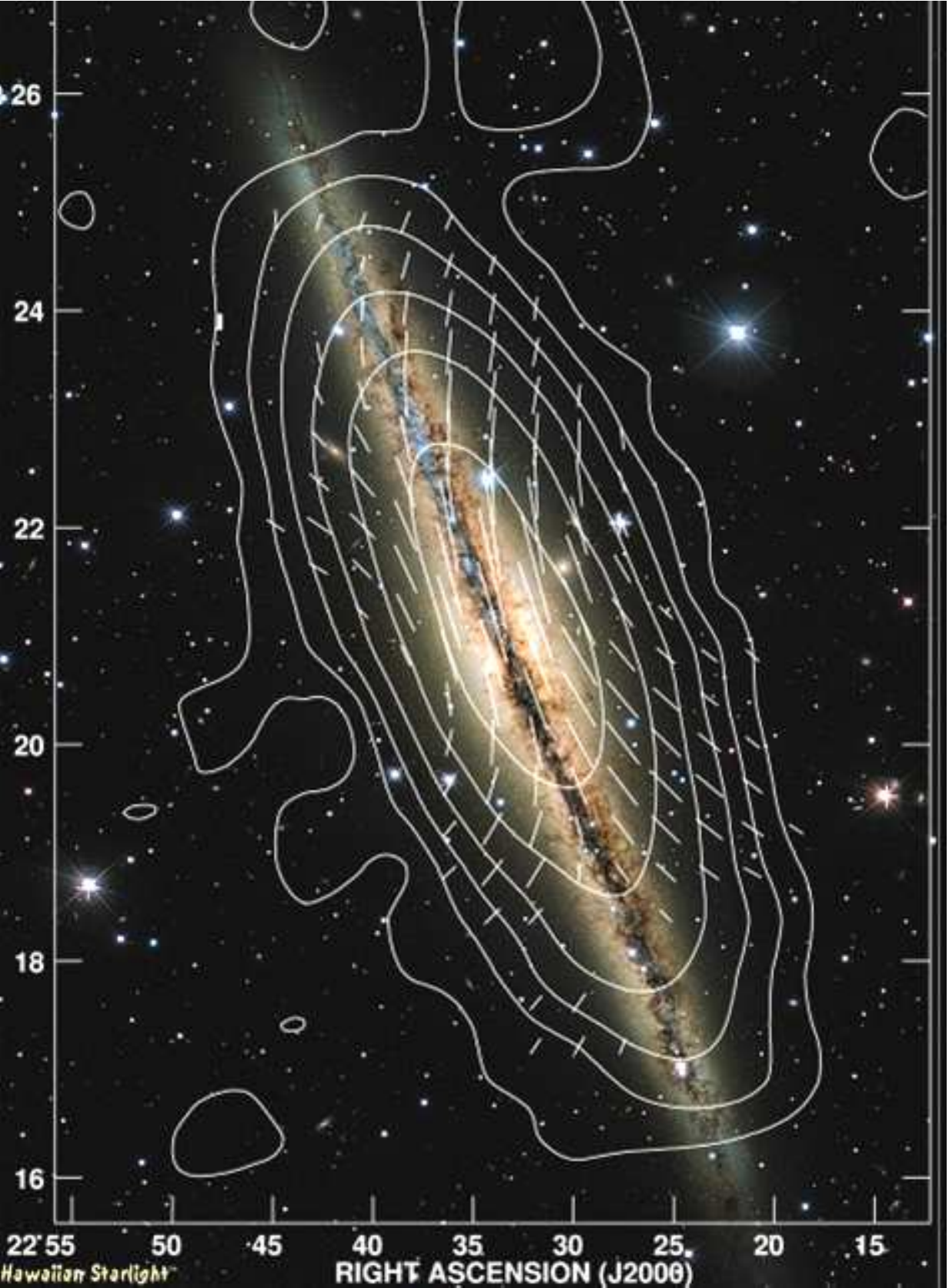}
\caption{Total radio emission and B--vectors of the edge-on galaxy NGC~891, a galaxy
similar to the Milky Way, observed at 3.6\,cm wavelength with the Effelsberg telescope at
$84^{\prime\prime}$ resolution (from \citet{krause09}). The background optical image is from the CFHT.
(Copyright: MPIfR Bonn)}
\label{fig:n891}
\end{center}
\end{minipage}\hfill
\begin{minipage}[t]{6.5cm}
\begin{center}
\includegraphics[width=6.25cm]{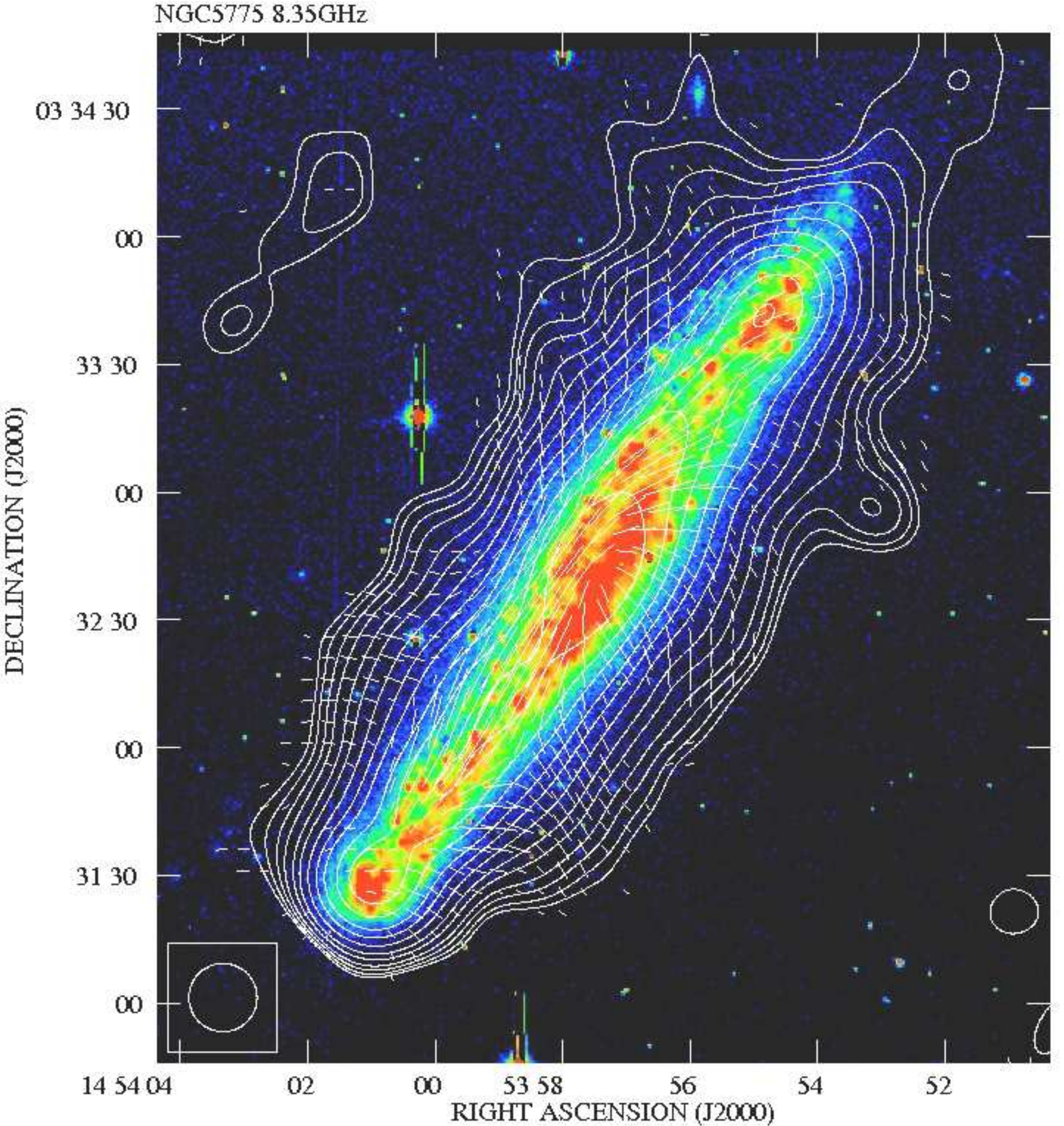}
\caption{Total radio intensity and B--vectors of the edge-on galaxy NGC~5775, combined from observations
at 3.6\,cm wavelength with the VLA and Effelsberg telescopes at $16^{\prime\prime}$ resolution
(from \citet{soida11}). The background H$\alpha$ image is from \citet{tuellmann00}. (Copyright: MPIfR Bonn)}
\label{fig:n5775}
\end{center}
\end{minipage}
\end{figure*}

\begin{figure*}[t]
\begin{center}
\includegraphics[width=10cm]{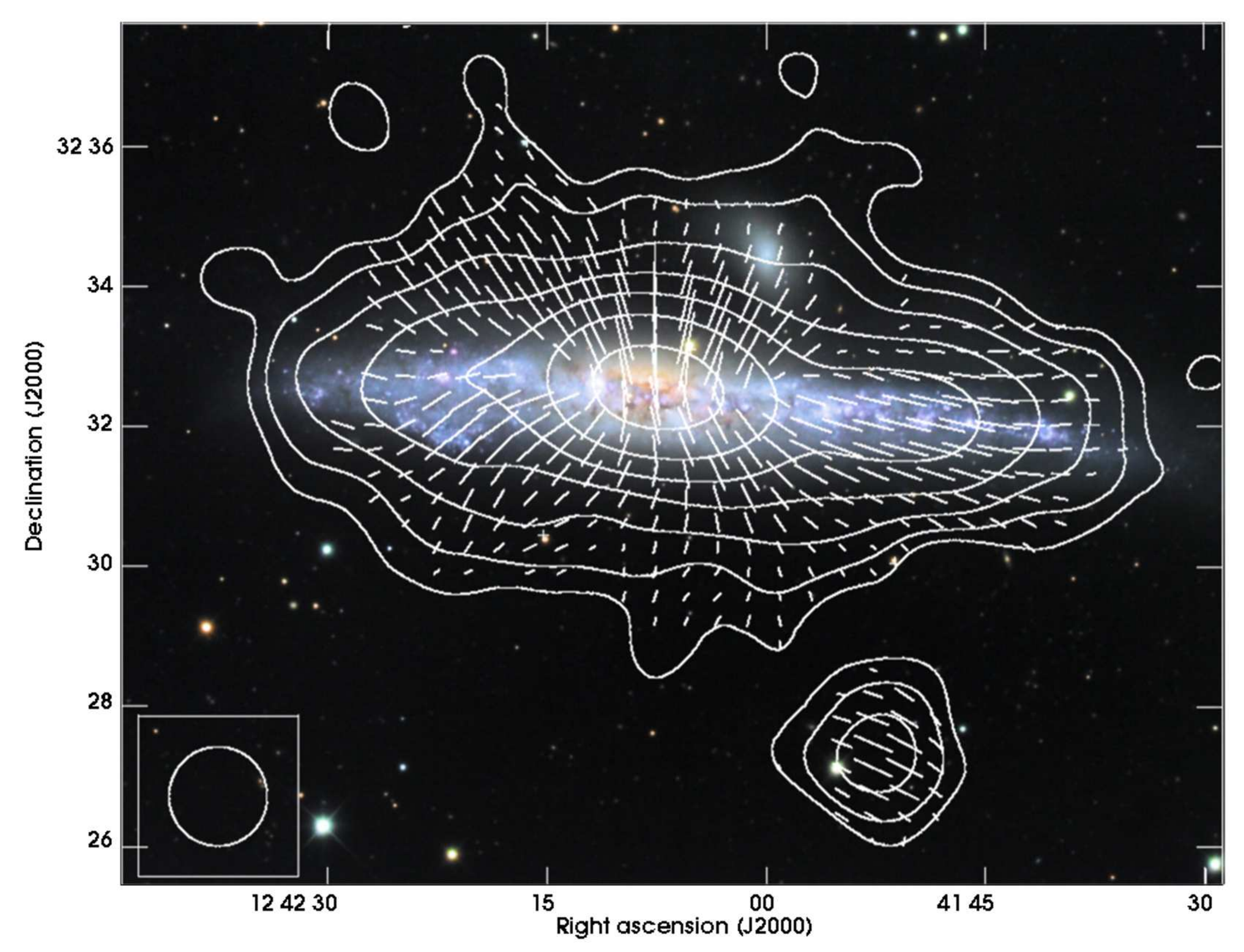}
\caption{Total radio emission (contours) and B--vectors of the edge-on galaxy NGC~4631, observed at
3.6\,cm wavelength with the Effelsberg telescope at $85^{\prime\prime}$ resolution (from \citet{mora13}).
The background optical image is from the Digital Sky Survey. (Copyright: MPIfR Bonn)}
\label{fig:n4631}
\end{center}
\end{figure*}

Hot gas, magnetic fields and cosmic rays in galaxy disks drive outflows that extend far beyond the galaxy
disks seen at optical wavelengths and form thick disks or halos at radio wavelengths. \footnote{As a definition
of how to distinguish ``thick disk'' from ``halo'' is missing, ``halo'' is used in the following.}
Radio polarization observations of several edge-on galaxies reveal
vertical field components in the halo forming an X-shaped pattern,
like in NGC~253 \citep{heesen09b}, NGC~891 (Fig.~\ref{fig:n891}), NGC~4631
(Fig.~\ref{fig:n4631}), and NGC~5775 (Fig.~\ref{fig:n5775}),
which may be related to outflows or to dynamo action \citep{moss10} driven by such outflows.

The vertical profile of the total radio continuum emission gives information about the outflow speed.
The profiles of five edge-on spiral galaxies observed
with high resolution can be described by two exponential scale heights $H_\mathrm{syn}$,
$300\pm50$\,pc for the thin disk and $1.8\pm0.2$\,kpc for the halo \citep{krause14}.
Neglecting the thermal contributions and assuming energy equipartition between total magnetic fields
and total cosmic rays (Sect.~\ref{eq}), the scale heights of the total magnetic field are typically
$H_\mathrm{B} \ge (3+\alpha)\,H_\mathrm{syn} \simeq 1.2$\,kpc for the thin disk and about 7\,kpc for
the halo, but probably larger due to energy losses of CREs.
Because the average field strengths and hence the synchrotron lifetimes of CREs are different
in different galaxies, the roughly constant halo scale heights indicate that outflow speeds increasing
with the average field strength (and hence with the star-formation rate) may balance the smaller
CRE lifetimes \citep{krause09}.

Stronger magnetic fields in the central regions cause higher synchrotron loss of CREs,
leading to the ``dumbbell'' shape of many radio
halos, e.g. in NGC~253 \citep{heesen09a} and NGC~4565 \citep{krause09}.
From the radio scale heights at several frequencies and the corresponding
lifetimes of CREs (depending on synchrotron, inverse Compton and adiabatic
losses) a wind speed of about 300~km/s was measured for the electrons in
the halo of NGC~253 \citep{heesen09a}.

Galaxies without a detectable radio halo are rare. UGC~10288 shows discrete high-latitude
radio features, but it does not have a global radio continuum halo \citep{irwin13}. No radio halo
was found in M~31 \citep{berkhuijsen13} and in NGC~7462 (Heesen et al., MNRAS, in press).
These three galaxies have a star-formation rate that is probably too low to drive outflows.

On the other hand, exceptionally large and almost spherical radio halos are observed around interacting
galaxies, namely NGC~4631 (Fig.~\ref{fig:n4631}) and the starburst galaxy M~82 \citep{adebahr13,reuter94}.
A few magnetic spurs could be resolved in NGC~4631, connected to star-forming regions \citep{golla94}.
These observations support the ideas of a strong galactic outflow ({\em galactic wind}),
driven by regions of star formation in the disk, possibly supported by the interaction with neighbouring
galaxies.

Above the central starburst region of NGC~253, an outwards-directed helical field of
about 20\,$\mu$G strength, extending to at least 1\,kpc height, could be identified
in the gas outflow cone with help of high-resolution RM mapping \citep{heesen11a}.
This field may help to confine the outflow.

In summary, the occurrence of radio halos is governed by the competition between
vertical transport of CREs by diffusion and/or convection (galactic wind) and energy
losses of CREs, dominated by synchrotron loss. CRE transport can be characterized by
the escape time $t_\mathrm{esc}$ needed to reach the scale height $H$ of the disk, beyond
which the synchrotron emission significantly weakens. In case of diffusive
transport, $t_\mathrm{esc} \propto (H^2/D)$, where $D$ is the diffusion coefficient,
while in case of convective transport $t_\mathrm{esc} \propto H/\mathrm{v_{wind}}$,
where $\mathrm{v_{wind}}$ is the average wind speed.

For strong magnetic fields or slow winds or slow diffusion ($t_\mathrm{syn} \ll  t_\mathrm{esc}$),
the CREs cannot leave the disk and no radio halo develops, as in the case of M~31 \citep{berkhuijsen13}.
For $t_\mathrm{syn} \lesssim t_\mathrm{esc}$,
the scale height $H$ of the synchrotron halo increases with distance from the center according to the increase in
synchrotron lifetime (``dumbbell'' halo) and with radio wavelength ($H \propto \lambda^{1/4}$
for diffusive or $H \propto \lambda^{1/2}$ for convective propagation).
For weak fields or fast winds ($t_\mathrm{syn} > t_\mathrm{esc}$), a large radio halo of elliptical
or spherical shape is formed with a wavelength-independent scale height that is determined by the
scale height of the total magnetic field and adiabatic losses of the expanding CRE flow.
\footnote{Detailed models of radio halos are presented in \citet{lerche82a,lerche82b}.}


\subsection{Rudimentary spirals: flocculent and irregular galaxies}
\label{irr}

\begin{figure*}[t]
\begin{center}
\includegraphics[width=9cm]{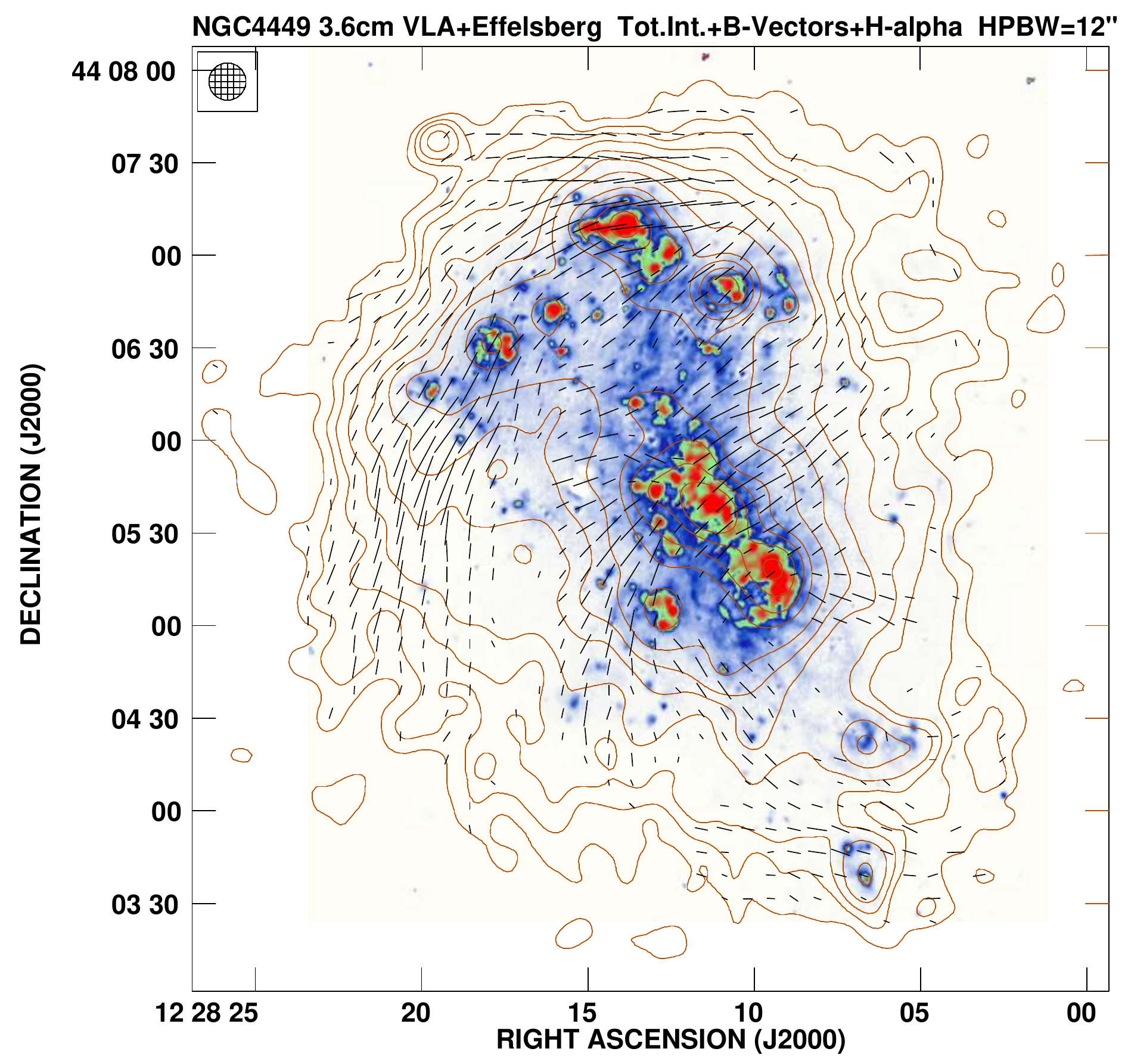}
\caption{Total radio emission (contours) and B--vectors of the
dwarf irregular galaxy NGC~4449, observed at 3.6\,cm wavelength with the VLA
at $12^{\prime\prime}$ resolution (from \citet{chyzy00}). The background
H$\alpha$ image is from Dominik Bomans (Bochum University).}
\label{fig:n4449}
\end{center}
\end{figure*}

Flocculent galaxies have disks but no gaseous spiral arms.
Nevertheless, magnetic spiral patterns have been observed in all
flocculent galaxies so far, indicating that the $\alpha-\Omega$ dynamo works
independently of spiral density waves. Ordered magnetic fields with
strengths similar to those in grand-design spiral galaxies have been
detected in the flocculent galaxies M~33 \citep{taba08}, NGC~3521 and NGC~5055
\citep{knapik00}, and in NGC~4414 \citep{soida02}. The mean degree of polarization
(corrected for the differences in spatial resolution) is also similar in
grand-design and flocculent galaxies.

Radio continuum images of irregular, slowly rotating galaxies may
reveal strong total magnetic fields, e.g. in the Magellanic-type
galaxy NGC~4449 (Fig.~\ref{fig:n4449}), with a partly ordered field
of about 7\,$\mu$G strength, a spiral pattern in the northeast and a
radial pattern in the central region \citep{chyzy00}.
Faraday rotation data shows that this ordered field is partly regular and
the $\alpha-\Omega$ dynamo is operating \citep{siejkowski14}.

The Large and the Small Magellanic Clouds are the closest irregular galaxies at distances of
50\,kpc and 60\,kpc, respectively, interacting with the Milky Way and each other.
Radio observations found the two galaxies to be weakly polarized \citep{haynes91,mao12},
with the exception of two highly polarized filaments in the southeast of the LMC.
A study based on RM grids suggests that both galaxies host large-scale
coherent fields, indicating that a large-scale dynamo also works under the less favourable
conditions of slow ordered rotation \citep{mao08,mao12}.
Due to their low potential well, low-mass galaxies are prone to outflows and
galactic winds driven by star formation.

In dwarf irregular galaxies, the strengths of the total (turbulent) field are generally smaller than
in spiral galaxies \citep{chyzy11}, except for starburst dwarfs, e.g. NGC~1569 with $10-15\,\mu$G
total field strength \citep{kepley10}, where star formation activity is sufficiently high
for the operation of the small-scale dynamo. Consequently, dwarf galaxies form the low-luminosity
tail of the radio--IR correlation for large galaxies \citep{chyzy11}.

Spots of faint polarized emission were detected in dwarf irregular galaxies, but no
large-scale regular fields so far \citep{heesen11b},
either due to a lack of telescope sensitivity or a lack of dynamo action.
In classical $\alpha-\Omega$ dynamo models, the dynamo number is too low for dynamo
action because rotation is almost chaotic, while models including support of dynamo action by
outflows predict $\alpha-\Omega$ dynamos even in dwarf galaxies \citep{rodrigues15}.

%
%

\subsection{Beyond spirals: interacting galaxies}
\label{inter}

\begin{figure*}[t]
\begin{center}
\includegraphics[width=7cm]{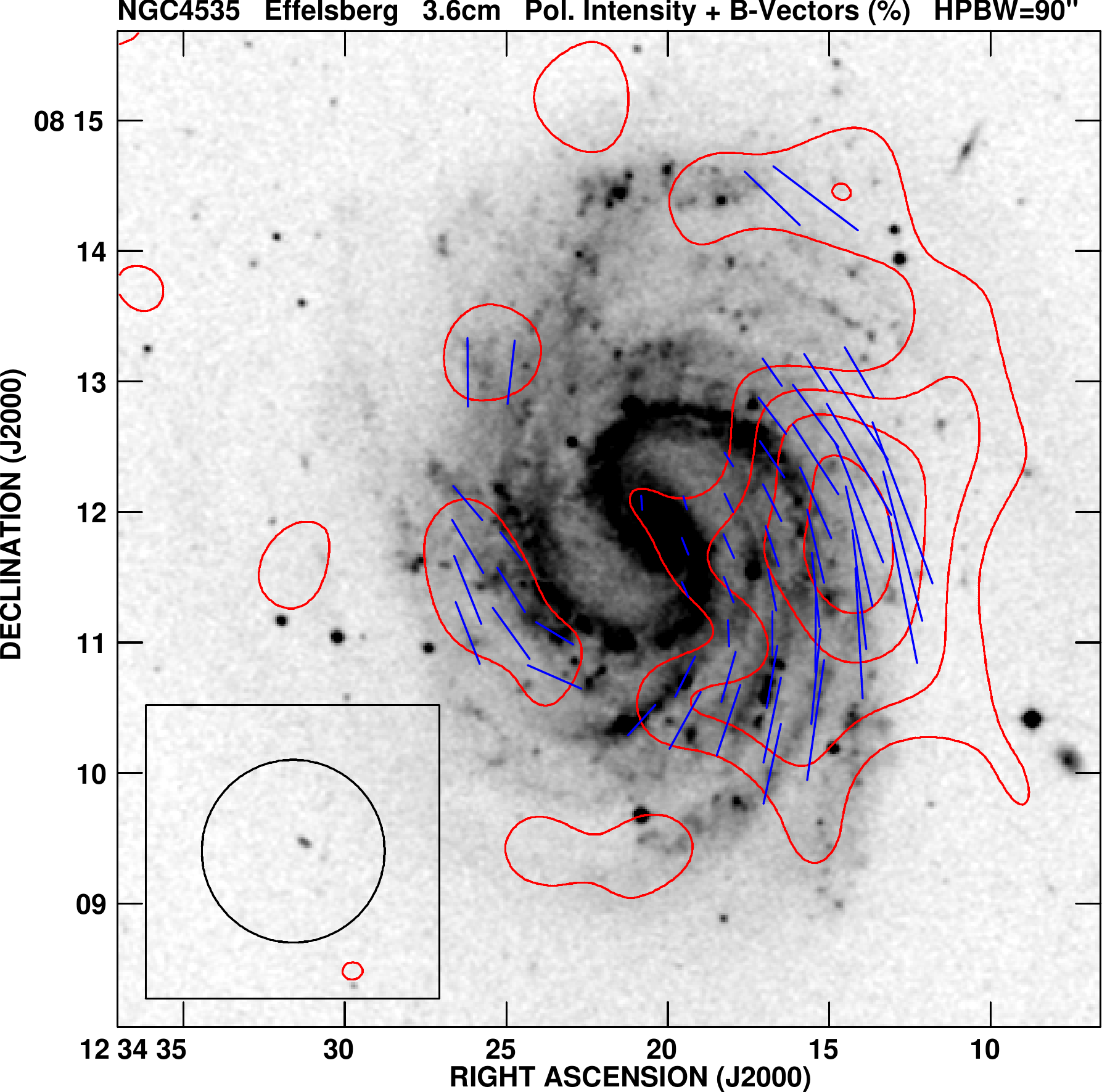}
\caption{Polarized radio emission (contours) and B--vectors of the
Virgo galaxy NGC~4535, observed at 3.6\,cm wavelength with the Effelsberg telescope at
$90^{\prime\prime}$ resolution (from \citet{wez12} and priv. comm.).
The background optical image is from the Digital Sky Survey.}
\label{fig:n4535}
\end{center}
\end{figure*}

Gravitational interaction between galaxies leads to asymmetric gas flows, compression, shear,
enhanced turbulence and outflows that can be traced by observing magnetic fields in radio polarization.
Magnetic fields can become aligned along the compression front or perpendicular to
the velocity gradients. Such gas flows make turbulent fields highly anisotropic.
Large-scale dynamos can be enhanced by ram pressure \citep{moss14}.

The classical interacting galaxy pair is NGC~4038/39, the
``Antennae'' \citep{chyzy04}. It shows bright, extended radio
emission filling the whole system. In the interaction region between
the galaxies, where star formation did not yet start, and at the
northeastern edge of the system, the magnetic field is partly
ordered, probably the result of compression and shearing motions
along the tidal tail. Particularly strong, almost unpolarized emission
comes from a region of violent star formation, hidden in dust.
The average total magnetic field is stronger than in normal spirals,
but the mean degree of polarization is unusually low, implying that
the fields are tangled.
The Antennae galaxies have been modeled by \citet{kotarba10}, who could
reproduce the main features observed in the radio image.

The total magnetic field in a sample of 16 systems in different interaction stages increases with
advancing interaction, which indicates enhanced production of random magnetic fields \citep{drzazga11}.
The strength of the ordered magnetic field are sensitive tools for revealing global galactic distortions.
The pattern of the ordered field traces the orientation of gas flow in tidally stretched spiral arms
and in tidal tails. Such outflows may have contributed to the magnetization of the intergalactic medium
in the early cosmological epoch.

Interaction with a dense intergalactic medium also imprints unique
signatures onto magnetic fields and thus onto the radio emission. The
Virgo cluster is a location of especially strong interaction effects
(Fig.~\ref{fig:n4535}), and almost all cluster galaxies observed so far show
asymmetries of their polarized emission because the outer magnetic fields were compressed
\citep{vollmer07,vollmer13,wez07,wez12}. Ordered fields are an excellent tracer of
past interactions between galaxies or with the intergalactic medium.

\subsection{Spiral galaxies with jets}
\label{jets}

\begin{figure*}[t]
\vspace*{2mm}
\begin{minipage}[t]{6.25cm}
\begin{center}
\includegraphics[width=6cm]{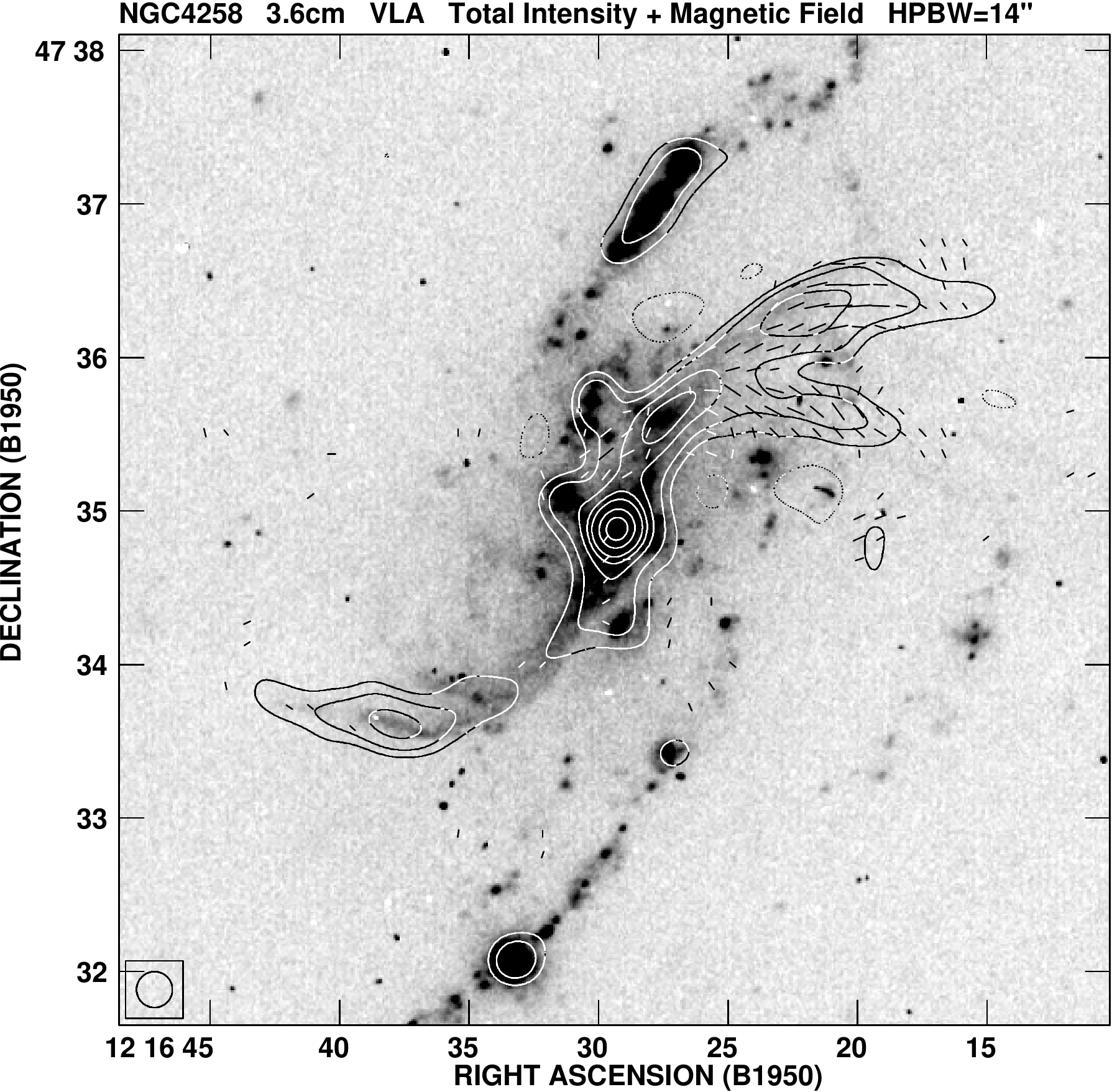}
\caption{Total radio intensity (contours) and B--vectors of the jets in NGC~4258, observed at 3.5\,cm
wavelength with the VLA at $14^{\prime\prime}$ resolution. The background H$\alpha$ image is from
the Hoher List Observatory of the University of Bonn (from \citet{krause04}).}
\label{fig:n4258}
\end{center}
\end{minipage}\hfill
\begin{minipage}[t]{5.25cm}
\begin{center}
\centerline{\includegraphics[width=5cm]{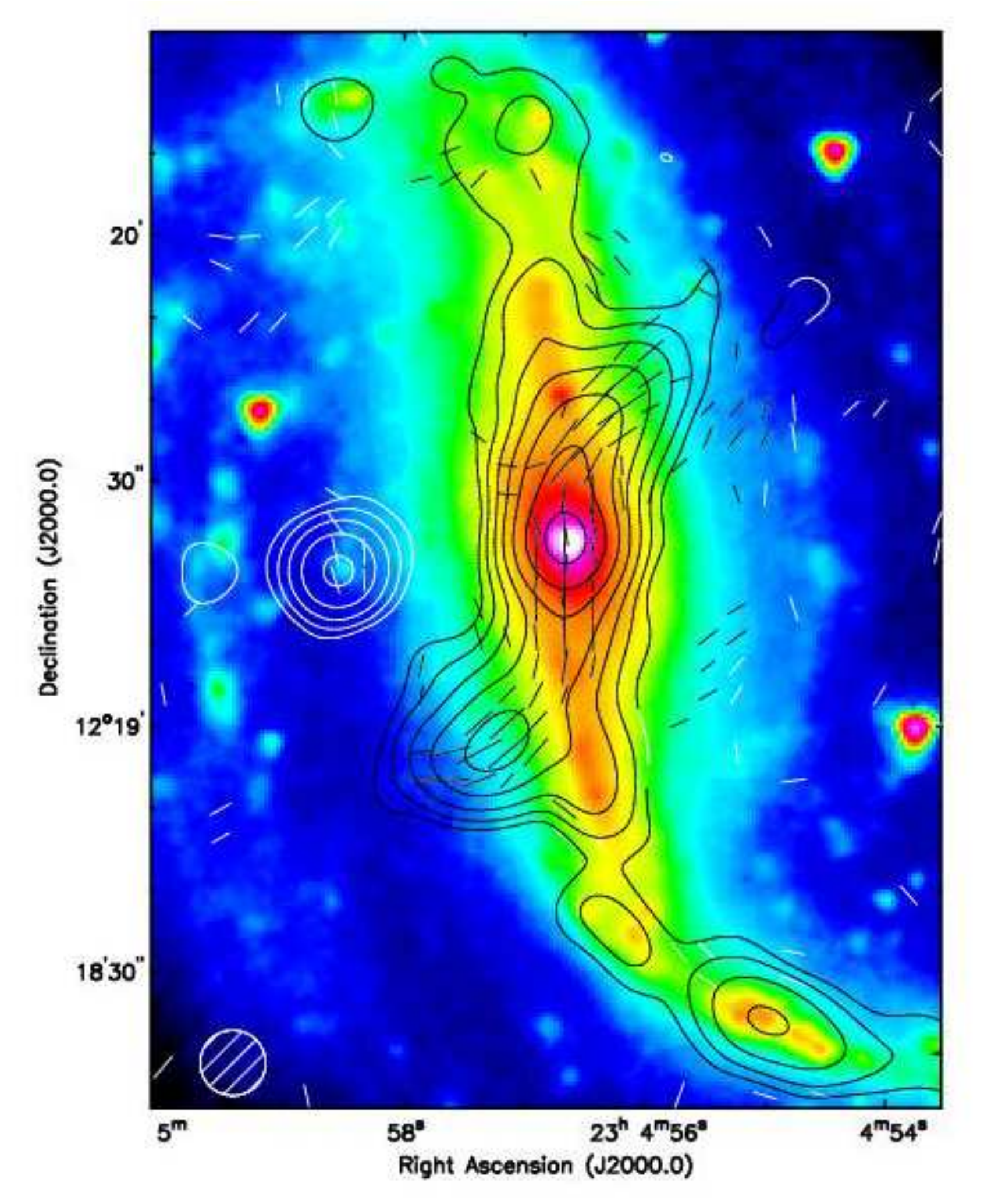}}
\caption{Total radio intensity (contours) and B--vectors of the jets in NGC~7479, observed at 3.5\,cm
wavelength with the VLA at $8^{\prime\prime}$ resolution. The background shows a Spitzer/IRAC image at
$3.6\,\mu$m (NASA/JPL-Caltech/Seppo Laine) (from \citet{laine08}.}
\label{fig:n7479}
\end{center}
\end{minipage}
\end{figure*}

Nuclear jets are observed in only a few spiral galaxies. These jets are weak and small compared to those
of radio galaxies and quasars. Detection is further hampered by the fact that they emerge at some
angle with respect to the disk, so that little interaction with the ISM occurs. If the nuclear
disk is oriented at a large angle to the disk, the jet hits a large amount of ISM matter,
CREs are accelerated in shocks, and the jet becomes radio-bright. However, not all jets are
consistent with this morphology.

NGC~4258 is one of the rare cases where large radio jets of at least 15\,kpc length are observed
(Fig.~\ref{fig:n4258}). A nuclear disk is observed in water maser emission, has an inner radius of
0.13\,pc and is seen almost edge-on \citep{greenhill95}.
As the jets emerge perpendicular to the nuclear disk, they have to pass the galactic disk at a rather
small angle. The magnetic field orientation is mainly along the jet direction.
The equipartition field strength is about $300\,\mu$G at the resolution of about 100\,pc
\citep{krause04}, which is a lower limit due to energy losses of the CREs and the limited resolution.

The barred galaxy NGC~7479 also shows remarkable jet-like radio continuum features: bright, narrow,
12\,kpc long in projection, and containing an aligned magnetic field (Fig.~\ref{fig:n7479}). The lack of
any optical or near-infrared emission associated with the jets suggests that at least the outer parts
of the jets are extraplanar features, although close to the disk plane. The equipartition strength is
$35-40\,\mu$G for the total magnetic field and about 10$\,\mu$G for the ordered magnetic field in the jets.
According to Faraday rotation measurements, the large-scale regular magnetic field along the bar points
towards the nucleus on both sides. Multiple reversals on scales of $1-2$\,kpc are detected, probably
occurring in the galaxy disk in front of the eastern jet by anisotropic fields in the shearing gas
flow in the bar potential.

Highly polarized radio emission from kpc--size jets has also been detected in NGC~3079
\citep{cecil01}, with the field orientations perpendicular to the jet's axis, and in the outflow
lobes of the Circinus Galaxy \citep{elmouttie95}. Two members of the Virgo cluster, NGC~4388 and
NGC~4438, have elongated radio features emerging from the active Seyfert--type nuclei and
extending roughly {\em perpendicular}\ to the planes of the disks \citep{hummel91}. Detailed images
including polarization are forthcoming as results of the CHANG-ES survey of edge-on galaxies
\citep{irwin12a}.

Jets in spiral galaxies may be more frequent than the available radio observations suggest.
Future low-frequency observations will help to search for weak synchrotron emission from
interface regions between the jets and the low-density halo gas.

\section{How does the Milky Way fit into the picture of nearby galaxies?}
\label{MW}

Direct measurements of the Voyager~2 spacecraft in the heliosheath indicate that the surrounding interstellar
magnetic field is $4-5\,\mu$G strong and oriented at an angle of about $30^\circ$ from the Galactic plane
\citep{opher09}, probably because the ISM field twists close to the heliosphere. Voyager~1 crossed into
interstellar space in 2012 and measured a smooth increase in field strength to $5.62\pm0.01\,\mu$G
\citep{burlaga13}. This value is very close to those obtained with other methods (see below) and to those
in nearby galaxies with a low star-formation rate (Sect.~\ref{strength}).

To deduce information about the strength and structure of the Galactic magnetic field,
pulsars are ideal objects because their RMs provide field directions at many distances from the
Sun \citep{noutsos12}. Since most pulsars are concentrated along the Galactic plane, they sample the
field in the disk. Combination of RM and DM data of pulsars (Eq.~\ref{dmrm}), assuming uncorrelated
fluctuations, gives an average strength of the local regular field along the line of sight of
$2.1\pm 0.3\,\mu$G and about 4\,$\mu$G at 3\,kpc Galacto-centric radius \citep{han06}. These are
upper limits in the case of correlated fluctuations \citep{beck03}, while lower limits in the case
of anticorrelated fluctuations or field reversals along the line of sight.

From the dispersion of pulsar RMs, the Galactic magnetic field was found to have a significant
turbulent component with a mean strength of $5-6\,\mu$G \citep{rand89,ohno93,han04}.
Magnetic turbulence occurs over a large spectrum of scales, with the largest scale
determined from pulsar RMs of $l_\mathrm{turb} \simeq 55$\,pc \citep{rand89}
or $l_\mathrm{turb} \simeq 10-100$\,pc \citep{ohno93}.
These values are consistent with the size of turbulent cells of $d \simeq 50$\,pc estimated from
beam depolarization and Faraday depolarization in external galaxies (Sects.~\ref{pol} and \ref{dp}).
However, turbulence scales of only a few parsecs in spiral arms were derived from RM structure
functions of polarized background sources \citep{haverkorn08}.

Modeling the surveys of the total synchrotron and $\gamma$--ray emission from the Milky Way yield field strengths
near the Sun of about 5\,$\mu$G of the isotropic turbulent field, about 2\,$\mu$G of the anisotropic turbulent
field and about 2\,$\mu$G of the regular field \citep{orlando13}, adding up to a total field strength of about
6\,$\mu$G. This is in excellent agreement with the Voyager and pulsar RM data (see above) and the Zeeman
splitting data of low-density gas clouds \citep{crutcher10}.
In the inner Galaxy the total field strength is about 10\,$\mu$G.
In the synchrotron filaments near the Galactic Center the total field strength is about 100\,$\mu$G
\citep{crocker10}.

\begin{figure*}[t]
\vspace*{2mm}
\begin{center}
\includegraphics[width=10cm]{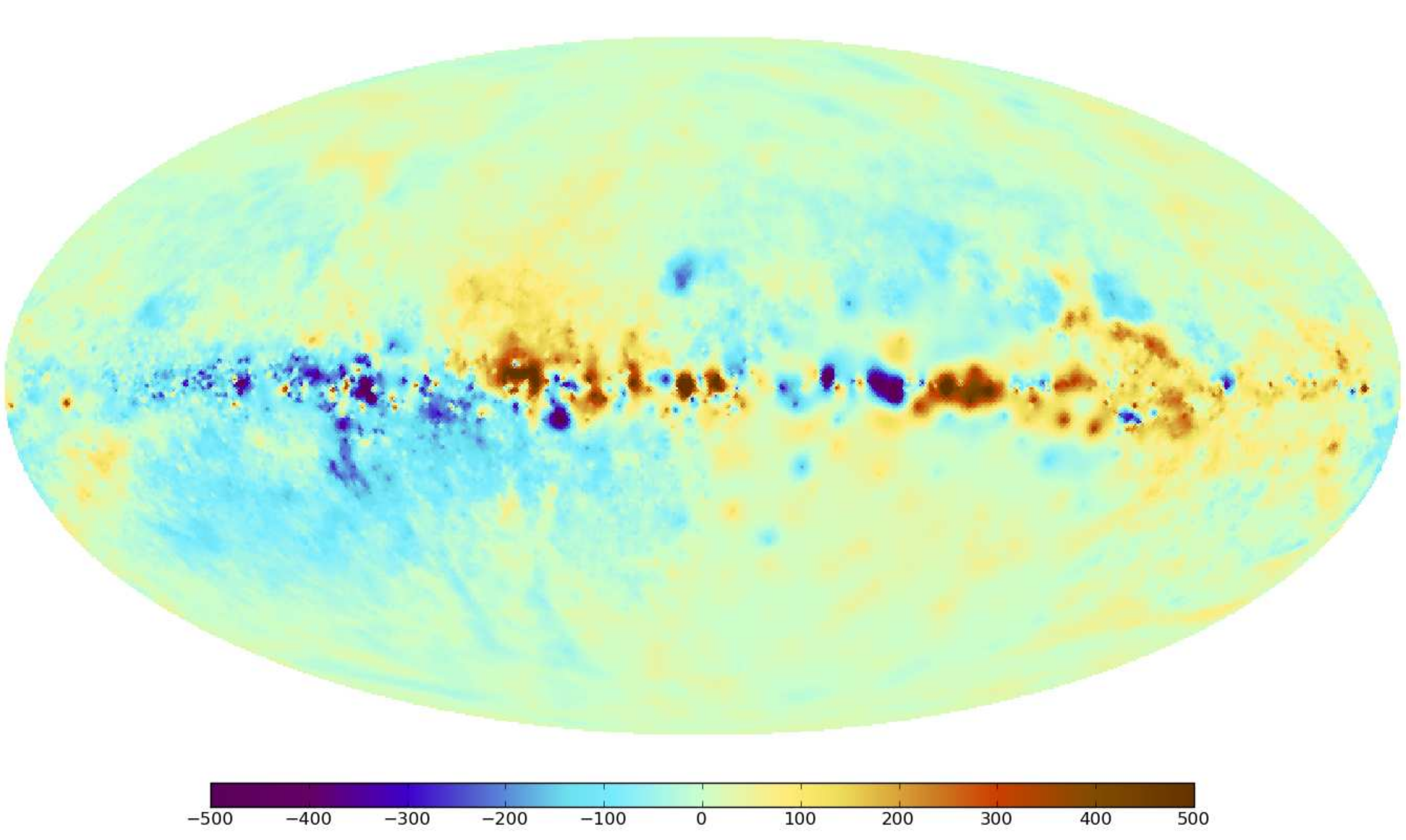}
\caption{All-sky map of rotation measures in the Milky
Way, constructed from the RM data of about 40000 polarized extragalactic
sources from the VLA NVSS survey \citep{taylor09} and other catalogs.
Red: positive RM, blue: negative RM (from \cite{oppermann12}).}
\label{fig:Galaxy1}
\end{center}
\end{figure*}

\begin{figure*}[t]
\vspace*{2mm}
\begin{center}
\includegraphics[width=7cm]{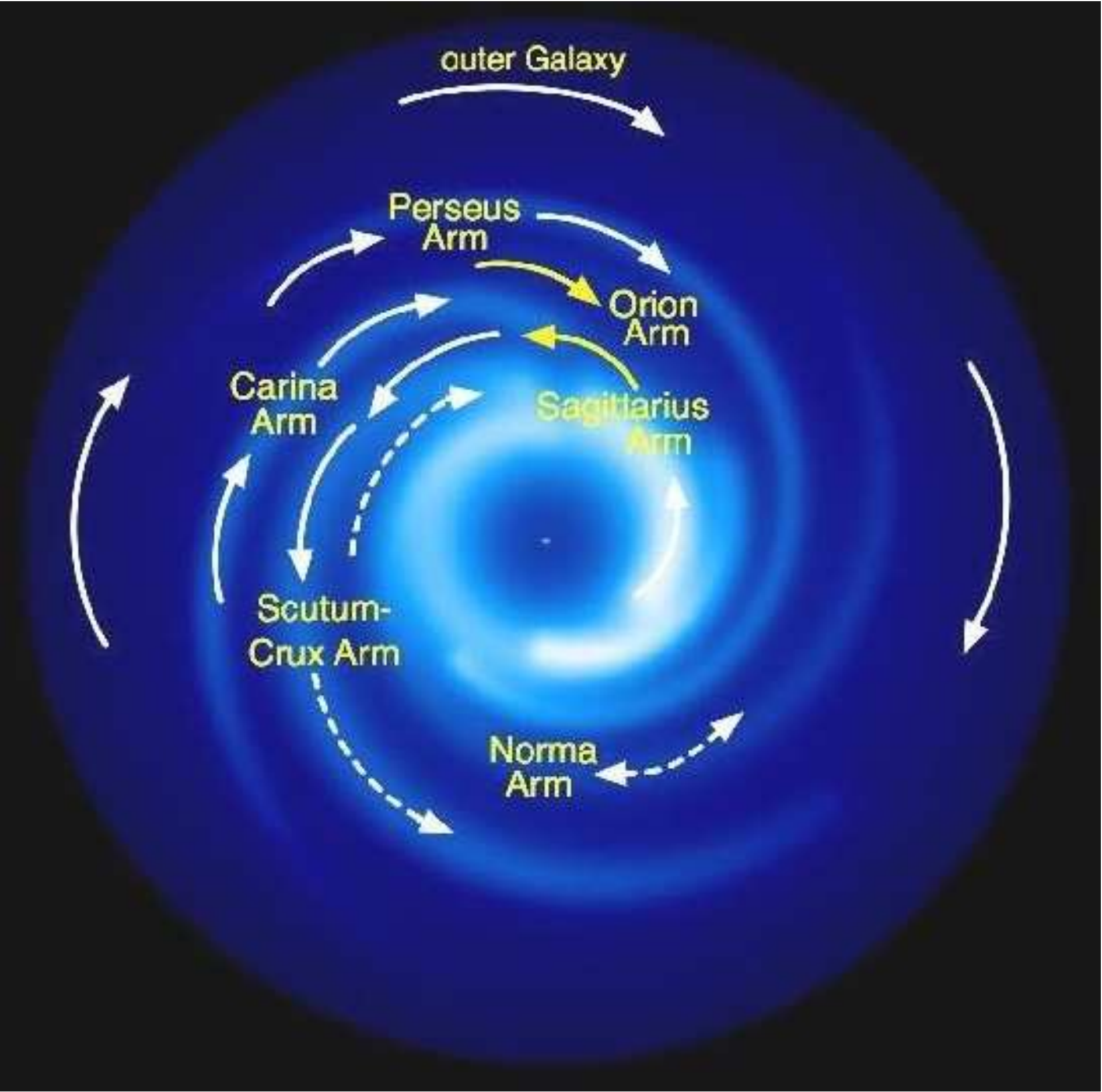}
\caption{Model of the magnetic field in the Milky Way (vectors), derived from Faraday rotation
measures of pulsars and extragalactic sources. Generally accepted results are indicated by yellow vectors,
while white vectors are not fully confirmed. The background image shows the distribution of ionized gas
delineating the spiral arms (from Jo-Anne Brown, Calgary).}
\label{fig:Galaxy2}
\end{center}
\end{figure*}

Optical polarization data of about 5500 selected stars in the Milky Way
yielded the orientation of the large-scale magnetic field near the Sun
\citep{fosalba02}, which is mostly parallel to the Galactic plane and
oriented along the local spiral arm.

The all-sky maps of polarized synchrotron emission at 1.4\,GHz from
the Milky Way from DRAO and Villa Elisa and at 22.8\,GHz from WMAP,
and the Effelsberg RM survey of polarized extragalactic sources, were
used to model the regular Galactic field \citep{sun08,sun10}. One
large-scale {\em field reversal}\ is required at about $1-2$\,kpc from
the Sun towards the Milky Way's center, consistent with pulsar data.

RM data from pulsars and extragalactic radio sources (Fig.~\ref{fig:Galaxy1})
was used to model the Galactic magnetic field \citep{nota10,vaneck11,jansson12a}.
A large-scale magnetic field reversal appears to be present between the Scutum-Crux-Sagittarius
arm and the Carina-Orion arm (Fig.~\ref{fig:Galaxy2}). As distances to most pulsars
are uncertain, this result should be taken with some caution.
The overall structure of the regular field in the disk of the Milky Way is still uncertain
\citep{noutsos09}. A larger sample of pulsar RM data and improved distance measurements to
pulsars are needed.
A satisfying explanation for the large-scale reversal in the Milky Way is still lacking.
So far no similar reversals have been detected in external galaxies (Sect.~\ref{reversals}).
Possible reasons are:\\
(1) The reversal in the Milky Way may be restricted to a thin layer near to the
plane and therefore hardly visible in the average RM data of
external galaxies along the line of sight.\\
(2) The reversal in the Milky Way may be of limited azimuthal extent and difficult
to observe in external galaxies with present-day telescopes.\\
(3) The reversal in the Milky Way may be part of a disturbed
field structure, e.g. due to interaction with the Magellanic clouds,
or a relic from seed fields \citep{moss12}.

Even less is known about the halo field of the Milky Way.
The vertical full equivalent thickness of the synchrotron emission from the
Milky Way is about 3\,kpc near the Sun \citep{beuermann85}, scaled to a distance
to the Galactic center of 8.5\,kpc, yielding an exponential scale height of
$H_\mathrm{syn} \simeq 1.5$\,kpc and at least 6\,kpc of the total field,
similar to that in external galaxies (Sect.~\ref{halo}).

The signs of RMs of extragalactic sources and of pulsars at Galactic longitudes
$l=90^\circ-270^\circ$ are mostly the same above and below the plane (Fig.~\ref{fig:Galaxy1}):
the magnetic field in the local disk is symmetric, while the RM signs towards the inner Galaxy
($l=270^\circ-90^\circ$) are {\em opposite}\ above and below the plane. This could be assigned
to an odd-symmetry (antisymmetric) halo field \citep{sun10}. However, according to RM data from
extragalactic sources, the local regular Galactic field has no significant vertical component
towards the northern Galactic pole and only a weak vertical component of $B_z\simeq0.3\,\mu$G towards
the south \citep{mao10}. This is neither consistent with an odd-symmetry nor with an even-symmetry
halo field as found in several external galaxies (Sect.~\ref{large}).
The halo field of the Milky Way has a more complicated structure than predicted by dynamo models.
Modeling the diffuse polarized emission and RMs gave evidence for an X-shaped vertical field component
\citep{jansson12a}, similar to that in external galaxies (Sect.~\ref{halo}).

In summary, strength, spiral pattern, and vertical extent of the Galactic magnetic field are similar
to those in nearby spiral galaxies. Two major differences still need to be understood: the large-scale field
reversal(s) and the antisymmetric pattern of the halo field, both of which are not observed in external
galaxies. Either our Milky Way is special, or the different observational methods deliver apparently
incompatible results.

\section{Summary and open questions}

Table~\ref{tab:summary} gives a concise description of the general structures of ordered
magnetic fields in various galaxy types as observed from radio polarization observations.
For most galaxy types only a few cases were observed so far. --
``Turbulent'' stands for ``isotropic turbulent or tangled'',
``ordered'' for ``anisotropic turbulent'' or ``regular''.
``Regular field'' refers to additional measurements of Faraday rotation.\\

For a list of radio polarization observations of galaxies, see \citet{beck+wielebinski13}
and updated versions on arXiv.
Radio images of total intensity and polarization are compiled on:
{\tt http://www.mpifr-bonn.mpg.de/atlasmag}.\\

\begin{table}
\caption{General field structures in nearby galaxies}
\begin{tabular}{lll}
\hline
{\bf Galaxy type} & {\bf Magnetic field structure} & {\bf Regular field} \\
\hline
Sc with strong & Spiral at inner arm edge and in & Strong,\\
\,\,\, density wave & \,\,\, interarm regions, turbulent in arms & \,\,\, fluctuating\\
Sb or Sc with weak & Spiral in interarm regions, & Moderate,\\
\,\,\, or moderate density wave & \,\,\, turbulent + ordered in arms & \,\,\, large-scale\\
Barred Sc & Ordered + turbulent along bar, & Moderate\\
 & \,\,\, spiral outside bar \\
Sa & Ordered + turbulent & Detected\\
S0 & Not detected & Not detected\\
Interacting spiral & Ordered, asymmetric & Weak\\
Spiral with nuclear jets & Ordered along jet & ?\\
Flocculent Sc or Sd & Spiral + turbulent in disk & Weak\\
Irregular & Turbulent in star-forming regions & Weak\\
 & \,\,\, + segments of ordered field\\
Starburst dwarf & Turbulent in star-forming regions & Not detected\\
Spheroidal dwarf & Not detected & Not detected\\
E without active nucleus & Not detected & Not detected\\
\hline
\label{tab:summary}
\end{tabular}
\end{table}

The open questions raised in this review may inspire future work:\\

\noindent (1) Morphology:

\begin{itemize}

\item How far do turbulent and regular magnetic fields extend radially in the disk and vertically above the disk?
\item How well are magnetic pitch angles aligned with the orientations of gaseous spiral arms?
\item How do magnetic pitch angles vary radially in the disk?
\item Do dominating bisymmetric spiral patterns (azimuthal mode $m=1$) of the regular field exist in galaxies?
\item Do azimuthal modes $m>2$ exist and how strong are these?
\item Which galaxy types host magnetic arms between gaseous spiral arms?
\item How common are field loops generated by the Parker instability?
\item Do regular fields in galaxies have large-scale reversals?
\item Are the X-shaped fields in galaxy halos related to outflows (anisotropic turbulent fields) or do they show
large-scale patterns in Faraday rotation (regular fields)?

\end{itemize}

\noindent (2) Origin and amplification:

\begin{itemize}

\item What was the main source of seed fields needed for the amplification of galactic fields in young galaxies?
\item Are small-scale fields generated by the small-scale dynamo or by tangling of regular fields?
\item What is the origin of magnetic arms between gaseous spiral arms?
\item What determines the efficiency and saturation level of the $\alpha-\Omega$ dynamo?
\item Is the mean helicity of the large-scale magnetic fields in galaxy disks non-zero, as predicted by
$\alpha-\Omega$ dynamo action?
\item Are small-scale fields with non-zero mean helicity expelled from the disk to keep the
$\alpha-\Omega$ dynamo alive?

\end{itemize}

\noindent (3) Dynamical importance:

\begin{itemize}

\item Where is energy equipartition between total magnetic field and total cosmic rays valid and where does it fail?
\item What is the dominant propagation mechanism of CREs in galaxy disks and halos and how does this depend on
magnetic field strength and structure?
\item How do magnetic fields interact with spiral density waves?
\item How does the relation between total magnetic field strength and gas density vary between galaxies
and within galaxies?
\item How do magnetic fields modify the star-formation rate and the star-formation efficiency?
\item Is amplification of small-scale fields in star-forming regions the reason for the radio--IR correlation?
\item Is the magnetic energy density larger than that of gas turbulence in the outer parts of galaxies?
\item Can magnetic fields affect the general rotation of gas?
\item Can galactic outflows magnetize the intergalactic medium?

\end{itemize}

\section{Outlook}
\label{future}

Future high-resolution, high-sensitivity observations at high radio frequencies with the
{\em Jansky Very Large Array}\ (VLA) and the planned {\em Square Kilometre Array}\ (SKA, see below),
combined with high-resolution CO observations with the {\em Atacama Large Millimeter Array}\ (ALMA),
will directly map the detailed field structure and the interaction with the molecular gas in external galaxies
\citep{beck+15}. Images of dust polarization of nearby galaxies in the submm range with ALMA will soon
become possible. Another exciting prospect is extragalactic starlight and infrared polarimetry with the
{\em European Extremely Large Telescope}\ (E-ELT) \citep{strassmeier09}.

Construction of the first phase of the SKA is planned to start in 2018.
The low-frequency radio telescopes {\em Low Frequency Array}\ (LOFAR)
\citep{vanhaarlem13} and the {\em Murchison Widefield Array}\ (MWA)
\citep{tingay13} are suitable instruments to search for extended synchrotron
radiation in outer galaxy disks and halos and the transition region to intergalactic space.
Several galaxies have already been observed with LOFAR \citep{beck13,mulcahy14}.
Low frequencies are also suited to search for small Faraday rotation
measures from weak fields in the local ISM of the Milky Way, in the ISM of external
galaxies and in intergalactic space, if Faraday depolarization is small.

Small-scale and large-scale magnetic fields may exist in S0 and elliptical galaxies without star
formation and without active nucleus because turbulence can be generated by the magneto-rotational
instability (MRI) \citep{sellwood99} or may result from star formation activity in the past. The
detection may become possible with the SKA via deep imaging or RM grids of background sources.

\begin{figure*}[t]\begin{center}
\includegraphics[width=9cm]{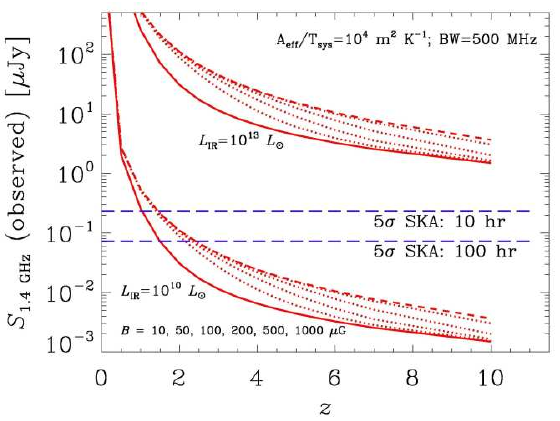}
\caption{Total synchrotron emission at 20\,cm wavelength as a function of
redshift $z$ and total magnetic field strength $B$, and the $5\sigma$
detection limits for 10\,h and 100\,h integration time with the full SKA
(from \citet{murphy09} and priv. comm.).}
\label{fig:murphy}
\end{center}
\end{figure*}

The detection of strong radio emission in distant galaxies (which is at
least partly of synchrotron origin) demonstrates that strong magnetic fields existed
already in young galaxies with strengths of several 100\,$\mu$G \citep{murphy09}.
Total synchrotron emission can be detected with the SKA out to even larger redshifts
in starburst galaxies, depending on luminosity and magnetic field strength
(Fig.~\ref{fig:murphy}).
As the amplification of turbulent fields is fast
and efficient (Sect.~\ref{dynamo}) and the star-formation rates were high
at high redshifts, the total fields can be expected to be very strong
already in young galaxies.
Then the energy range of the electrons emitting at a fixed observation frequency
shifts to low energies (Eq.~\ref{sync}), where ionization and bremsstrahlung
losses may become dominant, so that
there must be a maximum redshift until synchrotron emission is detectable.
Its measurement with the SKA will constrain models of the evolution of
magnetic fields in young galaxies \citep{schleicher13}.


Nearby galaxies seen edge-on generally show an ordered, disk-parallel
field near the disk plane \citep{dumke95}. As a result, polarized
emission can also be detected from distant, unresolved galaxies if
the inclination is larger than about $20^{\circ}$ \citep{stil09}.
This opens a new method to search for ordered fields in distant
galaxies with help of deep surveys.

If polarized emission from galaxies is too weak to be detected, the method of {\em RM grids}\ towards
background QSOs can still be applied. Here, the distance limit is given by the polarized flux of
the background QSO, which can be much higher than that of an intervening galaxy that is identified
optically by its Mg~II absorption system. With this method, significant regular fields
of several $\mu$G strength were detected in distant galaxies \citep{bernet08,bernet13,farnes14}.
Mean-field solutions of the $\alpha-\Omega$ dynamo theory predict patches of regular fields
observable as RMs in evolving regular fields already at $z\le3$, with increasing coherence
scale, until fully developed regular fields are formed at $z\simeq1-1.5$ \citep{arshakian09,rodrigues15}.
Note that RM values measured today have been reduced due to redshift by the dilution factor of
$(1+z)^{-2}$, so that observations at longer wavelengths are needed to detect RMs from regular fields
at high redshifts.

Faraday rotation in the direction of polarized lobes of radio galaxies allows to determine the
strength and pattern of a regular field in an intervening galaxy \citep{kronberg92,irwin13}. RM data
indicate that ionized gas may extend to several 10\,kpc distance from galaxies \citep{bernet13}.
A reliable model for the field structure in and around galaxies needs RM values from a large number of
polarized background sources, hence large sensitivity and high survey speed.


The SKA ``Cosmic Magnetism'' Key Science Project plans to observe a polarization
survey over the entire accessible sky with the SKA-MID Band\,2 (around 20\,cm wavelength)
\citep{johnston15}. Within 1\,h integration per field this will allow the detection of about
10 million discrete extragalactic sources and the measurement of their RMs,
about 300 RMs per square degree or a mean spacing between sources of about $3^{\prime}$.

Deep polarization observations with the SKA centered on a sample of nearby galaxies will allow studies
of the interaction between magnetic fields and the ISM gas \citep{beck+15}. The detailed
reconstruction of the 3D magnetic field structure will become possible with help of Faraday
tomography \citep{heald+15} or a dense grid of RM values from background sources
\citep{stepanov08}. More than 1000 RM values can be expected in the area of M~31.
Applying this method to more distant galaxies, simple patterns of regular fields can be
recognized out to distances of about 100\,Mpc \citep{stepanov08}.\\

Exciting prospects!

\newpage


%
%

\begin{acknowledgements}
The author would like to thank all colleagues who have pursued the studies of magnetic fields in the Milky Way
and in galaxies over the past 40 years, especially Richard Wielebinski, Elly M. Berkhuijsen, Marita Krause,
Patricia Reich, Wolfgang Reich, Sui Ann Mao and Aritra Basu at MPIfR, Andrew Fletcher and Anvar Shukurov at
Newcastle University, Chris Chy{\.z}y and Marek Urbanik at Krak{\'o}w University, Dmitry Sokoloff at Moscow
University, Peter Frick and Rodion Stepanov at ICMM Perm, Ralf-J{\"u}rgen Dettmar at Bochum University,
Detlef Elstner at AIP Potsdam, Katia Ferri{\`e}re at Toulouse University and Fatemeh Tabatabaei at IAC La Laguna.
-- Elly M. Berkhuijsen and Aritra Basu are acknowledged for careful reading of the manuscript and useful
comments.
-- Support from DFG Research Unit FOR1254 is acknowledged.
\end{acknowledgements}

\bibliographystyle{aa} 

\bibliography{aar}   

\begin{thebibliography}{259}
\expandafter\ifx\csname natexlab\endcsname\relax\def\natexlab#1{#1}\fi

\bibitem[{{Aab} {et~al.}(2015){Aab}, {Abreu}, {Aglietta}, {Ahn}, {Samarai},
  {Albuquerque}, {Allekotte}, {Allen}, {Allison}, {Almela}, \& et~al.}]{aab15}
{Aab}, A., {Abreu}, P., {Aglietta}, M., {et~al.} 2015, \apj, 804, 15

\bibitem[{{Adebahr} {et~al.}(2013){Adebahr}, {Krause}, {Klein},
  {We{\.z}gowiec}, {Bomans}, \& {Dettmar}}]{adebahr13}
{Adebahr}, B., {Krause}, M., {Klein}, U., {et~al.} 2013, \aap, 555, A23

\bibitem[{{Arbutina} {et~al.}(2012){Arbutina}, {Uro{\v s}evi{\'c}},
  {Andjeli{\'c}}, {Pavlovi{\'c}}, \& {Vukoti{\'c}}}]{arbutina12}
{Arbutina}, B., {Uro{\v s}evi{\'c}}, D., {Andjeli{\'c}}, M.~M., {Pavlovi{\'c}},
  M.~Z., \& {Vukoti{\'c}}, B. 2012, \apj, 746, 79

\bibitem[{{Arshakian} \& {Beck}(2011)}]{arshakian11}
{Arshakian}, T.~G. \& {Beck}, R. 2011, \mnras, 418, 2336

\bibitem[{{Arshakian} {et~al.}(2009){Arshakian}, {Beck}, {Krause}, \&
  {Sokoloff}}]{arshakian09}
{Arshakian}, T.~G., {Beck}, R., {Krause}, M., \& {Sokoloff}, D. 2009, \aap,
  494, 21

\bibitem[{{Bagetakos} {et~al.}(2011){Bagetakos}, {Brinks}, {Walter}, {de Blok},
  {Usero}, {Leroy}, {Rich}, \& {Kennicutt}}]{bagetakos11}
{Bagetakos}, I., {Brinks}, E., {Walter}, F., {et~al.} 2011, \aj, 141, 23

\bibitem[{{Balbus} \& {Hawley}(1998)}]{balbus98}
{Balbus}, S.~A. \& {Hawley}, J.~F. 1998, Reviews of Modern Physics, 70, 1

\bibitem[{{Basu} \& {Roy}(2013)}]{basu13}
{Basu}, A. \& {Roy}, S. 2013, \mnras, 433, 1675

\bibitem[{{Basu} {et~al.}(2012){Basu}, {Roy}, \& {Mitra}}]{basu12}
{Basu}, A., {Roy}, S., \& {Mitra}, D. 2012, \apj, 756, 141

\bibitem[{{Basu} {et~al.}(2015){Basu}, {Wadadekar}, {Beelen}, {Singh},
  {Archana}, {Sirothia}, \& {Ishwara-Chandra}}]{basu15}
{Basu}, A., {Wadadekar}, Y., {Beelen}, A., {et~al.} 2015, \apj, 803, 51

\bibitem[{{Battaner} \& {Florido}(2007)}]{battaner07}
{Battaner}, E. \& {Florido}, E. 2007, Astronomische Nachrichten, 328, 92

\bibitem[{{Beck} {et~al.}(2012{\natexlab{a}}){Beck}, {Lesch}, {Dolag},
  {Kotarba}, {Geng}, \& {Stasyszyn}}]{abeck12}
{Beck}, A.~M., {Lesch}, H., {Dolag}, K., {et~al.} 2012{\natexlab{a}}, \mnras,
  422, 2152

\bibitem[{{Beck}(2007)}]{beck07}
{Beck}, R. 2007, \aap, 470, 539

\bibitem[{{Beck}(2015)}]{beck15}
{Beck}, R. 2015, \aap, 578, A93

\bibitem[{{Beck} {et~al.}(2013){Beck}, {Anderson}, {Heald}, {Horneffer},
  {Iacobelli}, {K{\"o}hler}, {Mulcahy}, {Pizzo}, {Scaife}, {Wucknitz}, \&
  {LOFAR Magnetism Key Science Project Team}}]{beck13}
{Beck}, R., {Anderson}, J., {Heald}, G., {et~al.} 2013, Astronomische
  Nachrichten, 334, 548

\bibitem[{{Beck} {et~al.}(2015){Beck}, {Bomans}, {Colafrancesco}, {Dettmar},
  {Ferri{\`e}re}, {Fletcher}, {Heald}, {Heesen}, {Horellou}, {Krause}, {Lou},
  {Mao}, {Paladino}, {Schinnerer}, {Sokoloff}, {Stil}, \&
  {Tabatabaei}}]{beck+15}
{Beck}, R., {Bomans}, D., {Colafrancesco}, S., {et~al.} 2015, Advancing
  Astrophysics with the Square Kilometre Array (AASKA14), 94

\bibitem[{{Beck} {et~al.}(1996){Beck}, {Brandenburg}, {Moss}, {Shukurov}, \&
  {Sokoloff}}]{beck96}
{Beck}, R., {Brandenburg}, A., {Moss}, D., {Shukurov}, A., \& {Sokoloff}, D.
  1996, \araa, 34, 155

\bibitem[{{Beck} {et~al.}(2005){Beck}, {Fletcher}, {Shukurov}, {Snodin},
  {Sokoloff}, {Ehle}, {Moss}, \& {Shoutenkov}}]{beck05c}
{Beck}, R., {Fletcher}, A., {Shukurov}, A., {et~al.} 2005, \aap, 444, 739

\bibitem[{{Beck} {et~al.}(2012{\natexlab{b}}){Beck}, {Frick}, {Stepanov}, \&
  {Sokoloff}}]{beck12}
{Beck}, R., {Frick}, P., {Stepanov}, R., \& {Sokoloff}, D. 2012{\natexlab{b}},
  \aap, 543, A113

\bibitem[{{Beck} \& {Krause}(2005)}]{beck+krause05}
{Beck}, R. \& {Krause}, M. 2005, Astronomische Nachrichten, 326, 414

\bibitem[{{Beck} {et~al.}(1989){Beck}, {Loiseau}, {Hummel}, {Berkhuijsen},
  {Grave}, \& {Wielebinski}}]{beck89}
{Beck}, R., {Loiseau}, N., {Hummel}, E., {et~al.} 1989, \aap, 222, 58

\bibitem[{{Beck} {et~al.}(1994){Beck}, {Poezd}, {Shukurov}, \&
  {Sokoloff}}]{beck94}
{Beck}, R., {Poezd}, A.~D., {Shukurov}, A., \& {Sokoloff}, D.~D. 1994, \aap,
  289, 94

\bibitem[{{Beck} {et~al.}(2003){Beck}, {Shukurov}, {Sokoloff}, \&
  {Wielebinski}}]{beck03}
{Beck}, R., {Shukurov}, A., {Sokoloff}, D., \& {Wielebinski}, R. 2003, \aap,
  411, 99

\bibitem[{{Beck} \& {Wielebinski}(2013)}]{beck+wielebinski13}
{Beck}, R. \& {Wielebinski}, R. 2013, in Planets, Stars and Stellar
  Systems.~Volume 5: Galactic Structure and Stellar Populations, ed. T.~D.
  {Oswalt} \& G.~{Gilmore} (New York: Springer), 641

\bibitem[{{Bell}(1978)}]{bell78}
{Bell}, A.~R. 1978, \mnras, 182, 443

\bibitem[{{Bell}(2003)}]{bell03}
{Bell}, E.~F. 2003, \apj, 586, 794

\bibitem[{{Bell} {et~al.}(2011){Bell}, {Junklewitz}, \& {En{\ss}lin}}]{bell11}
{Bell}, M.~R., {Junklewitz}, H., \& {En{\ss}lin}, T.~A. 2011, \aap, 535, A85

\bibitem[{{Berkhuijsen} {et~al.}(2003){Berkhuijsen}, {Beck}, \&
  {Hoernes}}]{berkhuijsen03}
{Berkhuijsen}, E.~M., {Beck}, R., \& {Hoernes}, P. 2003, \aap, 398, 937

\bibitem[{{Berkhuijsen} {et~al.}(2013){Berkhuijsen}, {Beck}, \&
  {Tabatabaei}}]{berkhuijsen13}
{Berkhuijsen}, E.~M., {Beck}, R., \& {Tabatabaei}, F.~S. 2013, \mnras, 435,
  1598

\bibitem[{{Berkhuijsen} {et~al.}(1997){Berkhuijsen}, {Horellou}, {Krause},
  {Neininger}, {Poezd}, {Shukurov}, \& {Sokoloff}}]{berkhuijsen97}
{Berkhuijsen}, E.~M., {Horellou}, C., {Krause}, M., {et~al.} 1997, \aap, 318,
  700

\bibitem[{{Berkhuijsen} {et~al.}(2006){Berkhuijsen}, {Mitra}, \&
  {M\"uller}}]{berkhuijsen06}
{Berkhuijsen}, E.~M., {Mitra}, D., \& {M\"uller}, P. 2006, Astronomische
  Nachrichten, 327, 82

\bibitem[{{Bernet} {et~al.}(2012){Bernet}, {Miniati}, \& {Lilly}}]{bernet12}
{Bernet}, M.~L., {Miniati}, F., \& {Lilly}, S.~J. 2012, \apj, 761, 144

\bibitem[{{Bernet} {et~al.}(2013){Bernet}, {Miniati}, \& {Lilly}}]{bernet13}
{Bernet}, M.~L., {Miniati}, F., \& {Lilly}, S.~J. 2013, \apjl, 772, L28

\bibitem[{{Bernet} {et~al.}(2008){Bernet}, {Miniati}, {Lilly}, {Kronberg}, \&
  {Dessauges-Zavadsky}}]{bernet08}
{Bernet}, M.~L., {Miniati}, F., {Lilly}, S.~J., {Kronberg}, P.~P., \&
  {Dessauges-Zavadsky}, M. 2008, \nat, 454, 302

\bibitem[{{Bertone} {et~al.}(2006){Bertone}, {Vogt}, \&
  {En{\ss}lin}}]{bertone06}
{Bertone}, S., {Vogt}, C., \& {En{\ss}lin}, T. 2006, \mnras, 370, 319

\bibitem[{{Beuermann} {et~al.}(1985){Beuermann}, {Kanbach}, \&
  {Berkhuijsen}}]{beuermann85}
{Beuermann}, K., {Kanbach}, G., \& {Berkhuijsen}, E.~M. 1985, \aap, 153, 17

\bibitem[{{Bigiel} {et~al.}(2008){Bigiel}, {Leroy}, {Walter}, {Brinks}, {de
  Blok}, {Madore}, \& {Thornley}}]{bigiel08}
{Bigiel}, F., {Leroy}, A., {Walter}, F., {et~al.} 2008, \aj, 136, 2846

\bibitem[{{Birnboim} {et~al.}(2015){Birnboim}, {Balberg}, \&
  {Teyssier}}]{birnboim15}
{Birnboim}, Y., {Balberg}, S., \& {Teyssier}, R. 2015, \mnras, 447, 3678

\bibitem[{{Bisnovatyi-Kogan} {et~al.}(1973){Bisnovatyi-Kogan}, {Ruzmaikin}, \&
  {Syunyaev}}]{bisnovatyi73}
{Bisnovatyi-Kogan}, G.~S., {Ruzmaikin}, A.~A., \& {Syunyaev}, R.~A. 1973,
  Soviet Astronomy, 17, 137

\bibitem[{{Brandenburg} \& {Stepanov}(2014)}]{brandenburg14}
{Brandenburg}, A. \& {Stepanov}, R. 2014, \apj, 786, 91

\bibitem[{{Brandenburg} \& {Subramanian}(2005)}]{brandenburg05}
{Brandenburg}, A. \& {Subramanian}, K. 2005, \physrep, 417, 1

\bibitem[{{Braun} {et~al.}(2010){Braun}, {Heald}, \& {Beck}}]{braun10}
{Braun}, R., {Heald}, G., \& {Beck}, R. 2010, \aap, 514, A42

\bibitem[{{Brentjens} \& {de Bruyn}(2005)}]{brentjens05}
{Brentjens}, M.~A. \& {de Bruyn}, A.~G. 2005, \aap, 441, 1217

\bibitem[{{Broderick} {et~al.}(2012){Broderick}, {Chang}, \&
  {Pfrommer}}]{broderick12}
{Broderick}, A.~E., {Chang}, P., \& {Pfrommer}, C. 2012, \apj, 752, 22

\bibitem[{{Burlaga} {et~al.}(2013){Burlaga}, {Ness}, {Gurnett}, \&
  {Kurth}}]{burlaga13}
{Burlaga}, L.~F., {Ness}, N.~F., {Gurnett}, D.~A., \& {Kurth}, W.~S. 2013,
  \apjl, 778, L3

\bibitem[{{Burn}(1966)}]{burn66}
{Burn}, B.~J. 1966, \mnras, 133, 67

\bibitem[{{Buta} \& {McCall}(1999)}]{buta99}
{Buta}, R.~J. \& {McCall}, M.~L. 1999, \apjs, 124, 33

\bibitem[{{Calzetti} {et~al.}(2010){Calzetti}, {Wu}, {Hong}, {Kennicutt},
  {Lee}, {Dale}, {Engelbracht}, {van Zee}, {Draine}, {Hao}, {Gordon},
  {Moustakas}, {Murphy}, {Regan}, {Begum}, {Block}, {Dalcanton}, {Funes}, {Gil
  de Paz}, {Johnson}, {Sakai}, {Skillman}, {Walter}, {Weisz}, {Williams}, \&
  {Wu}}]{calzetti10}
{Calzetti}, D., {Wu}, S.-Y., {Hong}, S., {et~al.} 2010, \apj, 714, 1256

\bibitem[{{Cecil} {et~al.}(2001){Cecil}, {Bland-Hawthorn}, {Veilleux}, \&
  {Filippenko}}]{cecil01}
{Cecil}, G., {Bland-Hawthorn}, J., {Veilleux}, S., \& {Filippenko}, A.~V. 2001,
  \apj, 555, 338

\bibitem[{{Chamandy} {et~al.}(2015){Chamandy}, {Shukurov}, \&
  {Subramanian}}]{chamandy15}
{Chamandy}, L., {Shukurov}, A., \& {Subramanian}, K. 2015, \mnras, 446, L6

\bibitem[{{Chamandy} {et~al.}(2013{\natexlab{a}}){Chamandy}, {Subramanian}, \&
  {Shukurov}}]{chamandy13a}
{Chamandy}, L., {Subramanian}, K., \& {Shukurov}, A. 2013{\natexlab{a}},
  \mnras, 428, 3569

\bibitem[{{Chamandy} {et~al.}(2013{\natexlab{b}}){Chamandy}, {Subramanian}, \&
  {Shukurov}}]{chamandy13b}
{Chamandy}, L., {Subramanian}, K., \& {Shukurov}, A. 2013{\natexlab{b}},
  \mnras, 433, 3274

\bibitem[{{Chandrasekhar} \& {Fermi}(1953)}]{chandra53}
{Chandrasekhar}, S. \& {Fermi}, E. 1953, \apj, 118, 113

\bibitem[{{Cho} \& {Lazarian}(2005)}]{cho05}
{Cho}, J. \& {Lazarian}, A. 2005, \apj, 631, 361

\bibitem[{{Chy{\.z}y}(2008)}]{chyzy08}
{Chy{\.z}y}, K.~T. 2008, \aap, 482, 755

\bibitem[{{Chy{\.z}y} \& {Beck}(2004)}]{chyzy04}
{Chy{\.z}y}, K.~T. \& {Beck}, R. 2004, \aap, 417, 541

\bibitem[{{Chy{\.z}y} {et~al.}(2000){Chy{\.z}y}, {Beck}, {Kohle}, {Klein}, \&
  {Urbanik}}]{chyzy00}
{Chy{\.z}y}, K.~T., {Beck}, R., {Kohle}, S., {Klein}, U., \& {Urbanik}, M.
  2000, \aap, 355, 128

\bibitem[{{Chy{\.z}y} \& {Buta}(2008)}]{chyzy+buta08}
{Chy{\.z}y}, K.~T. \& {Buta}, R.~J. 2008, \apjl, 677, L17

\bibitem[{{Chy{\.z}y} {et~al.}(2011){Chy{\.z}y}, {We{\.z}gowiec}, {Beck}, \&
  {Bomans}}]{chyzy11}
{Chy{\.z}y}, K.~T., {We{\.z}gowiec}, M., {Beck}, R., \& {Bomans}, D.~J. 2011,
  \aap, 529, A94

\bibitem[{{Cox}(2005)}]{cox05}
{Cox}, D.~P. 2005, \araa, 43, 337

\bibitem[{{Crocker} {et~al.}(2010){Crocker}, {Jones}, {Melia}, {Ott}, \&
  {Protheroe}}]{crocker10}
{Crocker}, R.~M., {Jones}, D.~I., {Melia}, F., {Ott}, J., \& {Protheroe}, R.~J.
  2010, \nat, 463, 65

\bibitem[{{Crosthwaite} {et~al.}(2001){Crosthwaite}, {Turner}, {Hurt},
  {Levine}, {Martin}, \& {Ho}}]{crosthwaite01}
{Crosthwaite}, L.~P., {Turner}, J.~L., {Hurt}, R.~L., {et~al.} 2001, \aj, 122,
  797

\bibitem[{{Crutcher} {et~al.}(2010){Crutcher}, {Wandelt}, {Heiles},
  {Falgarone}, \& {Troland}}]{crutcher10}
{Crutcher}, R.~M., {Wandelt}, B., {Heiles}, C., {Falgarone}, E., \& {Troland},
  T.~H. 2010, \apj, 725, 466

\bibitem[{{de Avillez} \& {Breitschwerdt}(2005)}]{avillez05}
{de Avillez}, M.~A. \& {Breitschwerdt}, D. 2005, \aap, 436, 585

\bibitem[{{Dobbs} \& {Baba}(2014)}]{dobbs14}
{Dobbs}, C. \& {Baba}, J. 2014, \pasa, 31, 35

\bibitem[{{Dolag} {et~al.}(2011){Dolag}, {Kachelriess}, {Ostapchenko}, \&
  {Tom{\`a}s}}]{dolag11}
{Dolag}, K., {Kachelriess}, M., {Ostapchenko}, S., \& {Tom{\`a}s}, R. 2011,
  \apjl, 727, L4

\bibitem[{{Drzazga} {et~al.}(2011){Drzazga}, {Chy{\.z}y}, {Jurusik}, \&
  {Wi{\'o}rkiewicz}}]{drzazga11}
{Drzazga}, R.~T., {Chy{\.z}y}, K.~T., {Jurusik}, W., \& {Wi{\'o}rkiewicz}, K.
  2011, \aap, 533, A22

\bibitem[{{Dumas} {et~al.}(2011){Dumas}, {Schinnerer}, {Tabatabaei}, {Beck},
  {Velusamy}, \& {Murphy}}]{dumas11}
{Dumas}, G., {Schinnerer}, E., {Tabatabaei}, F.~S., {et~al.} 2011, \aj, 141, 41

\bibitem[{{Dumke} {et~al.}(1995){Dumke}, {Krause}, {Wielebinski}, \&
  {Klein}}]{dumke95}
{Dumke}, M., {Krause}, M., {Wielebinski}, R., \& {Klein}, U. 1995, \aap, 302,
  691

\bibitem[{{Durrer} \& {Neronov}(2013)}]{durrer13}
{Durrer}, R. \& {Neronov}, A. 2013, \aapr, 21, 62

\bibitem[{{Ehle} \& {Beck}(1993)}]{ehle93}
{Ehle}, M. \& {Beck}, R. 1993, \aap, 273, 45

\bibitem[{{Ehle} {et~al.}(1996){Ehle}, {Beck}, {Haynes}, {Vogler}, {Pietsch},
  {Elmouttie}, \& {Ryder}}]{ehle96}
{Ehle}, M., {Beck}, R., {Haynes}, R.~F., {et~al.} 1996, \aap, 306, 73

\bibitem[{{Ellis} \& {Axon}(1978)}]{ellis78}
{Ellis}, R.~S. \& {Axon}, D.~J. 1978, \apss, 54, 425

\bibitem[{{Elmouttie} {et~al.}(1995){Elmouttie}, {Haynes}, {Jones}, {Ehle},
  {Beck}, \& {Wielebinski}}]{elmouttie95}
{Elmouttie}, E., {Haynes}, R.~F., {Jones}, K.~L., {et~al.} 1995, \mnras, 275,
  L53

\bibitem[{{Elstner} {et~al.}(2014){Elstner}, {Beck}, \& {Gressel}}]{elstner14}
{Elstner}, D., {Beck}, R., \& {Gressel}, O. 2014, \aap, 568, A104

\bibitem[{{Farnes} {et~al.}(2014){Farnes}, {O'Sullivan}, {Corrigan}, \&
  {Gaensler}}]{farnes14}
{Farnes}, J.~S., {O'Sullivan}, S.~P., {Corrigan}, M.~E., \& {Gaensler}, B.~M.
  2014, \apj, 795, 63

\bibitem[{{Farnsworth} {et~al.}(2011){Farnsworth}, {Rudnick}, \&
  {Brown}}]{farnsworth11}
{Farnsworth}, D., {Rudnick}, L., \& {Brown}, S. 2011, \aj, 141, 191

\bibitem[{{Fathi} {et~al.}(2007){Fathi}, {Toonen}, {Falc{\'o}n-Barroso},
  {Beckman}, {Hernandez}, {Daigle}, {Carignan}, \& {de Zeeuw}}]{fathi07}
{Fathi}, K., {Toonen}, S., {Falc{\'o}n-Barroso}, J., {et~al.} 2007, \apjl, 667,
  L137

\bibitem[{{Fendt} {et~al.}(1998){Fendt}, {Beck}, \& {Neininger}}]{fendt98}
{Fendt}, C., {Beck}, R., \& {Neininger}, N. 1998, \aap, 335, 123

\bibitem[{{Ferri{\`e}re}(1996)}]{ferriere96}
{Ferri{\`e}re}, K. 1996, \aap, 310, 438

\bibitem[{{Ferri{\`e}re}(2001)}]{ferriere01}
{Ferri{\`e}re}, K.~M. 2001, Reviews of Modern Physics, 73, 1031

\bibitem[{{Fletcher}(2010)}]{fletcher10}
{Fletcher}, A. 2010, in Astronomical Society of the Pacific Conference Series,
  Vol. 438, Astronomical Society of the Pacific Conference Series, ed.
  R.~{Kothes}, T.~L. {Landecker}, \& A.~G. {Willis}, 197

\bibitem[{{Fletcher} {et~al.}(2011){Fletcher}, {Beck}, {Shukurov},
  {Berkhuijsen}, \& {Horellou}}]{fletcher11}
{Fletcher}, A., {Beck}, R., {Shukurov}, A., {Berkhuijsen}, E.~M., \&
  {Horellou}, C. 2011, \mnras, 412, 2396

\bibitem[{{Fletcher} {et~al.}(2004){Fletcher}, {Berkhuijsen}, {Beck}, \&
  {Shukurov}}]{fletcher04}
{Fletcher}, A., {Berkhuijsen}, E.~M., {Beck}, R., \& {Shukurov}, A. 2004, \aap,
  414, 53

\bibitem[{{Foglizzo} \& {Tagger}(1995)}]{foglizzo95}
{Foglizzo}, T. \& {Tagger}, M. 1995, \aap, 301, 293

\bibitem[{{Fosalba} {et~al.}(2002){Fosalba}, {Lazarian}, {Prunet}, \&
  {Tauber}}]{fosalba02}
{Fosalba}, P., {Lazarian}, A., {Prunet}, S., \& {Tauber}, J.~A. 2002, \apj,
  564, 762

\bibitem[{{Frick} {et~al.}(2001){Frick}, {Beck}, {Berkhuijsen}, \&
  {Patrickeyev}}]{frick01}
{Frick}, P., {Beck}, R., {Berkhuijsen}, E.~M., \& {Patrickeyev}, I. 2001,
  \mnras, 327, 1145

\bibitem[{{Frick} {et~al.}(2011){Frick}, {Sokoloff}, {Stepanov}, \&
  {Beck}}]{frick11}
{Frick}, P., {Sokoloff}, D., {Stepanov}, R., \& {Beck}, R. 2011, \mnras, 414,
  2540

\bibitem[{{Frick} {et~al.}(2016){Frick}, {Stepanov}, {Beck}, {Sokoloff},
  {Shukurov}, {Ehle}, \& {Lundgren}}]{frick16}
{Frick}, P., {Stepanov}, R., {Beck}, R., {et~al.} 2016, \aap, 585, A21

\bibitem[{{Gaensler} {et~al.}(2005){Gaensler}, {Haverkorn}, {Staveley-Smith},
  {Dickey}, {McClure-Griffiths}, {Dickel}, \& {Wolleben}}]{gaensler05}
{Gaensler}, B.~M., {Haverkorn}, M., {Staveley-Smith}, L., {et~al.} 2005,
  Science, 307, 1610

\bibitem[{{Gent} {et~al.}(2013){Gent}, {Shukurov}, {Sarson}, {Fletcher}, \&
  {Mantere}}]{gent13}
{Gent}, F.~A., {Shukurov}, A., {Sarson}, G.~R., {Fletcher}, A., \& {Mantere},
  M.~J. 2013, \mnras, 430, L40

\bibitem[{{Gie\ss\"ubel}(2012)}]{giess12}
{Gie\ss\"ubel}, R. 2012, {PhD} dissertation, Universit\"at zu K\"oln,
  {C}uvillier {V}erlag {G}\"ottingen, ISBN 978-3-95404-308-8

\bibitem[{{Gie{\ss}{\"u}bel} \& {Beck}(2014)}]{giess14}
{Gie{\ss}{\"u}bel}, R. \& {Beck}, R. 2014, \aap, 571, A61

\bibitem[{{Gie{\ss}{\"u}bel} {et~al.}(2013){Gie{\ss}{\"u}bel}, {Heald}, {Beck},
  \& {Arshakian}}]{giess13}
{Gie{\ss}{\"u}bel}, R., {Heald}, G., {Beck}, R., \& {Arshakian}, T.~G. 2013,
  \aap, 559, A27

\bibitem[{{Gissinger} {et~al.}(2009){Gissinger}, {Fromang}, \&
  {Dormy}}]{gissinger09}
{Gissinger}, C., {Fromang}, S., \& {Dormy}, E. 2009, \mnras, 394, L84

\bibitem[{{Goldreich} \& {Kylafis}(1981)}]{goldreich81}
{Goldreich}, P. \& {Kylafis}, N.~D. 1981, \apjl, 243, L75

\bibitem[{{Golla} \& {Hummel}(1994)}]{golla94}
{Golla}, G. \& {Hummel}, E. 1994, \aap, 284, 777

\bibitem[{{Greaves} {et~al.}(2000){Greaves}, {Holland}, {Jenness}, \&
  {Hawarden}}]{greaves00}
{Greaves}, J.~S., {Holland}, W.~S., {Jenness}, T., \& {Hawarden}, T.~G. 2000,
  \nat, 404, 732

\bibitem[{{Greenhill} {et~al.}(1995){Greenhill}, {Jiang}, {Moran}, {Reid},
  {Lo}, \& {Claussen}}]{greenhill95}
{Greenhill}, L.~J., {Jiang}, D.~R., {Moran}, J.~M., {et~al.} 1995, \apj, 440,
  619

\bibitem[{{Gressel} {et~al.}(2013){Gressel}, {Elstner}, \&
  {Ziegler}}]{gressel13}
{Gressel}, O., {Elstner}, D., \& {Ziegler}, U. 2013, \aap, 560, A93

\bibitem[{{Han} {et~al.}(1999){Han}, {Beck}, {Ehle}, {Haynes}, \&
  {Wielebinski}}]{han99}
{Han}, J.~L., {Beck}, R., {Ehle}, M., {Haynes}, R.~F., \& {Wielebinski}, R.
  1999, \aap, 348, 405

\bibitem[{{Han} {et~al.}(2004){Han}, {Ferriere}, \& {Manchester}}]{han04}
{Han}, J.~L., {Ferriere}, K., \& {Manchester}, R.~N. 2004, \apj, 610, 820

\bibitem[{{Han} {et~al.}(2006){Han}, {Manchester}, {Lyne}, {Qiao}, \& {van
  Straten}}]{han06}
{Han}, J.~L., {Manchester}, R.~N., {Lyne}, A.~G., {Qiao}, G.~J., \& {van
  Straten}, W. 2006, \apj, 642, 868

\bibitem[{{Hanasz} {et~al.}(2013){Hanasz}, {Lesch}, {Naab}, {Gawryszczak},
  {Kowalik}, \& {W{\'o}lta{\'n}ski}}]{hanasz13}
{Hanasz}, M., {Lesch}, H., {Naab}, T., {et~al.} 2013, \apjl, 777, L38

\bibitem[{{Hanasz} {et~al.}(2002){Hanasz}, {Otmianowska-Mazur}, \&
  {Lesch}}]{hanasz02}
{Hanasz}, M., {Otmianowska-Mazur}, K., \& {Lesch}, H. 2002, \aap, 386, 347

\bibitem[{{Hanasz} {et~al.}(2009){Hanasz}, {W{\'o}lta{\'n}ski}, \&
  {Kowalik}}]{hanasz09}
{Hanasz}, M., {W{\'o}lta{\'n}ski}, D., \& {Kowalik}, K. 2009, \apjl, 706, L155

\bibitem[{{Hanayama} {et~al.}(2005){Hanayama}, {Takahashi}, {Kotake}, {Oguri},
  {Ichiki}, \& {Ohno}}]{hanayama05}
{Hanayama}, H., {Takahashi}, K., {Kotake}, K., {et~al.} 2005, \apj, 633, 941

\bibitem[{{Harnett} {et~al.}(2004){Harnett}, {Ehle}, {Fletcher}, {Beck},
  {Haynes}, {Ryder}, {Thierbach}, \& {Wielebinski}}]{harnett04}
{Harnett}, J., {Ehle}, M., {Fletcher}, A., {et~al.} 2004, \aap, 421, 571

\bibitem[{{Haverkorn} {et~al.}(2008){Haverkorn}, {Brown}, {Gaensler}, \&
  {McClure-Griffiths}}]{haverkorn08}
{Haverkorn}, M., {Brown}, J.~C., {Gaensler}, B.~M., \& {McClure-Griffiths},
  N.~M. 2008, \apj, 680, 362

\bibitem[{{Haynes} {et~al.}(1991){Haynes}, {Klein}, {Wayte}, {Wielebinski},
  {Murray}, {Bajaja}, {Meinert}, {Buczilowski}, {Harnett}, {Hunt}, {Wark}, \&
  {Sciacca}}]{haynes91}
{Haynes}, R.~F., {Klein}, U., {Wayte}, S.~R., {et~al.} 1991, \aap, 252, 475

\bibitem[{{Heald}(2015)}]{heald15}
{Heald}, G. 2015, in Astrophysics and Space Science Library, Vol. 407, Magnetic
  Fields in Diffuse Media, ed. A.~{Lazarian}, E.~M. {de Gouveia Dal Pino}, \&
  C.~{Melioli}, 41

\bibitem[{{Heald} {et~al.}(2015){Heald}, {Beck}, {de Blok}, {Dettmar},
  {Fletcher}, {Gaensler}, {Haverkorn}, {Heesen}, {Horellou}, {Krause}, {Mao},
  {Oppermann}, {Scaife}, {Sokoloff}, {Stil}, {Tabatabaei}, {Takahashi},
  {Taylor}, \& {Williams}}]{heald+15}
{Heald}, G., {Beck}, R., {de Blok}, W.~J.~G., {et~al.} 2015, Advancing
  Astrophysics with the Square Kilometre Array (AASKA14), 106

\bibitem[{{Heald} {et~al.}(2009){Heald}, {Braun}, \& {Edmonds}}]{heald+09}
{Heald}, G., {Braun}, R., \& {Edmonds}, R. 2009, \aap, 503, 409

\bibitem[{{Heesen} {et~al.}(2009{\natexlab{a}}){Heesen}, {Beck}, {Krause}, \&
  {Dettmar}}]{heesen09a}
{Heesen}, V., {Beck}, R., {Krause}, M., \& {Dettmar}, R.-J. 2009{\natexlab{a}},
  \aap, 494, 563

\bibitem[{{Heesen} {et~al.}(2011{\natexlab{a}}){Heesen}, {Beck}, {Krause}, \&
  {Dettmar}}]{heesen11a}
{Heesen}, V., {Beck}, R., {Krause}, M., \& {Dettmar}, R.-J. 2011{\natexlab{a}},
  \aap, 535, A79

\bibitem[{{Heesen} {et~al.}(2014){Heesen}, {Brinks}, {Leroy}, {Heald}, {Braun},
  {Bigiel}, \& {Beck}}]{heesen14}
{Heesen}, V., {Brinks}, E., {Leroy}, A.~K., {et~al.} 2014, \aj, 147, 103

\bibitem[{{Heesen} {et~al.}(2009{\natexlab{b}}){Heesen}, {Krause}, {Beck}, \&
  {Dettmar}}]{heesen09b}
{Heesen}, V., {Krause}, M., {Beck}, R., \& {Dettmar}, R.-J. 2009{\natexlab{b}},
  \aap, 506, 1123

\bibitem[{{Heesen} {et~al.}(2011{\natexlab{b}}){Heesen}, {Rau}, {Rupen},
  {Brinks}, \& {Hunter}}]{heesen11b}
{Heesen}, V., {Rau}, U., {Rupen}, M.~P., {Brinks}, E., \& {Hunter}, D.~A.
  2011{\natexlab{b}}, \apjl, 739, L23

\bibitem[{{Heiles} \& {Crutcher}(2005)}]{heiles05}
{Heiles}, C. \& {Crutcher}, R. 2005, in Lecture Notes in Physics, Berlin
  Springer Verlag, Vol. 664, Cosmic Magnetic Fields, ed. R.~{Wielebinski} \&
  R.~{Beck}, 137

\bibitem[{{Helfer} {et~al.}(2003){Helfer}, {Thornley}, {Regan}, {Wong},
  {Sheth}, {Vogel}, {Blitz}, \& {Bock}}]{helfer03}
{Helfer}, T.~T., {Thornley}, M.~D., {Regan}, M.~W., {et~al.} 2003, \apjs, 145,
  259

\bibitem[{{Hildebrand} {et~al.}(2009){Hildebrand}, {Kirby}, {Dotson}, {Houde},
  \& {Vaillancourt}}]{hildebrand09}
{Hildebrand}, R.~H., {Kirby}, L., {Dotson}, J.~L., {Houde}, M., \&
  {Vaillancourt}, J.~E. 2009, \apj, 696, 567

\bibitem[{{Hill} {et~al.}(2012){Hill}, {Joung}, {Mac Low}, {Benjamin},
  {Haffner}, {Klingenberg}, \& {Waagan}}]{hill12}
{Hill}, A.~S., {Joung}, M.~R., {Mac Low}, M.-M., {et~al.} 2012, \apj, 750, 104

\bibitem[{{Hoang} \& {Lazarian}(2008)}]{hoang08}
{Hoang}, T. \& {Lazarian}, A. 2008, \mnras, 388, 117

\bibitem[{{Hoang} \& {Lazarian}(2014)}]{hoang14}
{Hoang}, T. \& {Lazarian}, A. 2014, \mnras, 438, 680

\bibitem[{{Horellou} \& {Fletcher}(2014)}]{horellou14}
{Horellou}, C. \& {Fletcher}, A. 2014, \mnras, 441, 2049

\bibitem[{{Houde} {et~al.}(2013){Houde}, {Fletcher}, {Beck}, {Hildebrand},
  {Vaillancourt}, \& {Stil}}]{houde13}
{Houde}, M., {Fletcher}, A., {Beck}, R., {et~al.} 2013, \apj, 766, 49

\bibitem[{{Houde} {et~al.}(2011){Houde}, {Rao}, {Vaillancourt}, \&
  {Hildebrand}}]{houde11}
{Houde}, M., {Rao}, R., {Vaillancourt}, J.~E., \& {Hildebrand}, R.~H. 2011,
  \apj, 733, 109

\bibitem[{{Houde} {et~al.}(2009){Houde}, {Vaillancourt}, {Hildebrand},
  {Chitsazzadeh}, \& {Kirby}}]{houde09}
{Houde}, M., {Vaillancourt}, J.~E., {Hildebrand}, R.~H., {Chitsazzadeh}, S., \&
  {Kirby}, L. 2009, \apj, 706, 1504

\bibitem[{{Hummel} \& {Saikia}(1991)}]{hummel91}
{Hummel}, E. \& {Saikia}, D.~J. 1991, \aap, 249, 43

\bibitem[{{Ideguchi} {et~al.}(2014){Ideguchi}, {Tashiro}, {Akahori},
  {Takahashi}, \& {Ryu}}]{ideguchi14}
{Ideguchi}, S., {Tashiro}, Y., {Akahori}, T., {Takahashi}, K., \& {Ryu}, D.
  2014, \apj, 792, 51

\bibitem[{{Irwin} {et~al.}(2012){Irwin}, {Beck}, {Benjamin}, {Dettmar},
  {English}, {Heald}, {Henriksen}, {Johnson}, {Krause}, {Li}, {Miskolczi},
  {Mora}, {Murphy}, {Oosterloo}, {Porter}, {Rand}, {Saikia}, {Schmidt},
  {Strong}, {Walterbos}, {Wang}, \& {Wiegert}}]{irwin12a}
{Irwin}, J., {Beck}, R., {Benjamin}, R.~A., {et~al.} 2012, \aj, 144, 43

\bibitem[{{Irwin} {et~al.}(2013){Irwin}, {Krause}, {English}, {Beck}, {Murphy},
  {Wiegert}, {Heald}, {Walterbos}, {Rand}, \& {Porter}}]{irwin13}
{Irwin}, J., {Krause}, M., {English}, J., {et~al.} 2013, \aj, 146, 164

\bibitem[{{Jansson} \& {Farrar}(2012)}]{jansson12a}
{Jansson}, R. \& {Farrar}, G.~R. 2012, \apj, 757, 14

\bibitem[{{Johnston-Hollitt} {et~al.}(2015){Johnston-Hollitt}, {Govoni},
  {Beck}, {Dehghan}, {Pratley}, {Akahori}, {Heald}, {Agudo}, {Bonafede},
  {Carretti}, {Clarke}, {Colafrancesco}, {Ensslin}, {Feretti}, {Gaensler},
  {Haverkorn}, {Mao}, {Oppermann}, {Rudnick}, {Scaife}, {Schnitzeler}, {Stil},
  {Taylor}, \& {Vacca}}]{johnston15}
{Johnston-Hollitt}, M., {Govoni}, F., {Beck}, R., {et~al.} 2015, Advancing
  Astrophysics with the Square Kilometre Array (AASKA14), 92

\bibitem[{{Kepley} {et~al.}(2010){Kepley}, {M{\"u}hle}, {Everett}, {Zweibel},
  {Wilcots}, \& {Klein}}]{kepley10}
{Kepley}, A.~A., {M{\"u}hle}, S., {Everett}, J., {et~al.} 2010, \apj, 712, 536

\bibitem[{{Kim} {et~al.}(2006){Kim}, {Kim}, \& {Ostriker}}]{kim06}
{Kim}, C.-G., {Kim}, W.-T., \& {Ostriker}, E.~C. 2006, \apjl, 649, L13

\bibitem[{{Kim} {et~al.}(2002){Kim}, {Ostriker}, \& {Stone}}]{kim02}
{Kim}, W.-T., {Ostriker}, E.~C., \& {Stone}, J.~M. 2002, \apj, 581, 1080

\bibitem[{{Kim} \& {Stone}(2012)}]{kim12}
{Kim}, W.-T. \& {Stone}, J.~M. 2012, \apj, 751, 124

\bibitem[{{Kleeorin} \& {Rogachevskii}(1990)}]{kleeorin90}
{Kleeorin}, N.~I. \& {Rogachevskii}, I.~V. 1990, in ESA Special Publication,
  Vol. 311, ESA Special Publication, ed. T.~D. {Guyenne} \& J.~J. {Hunt},
  21--23

\bibitem[{{Klein} \& {Fletcher}(2015)}]{klein15}
{Klein}, U. \& {Fletcher}, A. 2015, {Galactic and Intergalactic Magnetic
  Fields} (Heidelberg: Springer)

\bibitem[{{Knapik} {et~al.}(2000){Knapik}, {Soida}, {Dettmar}, {Beck}, \&
  {Urbanik}}]{knapik00}
{Knapik}, J., {Soida}, M., {Dettmar}, R.-J., {Beck}, R., \& {Urbanik}, M. 2000,
  \aap, 362, 910

\bibitem[{{Kotarba} {et~al.}(2010){Kotarba}, {Karl}, {Naab}, {Johansson},
  {Dolag}, {Lesch}, \& {Stasyszyn}}]{kotarba10}
{Kotarba}, H., {Karl}, S.~J., {Naab}, T., {et~al.} 2010, \apj, 716, 1438

\bibitem[{{Kotarba} {et~al.}(2009){Kotarba}, {Lesch}, {Dolag}, {Naab},
  {Johansson}, \& {Stasyszyn}}]{kotarba09}
{Kotarba}, H., {Lesch}, H., {Dolag}, K., {et~al.} 2009, \mnras, 397, 733

\bibitem[{{Krause} \& {Beck}(1998)}]{krause98}
{Krause}, F. \& {Beck}, R. 1998, \aap, 335, 789

\bibitem[{{Krause}(1990)}]{krause90}
{Krause}, M. 1990, in IAU Symposium, Vol. 140, Galactic and Intergalactic
  Magnetic Fields, ed. R.~{Beck}, R.~{Wielebinski}, \& P.~P. {Kronberg},
  187--196

\bibitem[{{Krause}(2009)}]{krause09}
{Krause}, M. 2009, in Revista Mexicana de Astronomia y Astrofisica Conference
  Series, Vol.~36, Revista Mexicana de Astronomia y Astrofisica Conference
  Series, 25--29

\bibitem[{{Krause}(2014)}]{krause14}
{Krause}, M. 2014, ArXiv e-prints, arXiv:1401.1317

\bibitem[{{Krause} {et~al.}(1989){Krause}, {Beck}, \& {Hummel}}]{krause89b}
{Krause}, M., {Beck}, R., \& {Hummel}, E. 1989, \aap, 217, 17

\bibitem[{{Krause} \& {L{\"o}hr}(2004)}]{krause04}
{Krause}, M. \& {L{\"o}hr}, A. 2004, \aap, 420, 115

\bibitem[{{Krause} {et~al.}(2006){Krause}, {Wielebinski}, \&
  {Dumke}}]{krause06}
{Krause}, M., {Wielebinski}, R., \& {Dumke}, M. 2006, \aap, 448, 133

\bibitem[{{Kronberg} {et~al.}(1992){Kronberg}, {Perry}, \&
  {Zukowski}}]{kronberg92}
{Kronberg}, P.~P., {Perry}, J.~J., \& {Zukowski}, E.~L.~H. 1992, \apj, 387, 528

\bibitem[{{Kulpa-Dybe{\l}} {et~al.}(2011){Kulpa-Dybe{\l}}, {Otmianowska-Mazur},
  {Kulesza-{\.Z}ydzik}, {Hanasz}, {Kowal}, {W{\'o}lta{\'n}ski}, \&
  {Kowalik}}]{kulpa11}
{Kulpa-Dybe{\l}}, K., {Otmianowska-Mazur}, K., {Kulesza-{\.Z}ydzik}, B.,
  {et~al.} 2011, \apjl, 733, L18

\bibitem[{{Kulsrud} {et~al.}(1997){Kulsrud}, {Cen}, {Ostriker}, \&
  {Ryu}}]{kulsrud97}
{Kulsrud}, R.~M., {Cen}, R., {Ostriker}, J.~P., \& {Ryu}, D. 1997, \apj, 480,
  481

\bibitem[{{Lacki} \& {Beck}(2013)}]{lacki13}
{Lacki}, B.~C. \& {Beck}, R. 2013, \mnras, 430, 3171

\bibitem[{{Lacki} {et~al.}(2010){Lacki}, {Thompson}, \& {Quataert}}]{lacki10}
{Lacki}, B.~C., {Thompson}, T.~A., \& {Quataert}, E. 2010, \apj, 717, 1

\bibitem[{{Laine} \& {Beck}(2008)}]{laine08}
{Laine}, S. \& {Beck}, R. 2008, \apj, 673, 128

\bibitem[{{Lazar} {et~al.}(2009){Lazar}, {Schlickeiser}, {Wielebinski}, \&
  {Poedts}}]{lazar09}
{Lazar}, M., {Schlickeiser}, R., {Wielebinski}, R., \& {Poedts}, S. 2009, \apj,
  693, 1133

\bibitem[{{Lerche} \& {Schlickeiser}(1982{\natexlab{a}})}]{lerche82a}
{Lerche}, I. \& {Schlickeiser}, R. 1982{\natexlab{a}}, \aap, 107, 148

\bibitem[{{Lerche} \& {Schlickeiser}(1982{\natexlab{b}})}]{lerche82b}
{Lerche}, I. \& {Schlickeiser}, R. 1982{\natexlab{b}}, \aap, 116, 10

\bibitem[{{Li} \& {Henning}(2011)}]{li11}
{Li}, H.-B. \& {Henning}, T. 2011, \nat, 479, 499

\bibitem[{{Liu} \& {Gao}(2010)}]{liu10}
{Liu}, F. \& {Gao}, Y. 2010, \apj, 713, 524

\bibitem[{{Lobo Gomes} {et~al.}(2015){Lobo Gomes}, {Magalh{\~a}es}, {Pereyra},
  \& {Rodrigues}}]{lobo15}
{Lobo Gomes}, A., {Magalh{\~a}es}, A.~M., {Pereyra}, A., \& {Rodrigues}, C.~V.
  2015, \apj, 806, 94

\bibitem[{{Lou} {et~al.}(1999){Lou}, {Han}, \& {Fan}}]{lou99}
{Lou}, Y.-Q., {Han}, J.~L., \& {Fan}, Z. 1999, \mnras, 308, L1

\bibitem[{{Machida} {et~al.}(2013){Machida}, {Nakamura}, {Kudoh}, {Akahori},
  {Sofue}, \& {Matsumoto}}]{machida13}
{Machida}, M., {Nakamura}, K.~E., {Kudoh}, T., {et~al.} 2013, \apj, 764, 81

\bibitem[{{Mao} {et~al.}(2010){Mao}, {Gaensler}, {Haverkorn}, {Zweibel},
  {Madsen}, {McClure-Griffiths}, {Shukurov}, \& {Kronberg}}]{mao10}
{Mao}, S.~A., {Gaensler}, B.~M., {Haverkorn}, M., {et~al.} 2010, \apj, 714,
  1170

\bibitem[{{Mao} {et~al.}(2008){Mao}, {Gaensler}, {Stanimirovi{\'c}},
  {Haverkorn}, {McClure-Griffiths}, {Staveley-Smith}, \& {Dickey}}]{mao08}
{Mao}, S.~A., {Gaensler}, B.~M., {Stanimirovi{\'c}}, S., {et~al.} 2008, \apj,
  688, 1029

\bibitem[{{Mao} {et~al.}(2012){Mao}, {McClure-Griffiths}, {Gaensler},
  {Haverkorn}, {Beck}, {McConnell}, {Wolleben}, {Stanimirovi{\'c}}, {Dickey},
  \& {Staveley-Smith}}]{mao12}
{Mao}, S.~A., {McClure-Griffiths}, N.~M., {Gaensler}, B.~M., {et~al.} 2012,
  \apj, 759, 25

\bibitem[{{Mao} {et~al.}(2015){Mao}, {Zweibel}, {Fletcher}, {Ott}, \&
  {Tabatabaei}}]{mao15}
{Mao}, S.~A., {Zweibel}, E., {Fletcher}, A., {Ott}, J., \& {Tabatabaei}, F.
  2015, \apj, 800, 92

\bibitem[{{Meidt} {et~al.}(2009){Meidt}, {Rand}, \& {Merrifield}}]{meidt09}
{Meidt}, S.~E., {Rand}, R.~J., \& {Merrifield}, M.~R. 2009, \apj, 702, 277

\bibitem[{{Meier}(2014)}]{meier14}
{Meier}, D.~S. 2014, in IAU Symposium, Vol. 303, IAU Symposium, ed. L.~O.
  {Sjouwerman}, C.~C. {Lang}, \& J.~{Ott}, 66--68

\bibitem[{{Mikhailov} {et~al.}(2014){Mikhailov}, {Kasparova}, {Moss}, {Beck},
  {Sokoloff}, \& {Zasov}}]{mikhailov14}
{Mikhailov}, E., {Kasparova}, A., {Moss}, D., {et~al.} 2014, \aap, 568, A66

\bibitem[{{Mora} \& {Krause}(2013)}]{mora13}
{Mora}, S.~C. \& {Krause}, M. 2013, \aap, 560, A42

\bibitem[{{Moss}(1998)}]{moss98a}
{Moss}, D. 1998, \mnras, 297, 860

\bibitem[{{Moss} {et~al.}(2013){Moss}, {Beck}, {Sokoloff}, {Stepanov},
  {Krause}, \& {Arshakian}}]{moss13}
{Moss}, D., {Beck}, R., {Sokoloff}, D., {et~al.} 2013, \aap, 556, A147

\bibitem[{{Moss} {et~al.}(1998){Moss}, {Shukurov}, \& {Sokoloff}}]{moss98c}
{Moss}, D., {Shukurov}, A., \& {Sokoloff}, D. 1998, Geophysical and
  Astrophysical Fluid Dynamics, 89, 285

\bibitem[{{Moss} {et~al.}(2010){Moss}, {Sokoloff}, {Beck}, \&
  {Krause}}]{moss10}
{Moss}, D., {Sokoloff}, D., {Beck}, R., \& {Krause}, M. 2010, \aap, 512, A61

\bibitem[{{Moss} {et~al.}(2014){Moss}, {Sokoloff}, {Beck}, \&
  {Krause}}]{moss14}
{Moss}, D., {Sokoloff}, D., {Beck}, R., \& {Krause}, M. 2014, \aap, 566, A40

\bibitem[{{Moss} {et~al.}(2012){Moss}, {Stepanov}, {Arshakian}, {Beck},
  {Krause}, \& {Sokoloff}}]{moss12}
{Moss}, D., {Stepanov}, R., {Arshakian}, T.~G., {et~al.} 2012, \aap, 537, A68

\bibitem[{{Moss} {et~al.}(2015){Moss}, {Stepanov}, {Krause}, {Beck}, \&
  {Sokoloff}}]{moss15}
{Moss}, D., {Stepanov}, R., {Krause}, M., {Beck}, R., \& {Sokoloff}, D. 2015,
  \aap, 578, A94

\bibitem[{{Mulcahy} {et~al.}(2014){Mulcahy}, {Horneffer}, {Beck}, {Heald},
  {Fletcher}, {Scaife}, {Adebahr}, {Anderson}, {Bonafede}, {Br{\"u}ggen},
  {Brunetti}, {Chy{\.z}y}, {Conway}, {Dettmar}, {En{\ss}lin}, {Haverkorn},
  {Horellou}, {Iacobelli}, {Israel}, {Junklewitz}, {Jurusik}, {K{\"o}hler},
  {Kuniyoshi}, {Orr{\'u}}, {Paladino}, {Pizzo}, {Reich}, \&
  {R{\"o}ttgering}}]{mulcahy14}
{Mulcahy}, D.~D., {Horneffer}, A., {Beck}, R., {et~al.} 2014, \aap, 568, A74

\bibitem[{{Murgia} {et~al.}(2005){Murgia}, {Helfer}, {Ekers}, {Blitz},
  {Moscadelli}, {Wong}, \& {Paladino}}]{murgia05}
{Murgia}, M., {Helfer}, T.~T., {Ekers}, R., {et~al.} 2005, \aap, 437, 389

\bibitem[{{Murphy}(2009)}]{murphy09}
{Murphy}, E.~J. 2009, \apj, 706, 482

\bibitem[{{Niklas}(1995)}]{niklas95}
{Niklas}, S. 1995, PhD thesis, University of Bonn

\bibitem[{{Niklas} \& {Beck}(1997)}]{niklas97}
{Niklas}, S. \& {Beck}, R. 1997, \aap, 320, 54

\bibitem[{{Nota} \& {Katgert}(2010)}]{nota10}
{Nota}, T. \& {Katgert}, P. 2010, \aap, 513, A65

\bibitem[{{Noutsos}(2009)}]{noutsos09}
{Noutsos}, A. 2009, in IAU Symposium, Vol. 259, IAU Symposium, ed. K.~G.
  {Strassmeier}, A.~G. {Kosovichev}, \& J.~E. {Beckman}, 15--24

\bibitem[{{Noutsos}(2012)}]{noutsos12}
{Noutsos}, A. 2012, \ssr, 166, 307

\bibitem[{{Ohno} \& {Shibata}(1993)}]{ohno93}
{Ohno}, H. \& {Shibata}, S. 1993, \mnras, 262, 953

\bibitem[{{Opher} {et~al.}(2009){Opher}, {Bibi}, {Toth}, {Richardson},
  {Izmodenov}, \& {Gombosi}}]{opher09}
{Opher}, M., {Bibi}, F.~A., {Toth}, G., {et~al.} 2009, \nat, 462, 1036

\bibitem[{{Oppermann} {et~al.}(2012){Oppermann}, {Junklewitz}, {Robbers},
  {Bell}, {En{\ss}lin}, {Bonafede}, {Braun}, {Brown}, {Clarke}, {Feain},
  {Gaensler}, {Hammond}, {Harvey-Smith}, {Heald}, {Johnston-Hollitt}, {Klein},
  {Kronberg}, {Mao}, {McClure-Griffiths}, {O'Sullivan}, {Pratley}, {Robishaw},
  {Roy}, {Schnitzeler}, {Sotomayor-Beltran}, {Stevens}, {Stil}, {Sunstrum},
  {Tanna}, {Taylor}, \& {Van Eck}}]{oppermann12}
{Oppermann}, N., {Junklewitz}, H., {Robbers}, G., {et~al.} 2012, \aap, 542, A93

\bibitem[{{Orlando} \& {Strong}(2013)}]{orlando13}
{Orlando}, E. \& {Strong}, A. 2013, \mnras, 436, 2127

\bibitem[{{Otmianowska-Mazur} {et~al.}(2002){Otmianowska-Mazur}, {Elstner},
  {Soida}, \& {Urbanik}}]{otmian02}
{Otmianowska-Mazur}, K., {Elstner}, D., {Soida}, M., \& {Urbanik}, M. 2002,
  \aap, 384, 48

\bibitem[{{Pakmor} {et~al.}(2014){Pakmor}, {Marinacci}, \&
  {Springel}}]{pakmor14}
{Pakmor}, R., {Marinacci}, F., \& {Springel}, V. 2014, \apjl, 783, L20

\bibitem[{{Pakmor} \& {Springel}(2013)}]{pakmor13}
{Pakmor}, R. \& {Springel}, V. 2013, \mnras, 432, 176

\bibitem[{{Patrikeev} {et~al.}(2006){Patrikeev}, {Fletcher}, {Stepanov},
  {Beck}, {Berkhuijsen}, {Frick}, \& {Horellou}}]{patrikeev06}
{Patrikeev}, I., {Fletcher}, A., {Stepanov}, R., {et~al.} 2006, \aap, 458, 441

\bibitem[{{Pierini} {et~al.}(2003){Pierini}, {Popescu}, {Tuffs}, \&
  {V{\"o}lk}}]{pierini03}
{Pierini}, D., {Popescu}, C.~C., {Tuffs}, R.~J., \& {V{\"o}lk}, H.~J. 2003,
  \aap, 409, 907

\bibitem[{{Planck Collaboration} {et~al.}(2015{\natexlab{a}}){Planck
  Collaboration}, {Ade}, {Aghanim}, {Alves}, {Arnaud}, {Arzoumanian},
  {Ashdown}, {Aumont}, {Baccigalupi}, {Banday}, \& et~al.}]{ade15b}
{Planck Collaboration}, {Ade}, P.~A.~R., {Aghanim}, N., {et~al.}
  2015{\natexlab{a}}, ArXiv e-prints, arXiv:1502.04123

\bibitem[{{Planck Collaboration} {et~al.}(2015{\natexlab{b}}){Planck
  Collaboration}, {Ade}, {Aghanim}, {Arnaud}, {Arroja}, {Ashdown}, {Aumont},
  {Baccigalupi}, {Ballardini}, {Banday}, \& et~al.}]{ade15a}
{Planck Collaboration}, {Ade}, P.~A.~R., {Aghanim}, N., {et~al.}
  2015{\natexlab{b}}, ArXiv e-prints, afXiv:1502.01594

\bibitem[{{Poedts} \& {Rogava}(2002)}]{poedts02}
{Poedts}, S. \& {Rogava}, A.~D. 2002, \aap, 385, 32

\bibitem[{{Price} \& {Bate}(2008)}]{price08}
{Price}, D.~J. \& {Bate}, M.~R. 2008, \mnras, 385, 1820

\bibitem[{{Rand} \& {Kulkarni}(1989)}]{rand89}
{Rand}, R.~J. \& {Kulkarni}, S.~R. 1989, \apj, 343, 760

\bibitem[{{Rees}(2005)}]{rees05}
{Rees}, M.~J. 2005, in Lecture Notes in Physics, Berlin Springer Verlag, Vol.
  664, Cosmic Magnetic Fields, ed. R.~{Wielebinski} \& R.~{Beck}, 1

\bibitem[{{Reuter} {et~al.}(1994){Reuter}, {Klein}, {Lesch}, {Wielebinski}, \&
  {Kronberg}}]{reuter94}
{Reuter}, H.-P., {Klein}, U., {Lesch}, H., {Wielebinski}, R., \& {Kronberg},
  P.~P. 1994, \aap, 282, 724

\bibitem[{{Rieder} \& {Teyssier}(2015)}]{rieder15}
{Rieder}, M. \& {Teyssier}, R. 2015, ArXiv e-prints, arXiv:1506.00849

\bibitem[{{Robishaw} {et~al.}(2008){Robishaw}, {Quataert}, \&
  {Heiles}}]{robishaw08}
{Robishaw}, T., {Quataert}, E., \& {Heiles}, C. 2008, \apj, 680, 981

\bibitem[{{Rodrigues} {et~al.}(2015){Rodrigues}, {Shukurov}, {Fletcher}, \&
  {Baugh}}]{rodrigues15}
{Rodrigues}, L.~F.~S., {Shukurov}, A., {Fletcher}, A., \& {Baugh}, C.~M. 2015,
  \mnras, 450, 3472

\bibitem[{{Rohde} {et~al.}(1999){Rohde}, {Beck}, \& {Elstner}}]{rohde99}
{Rohde}, R., {Beck}, R., \& {Elstner}, D. 1999, \aap, 350, 423

\bibitem[{{Scarrott} {et~al.}(1987){Scarrott}, {Ward-Thompson}, \&
  {Warren-Smith}}]{scarrott87}
{Scarrott}, S.~M., {Ward-Thompson}, D., \& {Warren-Smith}, R.~F. 1987, \mnras,
  224, 299

\bibitem[{{Schinnerer} {et~al.}(2013){Schinnerer}, {Meidt}, {Pety}, {Hughes},
  {Colombo}, {Garc{\'{\i}}a-Burillo}, {Schuster}, {Dumas}, {Dobbs}, {Leroy},
  {Kramer}, {Thompson}, \& {Regan}}]{schinnerer13}
{Schinnerer}, E., {Meidt}, S.~E., {Pety}, J., {et~al.} 2013, \apj, 779, 42

\bibitem[{{Schleicher} {et~al.}(2010){Schleicher}, {Banerjee}, {Sur},
  {Arshakian}, {Klessen}, {Beck}, \& {Spaans}}]{schleicher10}
{Schleicher}, D.~R.~G., {Banerjee}, R., {Sur}, S., {et~al.} 2010, \aap, 522,
  A115

\bibitem[{{Schleicher} \& {Beck}(2013)}]{schleicher13}
{Schleicher}, D.~R.~G. \& {Beck}, R. 2013, \aap, 556, A142

\bibitem[{{Schlickeiser}(2012)}]{schlickeiser12}
{Schlickeiser}, R. 2012, Physical Review Letters, 109, 261101

\bibitem[{{Schlickeiser} \& {Felten}(2013)}]{schlickeiser13}
{Schlickeiser}, R. \& {Felten}, T. 2013, \apj, 778, 39

\bibitem[{{Sellwood} \& {Balbus}(1999)}]{sellwood99}
{Sellwood}, J.~A. \& {Balbus}, S.~A. 1999, \apj, 511, 660

\bibitem[{{Seymour} {et~al.}(2008){Seymour}, {Dwelly}, {Moss}, {McHardy},
  {Zoghbi}, {Rieke}, {Page}, {Hopkins}, \& {Loaring}}]{seymour08}
{Seymour}, N., {Dwelly}, T., {Moss}, D., {et~al.} 2008, \mnras, 386, 1695

\bibitem[{{Shi} {et~al.}(2011){Shi}, {Helou}, {Yan}, {Armus}, {Wu}, {Papovich},
  \& {Stierwalt}}]{shi11}
{Shi}, Y., {Helou}, G., {Yan}, L., {et~al.} 2011, \apj, 733, 87

\bibitem[{{Shibata} \& {Matsumoto}(1991)}]{shibata91}
{Shibata}, K. \& {Matsumoto}, R. 1991, \nat, 353, 633

\bibitem[{{Shneider} {et~al.}(2014){Shneider}, {Haverkorn}, {Fletcher}, \&
  {Shukurov}}]{shneider14}
{Shneider}, C., {Haverkorn}, M., {Fletcher}, A., \& {Shukurov}, A. 2014, \aap,
  567, A82

\bibitem[{{Shukurov}(1998)}]{shukurov98}
{Shukurov}, A. 1998, \mnras, 299, L21

\bibitem[{{Shukurov}(2005)}]{shukurov05}
{Shukurov}, A. 2005, in Lecture Notes in Physics, Berlin Springer Verlag, Vol.
  664, Cosmic Magnetic Fields, ed. R.~{Wielebinski} \& R.~{Beck}, 113

\bibitem[{{Shukurov} {et~al.}(2006){Shukurov}, {Sokoloff}, {Subramanian}, \&
  {Brandenburg}}]{shukurov06}
{Shukurov}, A., {Sokoloff}, D., {Subramanian}, K., \& {Brandenburg}, A. 2006,
  \aap, 448, L33

\bibitem[{{Siejkowski} {et~al.}(2014){Siejkowski}, {Otmianowska-Mazur},
  {Soida}, {Bomans}, \& {Hanasz}}]{siejkowski14}
{Siejkowski}, H., {Otmianowska-Mazur}, K., {Soida}, M., {Bomans}, D.~J., \&
  {Hanasz}, M. 2014, \aap, 562, A136

\bibitem[{{Soida} {et~al.}(2002){Soida}, {Beck}, {Urbanik}, \&
  {Braine}}]{soida02}
{Soida}, M., {Beck}, R., {Urbanik}, M., \& {Braine}, J. 2002, \aap, 394, 47

\bibitem[{{Soida} {et~al.}(2011){Soida}, {Krause}, {Dettmar}, \&
  {Urbanik}}]{soida11}
{Soida}, M., {Krause}, M., {Dettmar}, R.-J., \& {Urbanik}, M. 2011, \aap, 531,
  A127

\bibitem[{{Sokoloff} {et~al.}(1992){Sokoloff}, {Shukurov}, \&
  {Krause}}]{sokoloff92}
{Sokoloff}, D., {Shukurov}, A., \& {Krause}, M. 1992, \aap, 264, 396

\bibitem[{{Sokoloff} {et~al.}(1998){Sokoloff}, {Bykov}, {Shukurov},
  {Berkhuijsen}, {Beck}, \& {Poezd}}]{sokoloff98}
{Sokoloff}, D.~D., {Bykov}, A.~A., {Shukurov}, A., {et~al.} 1998, \mnras, 299,
  189

\bibitem[{{Stenflo}(2012)}]{stenflo12}
{Stenflo}, J.~O. 2012, \aap, 547, A93

\bibitem[{{Stepanov} {et~al.}(2008){Stepanov}, {Arshakian}, {Beck}, {Frick}, \&
  {Krause}}]{stepanov08}
{Stepanov}, R., {Arshakian}, T.~G., {Beck}, R., {Frick}, P., \& {Krause}, M.
  2008, \aap, 480, 45

\bibitem[{{Stepanov} {et~al.}(2014){Stepanov}, {Shukurov}, {Fletcher}, {Beck},
  {La Porta}, \& {Tabatabaei}}]{stepanov14}
{Stepanov}, R., {Shukurov}, A., {Fletcher}, A., {et~al.} 2014, \mnras, 437,
  2201

\bibitem[{{Stil} {et~al.}(2009){Stil}, {Krause}, {Beck}, \& {Taylor}}]{stil09}
{Stil}, J.~M., {Krause}, M., {Beck}, R., \& {Taylor}, A.~R. 2009, \apj, 693,
  1392

\bibitem[{{Strassmeier} \& {Ilyin}(2009)}]{strassmeier09}
{Strassmeier}, K.~G. \& {Ilyin}, I.~V. 2009, in Science with the VLT in the ELT
  Era, ed. A.~{Moorwood}, 255

\bibitem[{{Strong} {et~al.}(2000){Strong}, {Moskalenko}, \&
  {Reimer}}]{strong00}
{Strong}, A.~W., {Moskalenko}, I.~V., \& {Reimer}, O. 2000, \apj, 537, 763

\bibitem[{{Sun} \& {Reich}(2010)}]{sun10}
{Sun}, X.-H. \& {Reich}, W. 2010, Research in Astronomy and Astrophysics, 10,
  1287

\bibitem[{{Sun} {et~al.}(2008){Sun}, {Reich}, {Waelkens}, \&
  {En{\ss}lin}}]{sun08}
{Sun}, X.~H., {Reich}, W., {Waelkens}, A., \& {En{\ss}lin}, T.~A. 2008, \aap,
  477, 573

\bibitem[{{Sun} {et~al.}(2015){Sun}, {Rudnick}, {Akahori}, {Anderson}, {Bell},
  {Bray}, {Farnes}, {Ideguchi}, {Kumazaki}, {O'Brien}, {O'Sullivan}, {Scaife},
  {Stepanov}, {Stil}, {Takahashi}, {van Weeren}, \& {Wolleben}}]{sun15}
{Sun}, X.~H., {Rudnick}, L., {Akahori}, T., {et~al.} 2015, \aj, 149, 60

\bibitem[{{Tabatabaei} {et~al.}(2007){Tabatabaei}, {Beck}, {Kr{\"u}gel},
  {Krause}, {Berkhuijsen}, {Gordon}, \& {Menten}}]{taba07}
{Tabatabaei}, F.~S., {Beck}, R., {Kr{\"u}gel}, E., {et~al.} 2007, \aap, 475,
  133

\bibitem[{{Tabatabaei} {et~al.}(2013{\natexlab{a}}){Tabatabaei}, {Berkhuijsen},
  {Frick}, {Beck}, \& {Schinnerer}}]{taba13b}
{Tabatabaei}, F.~S., {Berkhuijsen}, E.~M., {Frick}, P., {Beck}, R., \&
  {Schinnerer}, E. 2013{\natexlab{a}}, \aap, 557, A129

\bibitem[{{Tabatabaei} {et~al.}(2008){Tabatabaei}, {Krause}, {Fletcher}, \&
  {Beck}}]{taba08}
{Tabatabaei}, F.~S., {Krause}, M., {Fletcher}, A., \& {Beck}, R. 2008, \aap,
  490, 1005

\bibitem[{{Tabatabaei} {et~al.}(2013{\natexlab{b}}){Tabatabaei}, {Schinnerer},
  {Murphy}, {Beck}, {Groves}, {Meidt}, {Krause}, {Rix}, {Sandstrom}, {Crocker},
  {Galametz}, {Helou}, {Wilson}, {Kennicutt}, {Calzetti}, {Draine}, {Aniano},
  {Dale}, {Dumas}, {Engelbracht}, {Gordon}, {Hinz}, {Kreckel}, {Montiel}, \&
  {Roussel}}]{taba13a}
{Tabatabaei}, F.~S., {Schinnerer}, E., {Murphy}, E.~J., {et~al.}
  2013{\natexlab{b}}, \aap, 552, A19

\bibitem[{{Tamburro} {et~al.}(2009){Tamburro}, {Rix}, {Leroy}, {Mac Low},
  {Walter}, {Kennicutt}, {Brinks}, \& {de Blok}}]{tamburro09}
{Tamburro}, D., {Rix}, H.-W., {Leroy}, A.~K., {et~al.} 2009, \aj, 137, 4424

\bibitem[{{Tang} {et~al.}(2009){Tang}, {Ho}, {Koch}, {Girart}, {Lai}, \&
  {Rao}}]{tang09}
{Tang}, Y.-W., {Ho}, P.~T.~P., {Koch}, P.~M., {et~al.} 2009, \apj, 700, 251

\bibitem[{{Taylor} {et~al.}(2009){Taylor}, {Stil}, \& {Sunstrum}}]{taylor09}
{Taylor}, A.~R., {Stil}, J.~M., \& {Sunstrum}, C. 2009, \apj, 702, 1230

\bibitem[{{Tingay} {et~al.}(2013){Tingay}, {Kaplan}, {McKinley}, {Briggs},
  {Wayth}, {Hurley-Walker}, {Kennewell}, {Smith}, {Zhang}, {Arcus}, {Bhat},
  {Emrich}, {Herne}, {Kudryavtseva}, {Lynch}, {Ord}, {Waterson}, {Barnes},
  {Bell}, {Gaensler}, {Lenc}, {Bernardi}, {Greenhill}, {Kasper}, {Bowman},
  {Jacobs}, {Bunton}, {deSouza}, {Koenig}, {Pathikulangara}, {Stevens},
  {Cappallo}, {Corey}, {Kincaid}, {Kratzenberg}, {Lonsdale}, {McWhirter},
  {Rogers}, {Salah}, {Whitney}, {Deshpande}, {Prabu}, {Udaya Shankar},
  {Srivani}, {Subrahmanyan}, {Ewall-Wice}, {Feng}, {Goeke}, {Morgan},
  {Remillard}, {Williams}, {Hazelton}, {Morales}, {Johnston-Hollitt},
  {Mitchell}, {Procopio}, {Riding}, {Webster}, {Wyithe}, {Oberoi}, {Roshi},
  {Sault}, \& {Williams}}]{tingay13}
{Tingay}, S.~J., {Kaplan}, D.~L., {McKinley}, B., {et~al.} 2013, \aj, 146, 103

\bibitem[{{T{\"u}llmann} {et~al.}(2000){T{\"u}llmann}, {Dettmar}, {Soida},
  {Urbanik}, \& {Rossa}}]{tuellmann00}
{T{\"u}llmann}, R., {Dettmar}, R.-J., {Soida}, M., {Urbanik}, M., \& {Rossa},
  J. 2000, \aap, 364, L36

\bibitem[{{Urbanik} {et~al.}(1997){Urbanik}, {Elstner}, \& {Beck}}]{urbanik97}
{Urbanik}, M., {Elstner}, D., \& {Beck}, R. 1997, \aap, 326, 465

\bibitem[{{Van Eck} {et~al.}(2015){Van Eck}, {Brown}, {Shukurov}, \&
  {Fletcher}}]{vaneck15}
{Van Eck}, C.~L., {Brown}, J.~C., {Shukurov}, A., \& {Fletcher}, A. 2015, \apj,
  799, 35

\bibitem[{{Van Eck} {et~al.}(2011){Van Eck}, {Brown}, {Stil}, {Rae}, {Mao},
  {Gaensler}, {Shukurov}, {Taylor}, {Haverkorn}, {Kronberg}, \&
  {McClure-Griffiths}}]{vaneck11}
{Van Eck}, C.~L., {Brown}, J.~C., {Stil}, J.~M., {et~al.} 2011, \apj, 728, 97

\bibitem[{{van Haarlem} {et~al.}(2013){van Haarlem}, {Wise}, {Gunst}, {Heald},
  {McKean}, {Hessels}, {de Bruyn}, {Nijboer}, {Swinbank}, {Fallows},
  {Brentjens}, {Nelles}, {Beck}, {Falcke}, {Fender}, {H{\"o}randel},
  {Koopmans}, {Mann}, {Miley}, {R{\"o}ttgering}, {Stappers}, {Wijers},
  {Zaroubi}, {van den Akker}, {Alexov}, {Anderson}, {Anderson}, {van Ardenne},
  {Arts}, {Asgekar}, {Avruch}, {Batejat}, {B{\"a}hren}, {Bell}, {Bell}, {van
  Bemmel}, {Bennema}, {Bentum}, {Bernardi}, {Best}, {B{\^i}rzan}, {Bonafede},
  {Boonstra}, {Braun}, {Bregman}, {Breitling}, {van de Brink}, {Broderick},
  {Broekema}, {Brouw}, {Br{\"u}ggen}, {Butcher}, {van Cappellen}, {Ciardi},
  {Coenen}, {Conway}, {Coolen}, {Corstanje}, {Damstra}, {Davies}, {Deller},
  {Dettmar}, {van Diepen}, {Dijkstra}, {Donker}, {Doorduin}, {Dromer}, {Drost},
  {van Duin}, {Eisl{\"o}ffel}, {van Enst}, {Ferrari}, {Frieswijk}, {Gankema},
  {Garrett}, {de Gasperin}, {Gerbers}, {de Geus}, {Grie{\ss}meier}, {Grit},
  {Gruppen}, {Hamaker}, {Hassall}, {Hoeft}, {Holties}, {Horneffer}, {van der
  Horst}, {van Houwelingen}, {Huijgen}, {Iacobelli}, {Intema}, {Jackson},
  {Jelic}, {de Jong}, {Juette}, {Kant}, {Karastergiou}, {Koers}, {Kollen},
  {Kondratiev}, {Kooistra}, {Koopman}, {Koster}, {Kuniyoshi}, {Kramer},
  {Kuper}, {Lambropoulos}, {Law}, {van Leeuwen}, {Lemaitre}, {Loose}, {Maat},
  {Macario}, {Markoff}, {Masters}, {McFadden}, {McKay-Bukowski}, {Meijering},
  {Meulman}, {Mevius}, {Middelberg}, {Millenaar}, {Miller-Jones}, {Mohan},
  {Mol}, {Morawietz}, {Morganti}, {Mulcahy}, {Mulder}, {Munk}, {Nieuwenhuis},
  {van Nieuwpoort}, {Noordam}, {Norden}, {Noutsos}, {Offringa}, {Olofsson},
  {Omar}, {Orr{\'u}}, {Overeem}, {Paas}, {Pandey-Pommier}, {Pandey}, {Pizzo},
  {Polatidis}, {Rafferty}, {Rawlings}, {Reich}, {de Reijer}, {Reitsma},
  {Renting}, {Riemers}, {Rol}, {Romein}, {Roosjen}, {Ruiter}, {Scaife}, {van
  der Schaaf}, {Scheers}, {Schellart}, {Schoenmakers}, {Schoonderbeek},
  {Serylak}, {Shulevski}, {Sluman}, {Smirnov}, {Sobey}, {Spreeuw}, {Steinmetz},
  {Sterks}, {Stiepel}, {Stuurwold}, {Tagger}, {Tang}, {Tasse}, {Thomas},
  {Thoudam}, {Toribio}, {van der Tol}, {Usov}, {van Veelen}, {van der Veen},
  {ter Veen}, {Verbiest}, {Vermeulen}, {Vermaas}, {Vocks}, {Vogt}, {de Vos},
  {van der Wal}, {van Weeren}, {Weggemans}, {Weltevrede}, {White}, {Wijnholds},
  {Wilhelmsson}, {Wucknitz}, {Yatawatta}, {Zarka}, {Zensus}, \& {van
  Zwieten}}]{vanhaarlem13}
{van Haarlem}, M.~P., {Wise}, M.~W., {Gunst}, A.~W., {et~al.} 2013, \aap, 556,
  A2

\bibitem[{{V{\'a}zquez-Semadeni} {et~al.}(2005){V{\'a}zquez-Semadeni}, {Kim},
  \& {Ballesteros-Paredes}}]{vazquez05}
{V{\'a}zquez-Semadeni}, E., {Kim}, J., \& {Ballesteros-Paredes}, J. 2005,
  \apjl, 630, L49

\bibitem[{{V\"olk}(1989)}]{voelk89}
{V\"olk}, H.~J. 1989, \aap, 218, 67

\bibitem[{{Vollmer} {et~al.}(2013){Vollmer}, {Soida}, {Beck}, {Chung},
  {Urbanik}, {Chy{\.z}y}, {Otmianowska-Mazur}, \& {Kenney}}]{vollmer13}
{Vollmer}, B., {Soida}, M., {Beck}, R., {et~al.} 2013, \aap, 553, A116

\bibitem[{{Vollmer} {et~al.}(2007){Vollmer}, {Soida}, {Beck}, {Urbanik},
  {Chy{\.z}y}, {Otmianowska-Mazur}, {Kenney}, \& {van Gorkom}}]{vollmer07}
{Vollmer}, B., {Soida}, M., {Beck}, R., {et~al.} 2007, \aap, 464, L37

\bibitem[{{Wang} \& {Abel}(2009)}]{wang09}
{Wang}, P. \& {Abel}, T. 2009, \apj, 696, 96

\bibitem[{{Watson} {et~al.}(2001){Watson}, {Wiebe}, \& {Crutcher}}]{watson01}
{Watson}, W.~D., {Wiebe}, D.~S., \& {Crutcher}, R.~M. 2001, \apj, 549, 377

\bibitem[{{We{\.z}gowiec} {et~al.}(2012){We{\.z}gowiec}, {Urbanik}, {Beck},
  {Chy{\.z}y}, \& {Soida}}]{wez12}
{We{\.z}gowiec}, M., {Urbanik}, M., {Beck}, R., {Chy{\.z}y}, K.~T., \& {Soida},
  M. 2012, \aap, 545, A69

\bibitem[{{We{\.z}gowiec} {et~al.}(2007){We{\.z}gowiec}, {Urbanik}, {Vollmer},
  {Beck}, {Chy{\.z}y}, {Soida}, \& {Balkowski}}]{wez07}
{We{\.z}gowiec}, M., {Urbanik}, M., {Vollmer}, B., {et~al.} 2007, \aap, 471, 93

\bibitem[{{Wiener} {et~al.}(2013){Wiener}, {Zweibel}, \& {Oh}}]{wiener13}
{Wiener}, J., {Zweibel}, E.~G., \& {Oh}, S.~P. 2013, \apj, 767, 87

\bibitem[{{Yoast-Hull} {et~al.}(2013){Yoast-Hull}, {Everett}, {Gallagher}, \&
  {Zweibel}}]{yoast13}
{Yoast-Hull}, T.~M., {Everett}, J.~E., {Gallagher}, III, J.~S., \& {Zweibel},
  E.~G. 2013, \apj, 768, 53

\bibitem[{{Yoast-Hull} {et~al.}(2014){Yoast-Hull}, {Gallagher}, {Zweibel}, \&
  {Everett}}]{yoast14}
{Yoast-Hull}, T.~M., {Gallagher}, III, J.~S., {Zweibel}, E.~G., \& {Everett},
  J.~E. 2014, \apj, 780, 137

\end{thebibliography}

%
%

\end{document}